\documentclass[final,3p,times]{elsarticle}

\usepackage{amsmath,amssymb}
\usepackage{bm}
\usepackage{multirow}
\usepackage{subfigure}
\usepackage{lineno}
\biboptions{sort&compress}
\usepackage{tikz}
\usetikzlibrary{positioning}
\tikzset{>=stealth}
\usepackage{makecell}
\usepackage{url}

\journal{Journal of Sound and Vibration}

\begin{document}

\begin{frontmatter}

\title{Application of a Spectral Method to Simulate Quasi-Three-Dimensional Underwater Acoustic Fields}

		\author[1]{Houwang Tu}
\ead{tuhouwang@nudt.edu.cn}
		\author[2]{Yongxian Wang\corref{cor1}}
\cortext[cor1]{Corresponding author: \url{yxwang@nudt.edu.cn}}
		\author[2]{Wei Liu}
\ead{liuwei@nudt.edu.cn}
		\author[3]{Chunmei Yang}
\ead{ycm@fio.org.cn}
		\author[4]{Jixing Qin}
\ead{qjx@mail.ioa.ac.cn}
		\author[2]{Shuqing Ma}
		\author[1]{Xiaodong Wang}

		\address[1]{College of Computer, National University of Defense Technology, Changsha, 410073, China}
		\address[2]{College of Meteorology and Oceanography, National University of Defense Technology, Changsha, 410073, China}
		\address[3]{Key Laboratory of Marine Science and Numerical Modeling, First Institute of Oceanography, Ministry of Natural Resources, Qingdao, 266061, China}
		\address[4]{State Key Laboratory of Acoustics, Institute of Acoustics, Chinese Academy of Sciences, Beijing, 100190, China}

\begin{abstract}
The calculation of a three-dimensional underwater acoustic field has always been a key problem in computational ocean acoustics. Traditionally, this solution is usually obtained by directly solving the acoustic Helmholtz equation using a finite difference or finite element algorithm. Solving the three-dimensional Helmholtz equation directly is computationally expensive. For quasi-three-dimensional problems, the Helmholtz equation can be processed by the integral transformation approach, which can greatly reduce the computational cost. In this paper, a numerical algorithm for a quasi-three-dimensional sound field that combines an integral transformation technique, stepwise coupled modes and a spectral method is designed. The quasi-three-dimensional problem is transformed into a two-dimensional problem using an integral transformation strategy. A stepwise approximation is then used to discretize the range dependence of the two-dimensional problem; this approximation is essentially a physical discretization that further reduces the range-dependent two-dimensional problem to a one-dimensional problem. Finally, the Chebyshev--Tau spectral method is employed to accurately solve the one-dimensional problem. We provide the corresponding numerical program SPEC3D for the proposed algorithm and describe several representative numerical examples. In the numerical experiments, the consistency between SPEC3D and the analytical solution/high-precision finite difference program COACH verifies the reliability and capability of the proposed algorithm. A comparison of running times illustrates that the algorithm proposed in this paper is significantly faster than the full three-dimensional algorithm in terms of computational speed.
\end{abstract}



\end{frontmatter}

\newpage
\section{Introduction}
Accurately and efficiently identifying and locating underwater targets have always been the core pursuits of computational ocean acoustics \cite{Etter2018}. The most critical step in achieving these objectives is accurately simulating the acoustic field \cite{Jensen2011}. The acquisition of an accurate three-dimensional numerical sound field is computationally expensive, as it involves a numerical solution of the three-dimensional Helmholtz equation \cite{Lee1995,Linyt2019,Ivansson2021,Liuw2021}. A quasi-three-dimensional sound field (one of the two horizontal directions is range-independent) is also useful, as it is usually associated with continental slopes, continental shelves, or transitional seas.

Simplified solutions exist for quasi-three-dimensional sound fields that do not incur the high computational cost of full three-dimensional problems. Fawcett proposed using the integral transformation technique to transform the Helmholtz equation satisfied by the quasi-three-dimensional marine environment into a governing equation similar to the two-dimensional line source problem, which greatly reduces the amount of computation \cite{Fawcett1990}. The two-dimensional line source problems can be solved accurately using the coupled modes. When using the step approximation proposed by Evans to address the range dependence, the two-dimensional line source problems become one-dimensional modal equations \cite{Evans1983,Evans1986,Couple}. The modal equations need to be solved by discretization. The accuracy of the solution of the one-dimensional modal equations determines the accuracy of the quasi-three-dimensional sound field.

There are many numerical discretized methods in scientific and engineering computing, and the spectral method is a highly precise numerical technique similar to the finite difference method (FDM) and the finite element method (FEM) \cite{Boyd2001}. Compared with the FDM and FEM, the spectral method is a global approach, meaning its basis functions are defined in the entire solution domain, whereas the basis functions of the FDM and FEM are defined on the grid points and in the elements, respectively \cite{Jshen2011}. Strictly speaking, the spectral method is a weighted residual method, but the basis functions selected by the spectral method---unlike those of the general weighted residual method---are sets of orthogonal polynomials, and the orthogonality of these basis functions ensures that the spectral method converges quickly \cite{Canuto2006}. The research conducted by Orszag and Gottilieb illustrated that for smooth problems, the numerical error of the spectral method decreases exponentially with increasing truncation order \cite{Orszag1972,Gottlieb1977}. Consequently, benefiting from its high precision, the spectral method has been widely used in many numerical simulations of scientific and engineering problems \cite{Canuto1988,Guoby1998,Jekeli2011}.

Spectral methods have also been gradually applied to computational ocean acoustics in recent years. In 1993, Dzieciuch wrote a program to simulate the ideal fluid waveguide of the Munk sound speed profile using the Chebyshev--Tau spectral method \cite{Dzieciuch1993,aw}. However, this program could calculate only a single-layer water body in which the seawater density does not vary with depth and the seabed is a pressure-release boundary, and thus, it was not capable of solving higher-complexity acoustic propagation problems; nevertheless, this program represented the first use of the spectral method to solve the underwater acoustic propagation problem, and it demonstrated remarkable accuracy. More recently, in 2016, Evans proposed the Legendre--Galerkin spectral method to solve for the normal modes of underwater acoustic propagation in two-layer media and developed the program rimLG \cite{rimLG}. Because rimLG also requires the seabed to be a pressure-release boundary and applies Gaussian integration in each parameter layer, its computational speed slows when the acoustic profile data are dense, but it exhibits extremely high accuracy. Moreover, rimLG can simulate underwater acoustic waveguides with attenuation in the sediment, and the diversity of sound waveguides that can be synthesized is much greater than the capability of Dzieciuch's program. Subsequently, in 2019, Sabatini proposed a multidomain spectral collocation method that can accurately calculate the normal modes in the open ocean \cite{Sabatini2019}. This method can accurately compute the half-space boundary of the seafloor, although the size of the matrix eigenvalue problem is doubled. Since 2020, Tu et al. have performed a series of studies to design highly precise numerical algorithms based on the spectral method for solving underwater acoustic propagation problems. In March 2020, Tu et al. proposed an algorithm based on the Chebyshev--Tau spectral method to solve the underwater acoustic propagation problem in two-layer media \cite{Tuhw2020a,Tuhw2021a} and developed a corresponding simulation software named NM-CT \cite{NM-CT}. Numerical simulations have shown that the accuracy of NM-CT is comparable to that of rimLG and that the numerical errors are all on the order of $10^{-14}$, which is much smaller than the order of $10^{-6}$ of the classic FDM; in addition, the computational speed is much faster than that of rimLG. Then, in June 2021, Tu et al. proposed an algorithm for simulating underwater acoustic propagation in multilayer media based on the Legendre collocation spectral method \cite{Tuhw2021c,MultiLC}. The algorithm adopts an absorbing layer to simulate the acoustic half-space; in fact, there is a compromise between computational cost and computational accuracy. The application of an absorption layer does not double the matrix size but reduces the precision of eigenpairs. To date, the spectral method has shown remarkable success and potential in designing highly precise numerical algorithms for simulating underwater acoustic propagation. However, there has been no research on the application of spectral methods to simulate three-dimensional acoustic propagation. Therefore, the development of a high-precision quasi-three-dimensional underwater acoustic model based on spectral methods is a research topic worthy of gradual exploration.

In this paper, we used the spectral method combined with the integral transformation technique and stepwise coupled modes to develop a high-precision quasi-three-dimensional numerical algorithm and develop the corresponding numerical program. We have also carefully designed several representative numerical experiments to verify the accuracy and efficiency of the algorithm and program by comparing them with analytical solutions or high-precision full-three-dimensional models.

\section{Mathematical Modeling}
\subsection{Quasi-three-dimensional waveguide}
\label{section2.1}
\begin{figure}[htbp]
	\centering
\begin{tikzpicture}[node distance=2cm,scale = 0.5]
		\fill[orange,opacity=0.6](14*1.5,-4.5)--(12*1.5,-4.5)--(6.5*1.5-1,-7.5)--(4,-7.5)--(2,-5)--(4.5*1.5,-5)--(10*1.5,-2)--(12*1.5,-2)--cycle;
		\fill[cyan,opacity=0.6] (2,2)--(12*1.5,2)--(14*1.5,-0.5)--(14*1.5,-4.5)--(12*1.5,-4.5)--(6.5*1.5-1,-7.5)--(4,-7.5)--(2,-5)--cycle;		\fill[gray,opacity=0.2] (14*1.5,-10.5)--(12*1.5,-8)--(2,-8)--(4,-10.5)--cycle;
		\fill[orange,opacity=0.6] (2,-5)--(4,-7.5)--(6.5*1.5-1,-7.5)--(12*1.5,-4.5)--(14*1.5,-4.5)--(14*1.5,-10.5)--(4,-10.5)--(2,-8)--cycle;
		\fill[gray,opacity=0.3] (14*1.5,-10.5)--(12*1.5,-8)--(2,-8)--(4,-10.5)--cycle;
		\fill[brown] (2-0.04,-8)--(4,-10.5)--(14*1.5+0.04,-10.5)--(14*1.5+0.04,-12)--(4,-12)--(2-0.04,-9.5)--cycle;
		
		\draw[thick, ->](3,0.75)--(14*1.5,0.75) node[right]{$x$};
		\draw[thick, ->](3,0.75)--(3,-12) node[below]{$z$};       
		\draw[very thick](1.94,2)--(12.02*1.5,2);
		\draw[very thick](4,-0.5)--(14.02*1.5,-0.5);
		\draw[very thick](2,2)--(4,-0.5);
		\draw[thick, ->](2,2)--(5,-1.75) node[below]{$y$}; 
		\draw[very thick](12*1.5,2)--(14*1.5,-0.5);
		\draw[dashed,very thick](12*1.5,2.02)--(12*1.5,-9.5);
		\draw[very thick](2,2.02)--(2,-9.52);
		\draw[very thick](4,-0.5)--(4,-12);
		\draw[very thick](14*1.5,-0.5+0.04)--(14*1.5,-12.0);
	
		\draw[very thick](12*1.5,-2)--(14*1.5,-4.5);
		\draw[very thick](2,-5)--(4.5*1.5,-5);  
		\draw[very thick](2,-5)--(4,-7.5);
		\draw[very thick](4,-7.5)--(6.5*1.5-1,-7.5);
		\draw[very thick](4.5*1.5,-5)--(6.5*1.5-1,-7.5);
		\draw[very thick](6.49*1.5-1,-7.5)--(12*1.5,-4.5);
		\draw[very thick](4.5*1.5,-5)--(10.02*1.5,-2);
		\draw[very thick](12*1.5,-4.5)--(14*1.5,-4.5);
		\draw[very thick](10*1.5,-2)--(12*1.5+0.04,-2);
		\draw[very thick](10*1.5,-2)--(12*1.5,-4.5);
		
		\draw[very thick](2,-8)--(4,-10.5); 
		\draw[very thick](4-0.04,-10.5)--(14*1.5+0.04,-10.5);   
		\draw[dashed,very thick](14*1.5,-10.5)--(12*1.5,-8)--(2,-8); 
				
		\filldraw [red] (6*1.5,-2.7) circle [radius=4pt];
		\node at (2.7,0.6){$o$};
		\node at (9*1.5-1,-1.5){Water};
		\node at (9*1.5-1,-7.5){Sediment};
		\node at (9*1.5-1,-11.4){Acoustic half-space};
	\end{tikzpicture}
\caption{Schematic diagram of the quasi-three-dimensional waveguides.}
	\label{Figure1}
\end{figure}
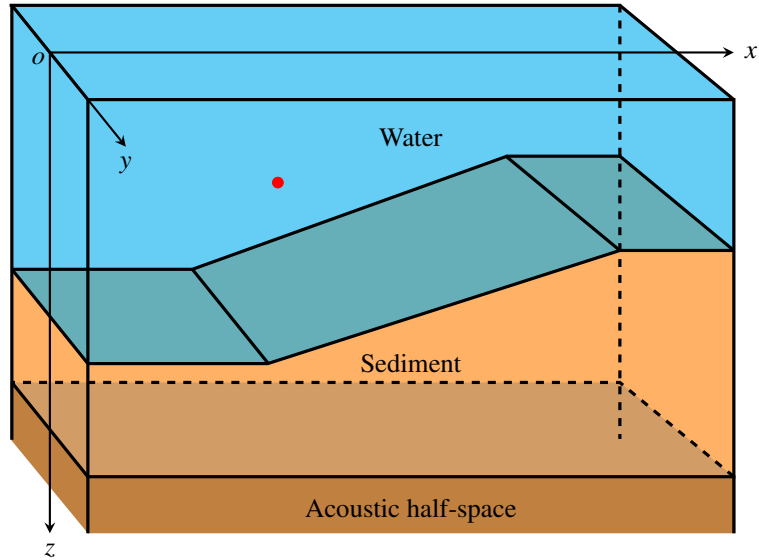

Quasi-three-dimensional waveguides can model common typical scenarios, including continental slopes, continental shelves, and transitional seas, and their structures are shown in Fig.~\ref{Figure1}. Terrain and acoustic parameters are range-dependent variables in the $xoz$-plane; in other words, parameters such as the sound speed profile, density and attenuation, as well as terrain, vary across the plane. In contrast, in the $yoz$--plane, the acoustic parameters are $y$-independent. Thus, this scenario is not a real three-dimensional problem but rather a quasi-three-dimensional problem. In the quasi-three-dimensional waveguide here, the sound speed $c\equiv c(x,z)$, density $\rho\equiv\rho(x,z)$, and attenuation coefficient $\alpha\equiv\alpha(x,z)$. To accurately simulate this problem physically, we can always place the sound source in the plane of $y=0$; thus, the governing equation of the problem can be written as \cite{Jensen2011}:
\begin{equation}
	\label{eq.1}
	\rho(x,z) \frac{\partial}{\partial x}\left[\frac{1}{\rho(x,z)} \frac{\partial p}{\partial x}\right]+\frac{\partial^{2} p}{\partial y^{2}}+\rho(x,z) \frac{\partial}{\partial z}\left[\frac{1}{\rho(x,z)} \frac{\partial p}{\partial z}\right]+k^{2}(x,z) p=-\delta\left(x-x_\mathrm{s}\right) \delta(y) \delta\left(z-z_\mathrm{s}\right)		
\end{equation}
where the harmonic point source is located at a horizontal range $x=x_\mathrm{s}$ and a depth $z=z_\mathrm{s}$, $k(x,z)=2\pi f/c(x,z)$ is the wavenumber, $f$ is the frequency of the sound source ($f$ arises from a temporal Fourier transform of the wave equation \cite{Jensen2011}), $p\equiv p(x,y,z)$, and $\delta(\cdot)$ is the Dirac function. Since the marine environment is $y$-independent, Eq.~\eqref{eq.1} can be transformed using the following Fourier transformations \cite{Fawcett1990}:

\begin{subequations}
	\begin{align}
	\label{eq.2a}
	\tilde{p}\left(x, k_{y}, z\right) &=\int_{-\infty}^{\infty} p(x, y, z) \mathrm{e}^{-\mathrm{i} y k_{y}} \mathrm{~d} y \\
	\label{eq.2b}
	p(x, y, z) &=\frac{1}{2 \pi} \int_{-\infty}^{\infty} \tilde{p}\left(x, k_{y}, z\right) \mathrm{e}^{\mathrm{i} y k_{y}} \mathrm{d} k_{y}
	\end{align}
\end{subequations}
This is accomplished by applying the following operators to both sides of Eq.~\eqref{eq.1}:
\[
	\int_{-\infty}^{\infty}(\cdot) \mathrm{e}^{-\mathrm{i} k_{y} y} \mathrm{d} y
\]
From Eq.~\eqref{eq.2a}, Eq.~\eqref{eq.1} can be transformed into the following form:
\begin{equation}
	\label{eq.3}
		\rho(x,z) \frac{\partial}{\partial x}\left[\frac{1}{\rho(x,z)} \frac{\partial \tilde{p}}{\partial x}\right] +\rho(x,z) \frac{\partial}{\partial z}\left[\frac{1}{\rho(x,z)} \frac{\partial \tilde{p}}{\partial z}\right]+ \left(k^{2}-k_{y}^{2}\right) \tilde{p}=-\delta\left(x-x_\mathrm{s}\right) \delta\left(z-z_\mathrm{s}\right)
\end{equation}
For each particular $k_y$, Eq.~\eqref{eq.3} has almost the same form as the governing equation for an infinitely long line source sound field in a two-dimensional plane. Eq.~\eqref{eq.3} is solved by subdividing the environment into segments along the $x$-axis; within each segment, the density and speed of sound are assumed to vary only with depth. Details are described in Sec.~\ref{section2.2}. After calculating the sound pressure $\tilde{p}(x,k_y,z)$ in the $k_y$--domain, the sound pressure $p(x,y,z)$ in the three-dimensional domain can be easily calculated through the inverse Fourier transform of Eq.~\eqref{eq.2b}.

\subsection{Stepwise approximation of a two-dimensional marine environment}
\label{section2.2}
The previous section demonstrates that Eq.~\eqref{eq.3} is highly similar in form to the governing equation of the sound field of an infinitely long line source in a two-dimensional plane.
\begin{figure}[htbp]
\centering
\begin{tikzpicture}[node distance=2cm,scale = 0.8]
	\node at (1.8,0){$0$};
	\fill[brown] (14,-8) rectangle (2,-6.7);
	\fill[cyan,opacity=0.6] (14,0)--(14,-1.7)--(12,-1.7)--(4,-5.7)--(2,-5.7)--(2,0)--cycle;
	\fill[orange,opacity=0.6] (14,-1.7)--(12,-1.7)--(4,-5.7)--(2,-5.7)--(2,-6.7)--(14,-6.7)--cycle;
	\draw[very thick, ->](2,0)--(14.5,0) node[right]{$x$};
	\draw[very thick, ->](2.02,0)--(2.02,-8.5) node[below]{$z$};
	\draw[very thick](2,-5.7)--(4,-5.7);
	\draw[dashed, very thick](4,-5.7)--(5,-5.7);
	\draw[dashed, very thick](5,-5.7)--(5,-4.7);
	\draw[dashed, very thick](5,-4.7)--(6.8,-4.7);
	\draw[dashed, very thick](6.8,-4.7)--(6.8,-3.7);
	\draw[dashed, very thick](6.8,-3.7)--(9.2,-3.7);
	\draw[dashed, very thick](8,0)--(8,-3.7);		
	\draw[dashed, very thick](9.2,-3.7)--(9.2,-2.7);  
	\draw[dashed, very thick](9.2,-2.7)--(11,-2.7);
	\draw[dashed, very thick](11,-2.7)--(11,-1.7); 
	\draw[dashed, very thick](11,-1.7)--(12,-1.7);		    		
	\draw[very thick](4,-5.7)--(12,-1.7);
	\draw[very thick](12,-1.7)--(14,-1.7);
	\draw[dashed, very thick](5,0)--(5,-4.7);
	\draw[dashed, very thick](6.8,0)--(6.8,-3.7);
	\draw[dashed, very thick](9.2,0)--(9.2,-2.7); 
	\draw[dashed, very thick](11,0)--(11,-1.7); 
	\draw[dashed, very thick](2.02,-6.7)--(14,-6.7);
	\filldraw [red] (8,-2) circle [radius=2.5pt];
	\draw[very thick](2,-2)--(2.2,-2);
	\node at (1.75,-2){$z_\mathrm{s}$};
	\node at (7.4,-0.5){$j_\mathrm{s}$};
	\node at (8.6,-0.5){$j_\mathrm{s}+1$};
	\draw[very thick, ->](7,-1.5)--(7.8,-1.5);
	\node at (7.4,-1.9){$a^{j_\mathrm{s}}$};
	\draw[very thick, ->](7.8,-2.5)--(7,-2.5);
	\node at (7.4,-2.9){$b^{j_\mathrm{s}}$};
	\draw[very thick, ->](8.2,-1.5)--(9,-1.5);
	\node at (8.6,-1.9){$a^{j_\mathrm{s}+1}$};
	\draw[very thick, ->](9,-2.5)--(8.2,-2.5);
	\node at (8.6,-2.9){$b^{j_\mathrm{s}+1}$};
	\draw[very thick, ->](3.8,-1)--(3,-1);
	\node at (3.4,-1.4){$b^{1}$};
	\draw[very thick, ->](12,-1)--(12.8,-1);
	\node at (12.4,-1.4){$a^{J}$};							
	\node at (8,0.2){$x_\mathrm{s}$};
	\node at (7,0.2){$x_{j_\mathrm{s}-1}$};
	\node at (9,0.2){$x_{j_\mathrm{s}+1}$};
	\node at (4.5,-4.7){$h_{j-1}$};			
	\node at (6.5,-3.7){$h_{j}$};
	\node at (11.5,-2.7){$h_{j+1}$};
	\node at (11,0.2){$x_{j+1}$};
	\node at (6,-1.0){$c_w^{j-1}$};
	\node at (6,-1.7){$\rho_w^{j-1}$}; 
	\node at (6,-2.4){$\alpha_w^{j-1}$};    		 	
	\node at (10.2,-0.5){$c_w^{j+1}$};
	\node at (10.2,-1.2){$\rho_w^{j+1}$};
	\node at (10.2,-1.9){$\alpha_w^{j+1}$};
	\node at (6.1,-5.5){$c_b^{j-1},\rho_b^{j-1}$};
	\node at (6,-6.3){$\alpha_b^{j-1}$};    		
	\node at (8,-4.5){$c_b^j,\rho_b^j,\alpha_b^j$}; \node at (10.3,-3.5){$c_b^{j+1},\rho_b^{j+1}$};
	\node at (10.2,-4.3){$\alpha_b^{j+1}$};
	\node at (8,-7.4){$c_\infty,\quad \rho_\infty,\quad \alpha_\infty$};    		
\end{tikzpicture}
\caption{Stepwise coupled modes of a line source \cite{Tuhw2022b}.}
\label{Figure2}
\end{figure}
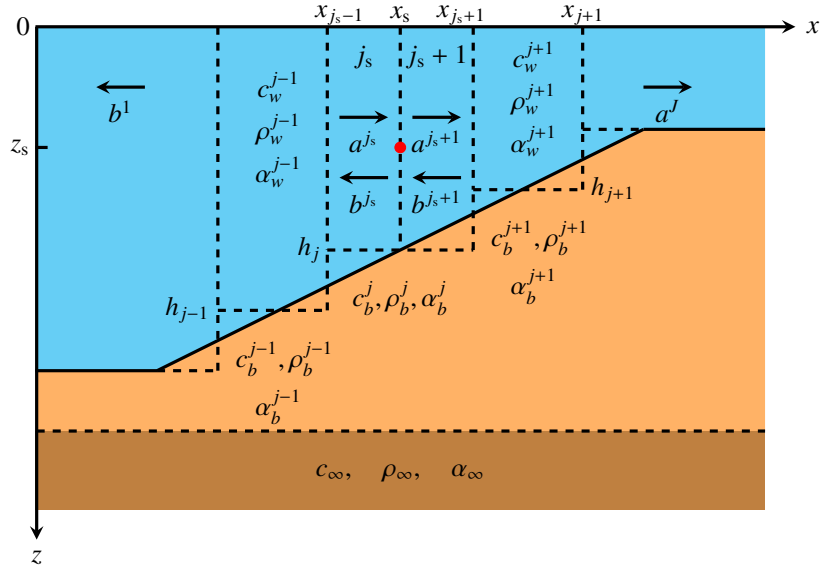
As shown in Fig.~\ref{Figure2}, for an $x$-dependent marine environment, the classic solution is to divide the environment into a sufficiently large number $J$ of narrow segments along the $x$-direction \cite{Evans1983}. In the $j$-th flat segment, the sound field is approximated as \cite{Jensen2011,Tuhw2022b}:

\begin{subequations}
	\label{eq.4}
	\begin{gather}
		\label{eq.4a}
		\tilde{p}^{j}(x, z)\approx \sum_{m=1}^{M}\left[a_{m}^{j} E_{m}^{j}(x)+b_{m}^{j} F_{m}^{j}(x)\right] \Psi_{m}^{j}(z) \\
		E_{m}^{j}(x)=\exp \left[\mathrm{i} \mu_ m^{j}\left(x-x_{j-1}\right)\right]\\
		F_{m}^{j}(x)=\exp \left[-\mathrm{i} \mu_ m^{j}\left(x-x_{j}\right)\right]
	\end{gather}
\end{subequations}
where $j=1,2,\cdots,J$; $M$ is the total number of normal modes needed to synthesize the sound field; $\{a_{m}^{j}\}_{m=1}^M$ and $\{b_{m}^{j}\}_{m=1}^M$ indicate the amplitudes of the forward and backward propagation modes, respectively, which are also called the coupling coefficients of the $j$-th segment \cite{Tuhw2022b}; $E_{m}^{j}(x)$ and $F_{m}^{j}(x)$ are the normalized range functions for special cases when $j=1$, $x_{j-1}=x_1$; and $\{\mu_m^j,\Psi_{m}^{j}(z)\}$ is the $m$-th eigensolution of the $j$-th segment and satisfies the following modal equation:
\begin{equation}
	\label{eq.5}
	\begin{gathered}
		\rho(z) \frac{\mathrm{d}}{\mathrm{d} z}\left[\frac{1}{\rho(z)} \frac{\mathrm{d} \Psi(z)}{\mathrm{d} z}\right]+\left[k^{2}(z)-k_y^2-\mu^{2}\right]\Psi(z)=0\\
		k(z)=2\pi f(1+\mathrm{i}\eta\alpha)/c(z),\quad \eta=(40\pi \log_{10}{\mathrm{e}})^{-1}
	\end{gathered}
\end{equation}
We redefine $k(z)$ in Eq.~\eqref{eq.1} to account for the effect of attenuation on acoustic propagation. Here, $\mu^2$, the constant obtained by the separation of variables, physically represents the horizontal wavenumber, and $\alpha$ is the attenuation coefficient in units of dB$/\lambda$, where $\lambda$ is the wavelength.

Ignoring the contribution of the continuous spectrum and adding appropriate boundary conditions, Eq.~\eqref{eq.5} has a set of eigensolutions $\{(\mu_{m},\Psi_m)\}_{m=1}^\infty$, where $\Psi_m$ is also called the eigenmode. The eigenmodes of Eq.~\eqref{eq.5} should be normalized as:
\begin{equation}
	\label{eq.6}
	\int_{0}^{H} \frac{\Psi^2_m(z)}{\rho(z)}\mathrm {d} z=1\quad m = 1, 2, \dots
\end{equation}
or:
\begin{equation}
	\label{eq.7}
	\begin{gathered}
		\int_{0}^{\infty} \frac{\Psi_{m}^{2}(z)}{\rho(z)} \mathrm{d} z = \int_{0}^{H} \frac{\Psi_{m}^{2}(z)}{\rho(z)} \mathrm{d} z +\frac{\Psi_{m}^{2}(H)}{2 \rho_\infty\gamma_{\infty}} =1,\quad m=1,2,\dots\\
		\gamma_{\infty}=\sqrt{\mu^{2}-k_\infty^2},\quad k_\infty=2\pi f(1+\mathrm{i}\eta\alpha_\infty)/c_{\infty}
	\end{gathered}
\end{equation}
where $H$ is the ocean depth. Eq.~\eqref{eq.6} applies to the case where the upper and lower boundaries are ideal boundaries, while Eq.~\eqref{eq.7} applies to the case where the upper boundary is an ideal boundary but the lower boundary is an acoustic half-space.

Segmenting the environment means that we solve the sound field in $J$ segments. Subsequently, the complete sound field can be obtained by simply imposing the continuity of the pressure perturbation and the continuity of the horizontal velocity perturbation. The method of coupling segments explicitly imposes two continuity conditions on the sides of the segments. In other words, the coefficients $a_m^j$ and $b_m^j$ are computed by enforcing the continuity of pressure and the continuity of horizontal velocity at the interface between regular segments. The first segment condition requires that the acoustic pressure be continuous at the $j$-th side, and the second segment condition requires that the horizontal velocity be continuous at the $j$-th side:

\begin{subequations}
	\label{eq.8}
	\begin{align}
		\tilde{p}^{j+1}\left(x_{j}, z\right)&=\tilde{p}^{j}\left(x_{j}, z\right)\\
		\frac{1}{\rho_{j+1}(z)} \frac{\partial \tilde{p}^{j+1}\left(x_{j}, z\right)}{\partial x}&=\frac{1}{\rho_{j}(z)} \frac{\partial \tilde{p}^{j}\left(x_{j}, z\right)}{\partial x}
	\end{align}
\end{subequations}
For the first segment condition, we have:
\begin{equation}
	\label{eq.9}
	\sum_{m=1}^{M}\left[a_{m}^{j+1}E_{m}^{j+1}(x_j)+b_{m}^{j+1}F_{m}^{j+1}(x_j) \right] \Psi_{m}^{j+1}(z)=\sum_{m=1}^{M}\left[a_{m}^{j} E_{m}^{j}\left(x_{j}\right)+b_{m}^{j}F_{m}^{j}\left(x_{j}\right)\right] \Psi_{m}^{j}(z)
\end{equation}
Next, we apply the following operator to the above equation:
\[
	\int (\cdot) \frac{\Psi_{\ell}^{j+1}(z)}{\rho_{j+1}(z)} \mathrm{d} z
\]
Furthermore, we apply the orthogonal normalization formula in Eq.~\eqref{eq.6} or \eqref{eq.7} to the eigenmodes of the $(j+1)$-th segment:

\begin{subequations}
	\label{eq.10}
	\begin{gather}
		\label{eq.10a}
		a_{\ell}^{j+1}+b_{\ell}^{j+1}F_{\ell}^{j+1}\left(x_{j}\right) =\sum_{m=1}^{M}\left[a_{m}^{j} E_{m}^{j}\left(x_{j}\right)+b_{m}^{j}\right] \tilde{c}_{\ell m}\\
		\tilde{c}_{\ell m}=\int_0^H \frac{\Psi_{\ell}^{j+1}(z) \Psi_{m}^{j}(z)}{\rho_{j+1}(z)} \mathrm{d} z,\quad \ell=1,2, \ldots, M\\
		\label{eq.10c}
		\tilde{c}_{\ell m}=\int_0^\infty \frac{\Psi_{\ell}^{j+1}(z) \Psi_{m}^{j}(z)}{\rho_{j+1}(z)} \mathrm{d} z=\int_0^H \frac{\Psi_{\ell}^{j+1}(z) \Psi_{m}^{j}(z)}{\rho_{j+1}(z)} \mathrm{d} z+\frac{\Psi_{\ell}^{j+1}(H) \Psi_{m}^{j}(H)}{\rho_\infty \left(\gamma_\ell^{j+1}+\gamma_m^j \right)}
	\end{gather}
\end{subequations}
Next, we find the partial derivatives on both sides of Eq.~\eqref{eq.4a}:
\begin{equation}
	\label{eq.11}
	\frac{1}{\rho_{j}(z)} \frac{\partial \tilde{p}^{j}(x, z)}{\partial x} \simeq \frac{1}{\rho_{j}(z)} \sum_{m=1}^{M} \mu_m^j\left[a_{m}^{j} E_{m}^{j}(x)-b_{m}^{j} F_{m}^{j}(x)\right] \Psi_{m}^{j}(z)
\end{equation}
For the second segment condition, we have:
\begin{equation}
	\label{eq.12}
	\frac{1}{\rho_{j+1}(z)}\sum_{m=1}^{M}\mu_m^{j+1}\left[a_{m}^{j+1}-b_{m}^{j+1}F_{m}^{j+1}\left(x_{j}\right)\right]\Psi_{m}^{j+1}(z)=\frac{1}{\rho_{j}(z)}\sum_{m=1}^{M}\mu_m^j\left[a_{m}^{j} E_{m}^{j}\left(x_{j}\right)-b_{m}^{j}\right] \Psi_{m}^{j}(z)
\end{equation}
Similarly, we apply the following operator to the above equation:
\[
	\int(\cdot) \Psi_{\ell}^{j+1}(z) \mathrm{d} z
\]
Next, we apply the orthogonal normalization formula in Eq.~\eqref{eq.6} or \eqref{eq.7} to the eigenmodes of the $(j+1)$-th segment, yielding the following:

\begin{subequations}
	\label{eq.13}
	\begin{gather}
		\label{eq.13a}
		a_{\ell}^{j+1}-b_{\ell}^{j+1}F_{\ell}^{j+1} =\sum_{m=1}^{M}\left[a_{m}^{j} E_{m}^{j}\left(x_{j}\right)-b_{m}^{j}\right] \hat{c}_{\ell m}\\
		\hat{c}_{\ell m}=\frac{\mu_m^{j}}{\mu_\ell^{j+1}} \int_0^H \frac{\Psi_{\ell}^{j+1}(z) \Psi_{m}^{j}(z)}{\rho_{j}(z)} \mathrm{d} z, \quad \ell=1,2, \ldots, M\\
		\label{eq.13c}
		\hat{c}_{\ell m}=\frac{\mu_m^{j}}{\mu_\ell^{j+1}}\int_0^\infty \frac{\Psi_{\ell}^{j+1}(z) \Psi_{m}^{j}(z)}{\rho_{j+1}(z)} \mathrm{d} z=\frac{\mu_m^{j}}{\mu_\ell^{j+1}}\left[\int_0^H \frac{\Psi_{\ell}^{j+1}(z) \Psi_{m}^{j}(z)}{\rho_{j+1}(z)} \mathrm{d} z+\frac{\Psi_{\ell}^{j+1}(H) \Psi_{m}^{j}(H)}{\rho_\infty \left(\gamma_\ell^{j+1}+\gamma_m^j \right)}\right]
	\end{gather}
\end{subequations}
The coupling integrals in Eqs.~\eqref{eq.10} and \eqref{eq.13} can be obtained using general numerical integration. In the numerical model developed later in this paper, the coupling integrals are evaluated by the trapezoidal rule.

Then, the above Eqs.~\eqref{eq.10a} and \eqref{eq.13a} can be naturally written in the following matrix-vector form:

\begin{subequations}
	\label{eq.14}
	\begin{gather}
		\mathbf{a}^{j+1}-\mathbf{F}^{j+1}\mathbf{b}^{j+1}=\widehat{\mathbf{C}}^{j}\left(\mathbf{E}^{j} \mathbf{a}^{j}-\mathbf{b}^{j}\right)\\
		\mathbf{a}^{j+1}+\mathbf{F}^{j+1}\mathbf{b}^{j+1}=\widetilde{\mathbf{C}}^{j}\left(\mathbf{E}^{j} \mathbf{a}^{j}+\mathbf{b}^{j}\right) 
	\end{gather}	 
\end{subequations}
and can be combined into the following form:

\begin{subequations}
	\label{eq.15}
	\begin{align}
			\left[\begin{array}{l}
				\mathbf{a}^{j+1} \\
				\mathbf{b}^{j+1}
			\end{array}\right]&=\left[\begin{array}{ll}
				\mathbf{R}_{1}^{j} & \mathbf{R}_{2}^{j} \\
				\mathbf{R}_{3}^{j} & \mathbf{R}_{4}^{j}
			\end{array}\right]\left[\begin{array}{c}
				\mathbf{a}^{j} \\
				\mathbf{b}^{j}
			\end{array}\right]\\
			\mathbf{R}_{1}^{j} &=\frac{1}{2}\left(\widetilde{\mathbf{C}}^{j}+\widehat{\mathbf{C}}^{j}\right) \mathbf{E}^{j} \\
			\mathbf{R}_{2}^{j} &=\frac{1}{2}\left(\widetilde{\mathbf{C}}^{j}-\widehat{\mathbf{C}}^{j}\right) \\
			\mathbf{R}_{3}^{j} &=\frac{1}{2}\left(\mathbf{F}^{j+1}\right)^{-1}\left(\widetilde{\mathbf{C}}^{j}-\widehat{\mathbf{C}}^{j}\right) \mathbf{E}^{j} \\
			\mathbf{R}_{4}^{j} &=\frac{1}{2}\left(\mathbf{F}^{j+1}\right)^{-1}\left(\widetilde{\mathbf{C}}^{j}+\widehat{\mathbf{C}}^{j}\right)
	\end{align}
\end{subequations}

Note that since the sound source is located at $(x_\mathrm{s}, z_\mathrm{s})$, a virtual boundary $j_\mathrm{s}$ needs to be introduced at the horizontal distance $x_\mathrm{s}$. The introduction of this virtual boundary at the sound source increases the number of segments to $J+1$. All boundaries except $j_\mathrm{s}$ still satisfy the boundary conditions of Eq.~\eqref{eq.8}. However, due to the existence of the sound source, the boundary conditions at the virtual boundary $j_\mathrm{s}$ need to be modified. We can obtain the relationship between the coupling coefficients of the $j_\mathrm{s}$-th and $(j_\mathrm{s}+1)$-th segments \cite{Tuhw2022b}:
\begin{equation}
	\label{eq:16}
	\begin{gathered}
		\left[\begin{array}{l}
			\mathbf{a}^{j_\mathrm{s}+1} \\
			\mathbf{b}^{j_\mathrm{s}+1}
		\end{array}\right]=\left[\begin{array}{ll}
			\mathbf{E}^{j_\mathrm{s}} & \mathbf{0} \\
			\mathbf{0} & (\mathbf{F}^{j_\mathrm{s}+1})^{-1}
		\end{array}\right]\left[\begin{array}{c}
			\mathbf{a}^{j_\mathrm{s}} \\
			\mathbf{b}^{j_\mathrm{s}}
		\end{array}\right]+\left[\begin{array}{c}
			-\frac{1}{2}\mathbf{s} \\
			\frac{1}{2}\left(\mathbf{F}^{j_\mathrm{s}+1}\right)^{-1}\mathbf{s}
		\end{array}\right]\\
		s_{m}=-\frac{\mathrm{i}}{\rho\left(z_\mathrm{s}\right)} \frac{\Psi_{m}^{j_\mathrm{s}}\left(z_\mathrm{s}\right)}{\mu_{m}^{j_\mathrm{s}}}, \quad m=1,2, \cdots, M
	\end{gathered}
\end{equation}

Since the sound source is located at $x_\mathrm{s} \neq 0$, the segment conditions at $x=0$ and $x=\infty$ are the radiation conditions, i.e., $\mathbf{a}^1=0$ and $\mathbf{b}^J=0$. The global matrix used to determine the coupling coefficients can be assembled as follows:
\begin{equation}
		\label{eq.17}
		\left[\begin{array}{ccccccccccccc}
			\mathbf{R}_{2}^{1} & -\mathbf{I} & \mathbf{0} & & & & & \\
			\mathbf{R}_{4}^{1} & \mathbf{0} & -\mathbf{I} & & & & & \\
			& \mathbf{R}_{1}^{2} &\mathbf{R}_{2}^{2} &-\mathbf{I} &\mathbf{0} & & & & \\
			& \mathbf{R}_{3}^{2} &\mathbf{R}_{4}^{2} &\mathbf{0}  &-\mathbf{I} & & & & \\
			& & &\ddots & \ddots & \ddots & \ddots& & \\
			& & & & & \mathbf{E}^{j_\mathrm{s}} & \mathbf{0} & -\mathbf{I} & \mathbf{0} \\
			& & & & & \mathbf{0} & (\mathbf{F}^{j_\mathrm{s}+1})^{-1} & \mathbf{0} & -\mathbf{I}\\
			& & & & & & &  \ddots & \ddots & \ddots & \ddots&  \\
			& & & & & & & & &\mathbf{R}_{1}^{J-1} & \mathbf{R}_{2}^{J-1} & -\mathbf{I} \\
			& & & & & & & & &\mathbf{R}_{3}^{J-1} & \mathbf{R}_{4}^{J-1} & \mathbf{0}
			\end{array}\right]\left[\begin{array}{c}
			\mathbf{b}^{1} \\
			\mathbf{a}^{2} \\
			\mathbf{b}^{2} \\
			\mathbf{a}^{3} \\
			\vdots \\
			\mathbf{a}^{j_\mathrm{s}} \\
			\mathbf{b}^{j_\mathrm{s}} \\
			\vdots \\
			\mathbf{b}^{J-1}\\
			\mathbf{a}^{J}
			\end{array}\right]=\left[\begin{array}{c}
			\mathbf{0} \\
			\mathbf{0} \\
			\mathbf{0} \\
			\mathbf{0} \\
			\vdots \\
			\frac{1}{2}\mathbf{s} \\
			-\frac{1}{2}(\mathbf{F}^{j_\mathrm{s}+1})^{-1}\mathbf{s} \\
			\vdots \\
			\mathbf{0}\\
			\mathbf{0} \\
			\end{array}\right]
\end{equation}
Note that the global matrix is a sparse band matrix of order $(2J-2) \times M$; the upper bandwidth is $(2M-1)$, and the lower bandwidth is $(3M-1)$. The linear algebraic equations are then solved to obtain the coupling coefficients, and Eq.~\eqref{eq.4} is implemented to synthesize the sound pressure field of the entire waveguide.

Specifically, when $x_\mathrm{s}=0$, the segment condition at the acoustic source $x_\mathrm{s}=0$ is:
\begin{equation}
		\label{eq.18}
		\begin{gathered}
			\mathbf{a}^{1}=\mathbf{s}\\
			s_{m}=\frac{\mathrm{i}}{2 \rho\left(z_\mathrm{s}\right)} \Psi_{m}^{1}\left(z_\mathrm{s}\right) \frac{\mathrm{e}^{\mathrm{i} \mu_{m}^{1} x_{1}}}{\mu_{m}^{1}}, \quad m=1,2, \cdots, M  
	\end{gathered}
\end{equation}
Therefore, the global matrix becomes:
\begin{equation}
		\label{eq.19}
		\left[\begin{array}{ccccccccc}
			\mathbf{I} & \mathbf{0} & \mathbf{0} & & & & & \\
			\mathbf{R}_{1}^{1} & \mathbf{R}_{2}^{1} & -\mathbf{I} & \mathbf{0} & & & & \\
			\mathbf{R}_{3}^{1} & \mathbf{R}_{4}^{1} & \mathbf{0} & -\mathbf{I} & & & & \\
			& & \ddots & \ddots & \ddots & \ddots & & \\
			& & & &\mathbf{R}_{1}^{J-2} & \mathbf{R}_{2}^{J-2} & -\mathbf{I} & \mathbf{0} & \\
			& & & &\mathbf{R}_{3}^{J-2} & \mathbf{R}_{4}^{J-2} & \mathbf{0} & -\mathbf{I} & \\
			& & & & & &\mathbf{R}_{1}^{J-1} & \mathbf{R}_{2}^{J-1} & -\mathbf{I} \\
			& & & & & &\mathbf{R}_{3}^{J-1} & \mathbf{R}_{4}^{J-1} & \mathbf{0}
			\end{array}\right]\left[\begin{array}{c}
			\mathbf{a}^{1} \\
			\mathbf{b}^{1} \\
			\mathbf{a}^{2} \\
			\vdots \\
			\mathbf{b}^{J-2} \\
			\mathbf{a}^{J-1} \\
			\mathbf{b}^{J-1} \\
			\mathbf{a}^{J}
			\end{array}\right]=\left[\begin{array}{c}
			\mathbf{s} \\
			\mathbf{0} \\
			\mathbf{0} \\
			\vdots \\    
			\mathbf{0} \\
			\mathbf{0} \\
			\mathbf{0} \\
			\mathbf{0}
		\end{array}\right]
\end{equation}

\section{Numerical Discretization}
\subsection{Range-independent marine environment}
In the previous section, we fully derived the solution of a quasi-three-dimensional waveguide. Here, we solve for the eigensolutions in each segment, that is, the solutions $\{(\mu_{m}^j,\Psi_m^j)\}_{m=1}^M$ of Eq.~\eqref{eq.5}. For the sediment-covered marine environment considered in Figs.~\ref{Figure1} and \ref{Figure2}, within range-independent segments, $\rho(z)$, $c(z)$ and $\alpha(z)$ are discontinuous at the interface $z=h$; thus, the ocean is divided into a discontinuous water column and a bottom sediment. The marine environmental parameters are defined as follows:
\begin{equation}
		\label{eq.20}
			c(z)= \begin{cases}
				c_w(z),&0 \leq z \leq h\\
				c_b(z),&h \leq z \leq H\\
				c_\infty,&z\geq H
			\end{cases},\quad \rho(z)= \begin{cases}
				\rho_w(z),&0 \leq z \leq h\\
				\rho_b(z),&h \leq z \leq H\\
				\rho_\infty,&z\geq H
			\end{cases},\quad \alpha(z)= \begin{cases}
				\alpha_w(z),&0 \leq z \leq h\\
				\alpha_b(z),&h \leq z \leq H\\
				\alpha_\infty,&z\geq H
			\end{cases}
\end{equation}

To solve Eq.~\eqref{eq.5}, it is necessary to impose boundary conditions at the sea surface ($z=0$) and the seabed ($z=H$) and interface conditions at the interface ($z=h$). The sea surface is usually set as a pressure-release boundary:
\begin{equation}
	\label{eq.21}
	\Psi(z=0)=0
\end{equation}
In contrast, the seabed can be either a pressure-release boundary or a rigid seabed:

\begin{subequations}
	\label{eq.22}
	\begin{gather}
		\Psi(z=H)=0\\
		\Psi'(z=H)=0  
	\end{gather}
\end{subequations}
Furthermore, for an acoustic half-space, the lower boundary should generally satisfy the following \cite{Jensen2011}:
\begin{equation}
	\label{eq.23}
	\Psi(H)+\frac{\rho_\infty}{\rho_b(H)\gamma_\infty} \Psi'(H)=0
\end{equation}

The interface conditions are defined as follows:

\begin{subequations}
	\label{eq.24}
	\begin{gather}
			\Psi(h^{-})=\Psi(h^{+}) \\
			\frac{1}{\rho(z=h^{-})} \frac{\mathrm{d}\Psi(h^{-} )}{\mathrm {d}z}= \frac{1}{\rho(z=h^{+})}\frac{\mathrm{d}\Psi(h^{+} )}{\mathrm {d}z}
	\end{gather}
\end{subequations}
where the superscripts $h^{-}$ and $h^{+}$ denote above and below $h$, respectively.

\subsection{Chebyshev--Tau spectral method}
\label{section3.2}
Traditionally, two numerical methods are used to solve for the local normal modes: the FDM, which is implemented by the KRAKEN/KRAKENC programs \cite{Kraken2001}, and the Galerkin method, which is implemented by COUPLE \cite{Couple}. The spectral method is a new type of numerical method known for its high precision that has been introduced into computational ocean acoustics in recent years. Generally, the spectral method includes both the Galerkin method (with orthogonal polynomials as the basis functions) and the collocation method (with the extreme points of orthogonal polynomials as the nodes). Numerical experiments have shown that the former method is slightly more accurate than the latter \cite{Jshen2011}. This paper employs the Chebyshev--Tau spectral method. Different from the classic Galerkin-type spectral method, the Chebyshev--Tau spectral method uses the original Chebyshev polynomial as the basis function and does not require the basis function to satisfy the boundary conditions but instead transforms the boundary conditions into the spectral space to enforce the requirement \cite{Lanczos1938}.

Here, we employ the Chebyshev--Tau spectral method to solve for the local modes (Eq.~\eqref{eq.5}) in the $J$ range-independent segments illustrated in Fig.~\ref{Figure2}. Since the Chebyshev polynomial $\{T_i(t)\}$, that is, the basis function, is defined in $t\in[-1,1]$, the equation to be solved, Eq.~\eqref{eq.5}, must first be scaled to $t\in[-1,1]$ as:
\begin{equation}
	\label{eq.25}
	\frac{4}{\vert \Delta h\vert^2}\rho(t)\frac{\mathrm{d}}{\mathrm{d}t}\left(\frac{1}{\rho(t)}\frac{\mathrm{d}\Psi(t)}{\mathrm {d}t}\right) +\left[k^{2}(t)-k_y^2\right]\Psi(t) = \mu^2 \Psi(t)
\end{equation}
where $\Delta h$ denotes the thickness of the medium. The variable to be determined, $\Psi(t)$, is transformed into the spectral space spanned by the basis functions $\{T_i(t)\}_{i=0}^N$:
\begin{equation}
	\label{eq.26}
	\Psi(t) \approx \sum_{i=0}^{N}\hat{\Psi}_{i}T_i(t)
\end{equation}
where $\{\hat{\Psi}_i\}_{i=0}^N$ represents the spectral coefficients of $\Psi(t)$. This function approximation becomes increasingly accurate as $N$ increases. Due to the advantageous properties of the Chebyshev polynomial, the following is easy to prove \cite{Canuto2006}:

\begin{subequations}
		\label{eq.27}
		\begin{gather}
			\label{eq.27a}		
			\hat{\Psi}'_i \approx \frac{2}{c_i}
			\sum_{\substack{j=i+1,\\ 
					j+i=\text{odd}
			}}^{N} j \hat{\Psi}_j, \quad c_0=2,c_{i>1}=1
			\Longleftrightarrow \bm{\hat{\Psi}}' \approx \mathbf{D}_N \bm{\hat{\Psi}} \\
			\label{eq.27b}
			\widehat{(v\Psi)}_i \approx 
			\frac{1}{2} \sum_{m+n=i}^{N} \hat{\Psi}_m\hat{v}_n +
			\frac{1}{2} \sum_{\vert m-n\vert=i}^{N} \hat{\Psi}_m\hat{v}_n  \Longleftrightarrow  \widehat{\bm{(v\Psi)}} \approx \mathbf{C}_v \bm{\hat{\Psi}}
	\end{gather}
\end{subequations}
where $\mathbf{\hat{\Psi}}$ is a column vector composed of $\{\hat{\Psi}_i\}_{i=1}^N$. Eq.~\eqref{eq.27a} describes the relationship between the spectral coefficients of a function and the spectral coefficients of its derivative. Similarly, Eq.~\eqref{eq.27b} represents the relationship between the spectral coefficients of the product of two functions and the spectral coefficients of the individual functions, where the right-hand side is the matrix-vector representation of the relationship.

As shown in Eqs.~\eqref{eq.26} and \eqref{eq.27}, Eq.~\eqref{eq.5} can be discretized into the following matrix-vector form:
\begin{equation}
	\label{eq.28}
	\left(\frac{4}{\vert\Delta h\vert^2}\mathbf{C}_{\rho}\mathbf{D}_{N}\mathbf{C}_{1/\rho}\mathbf{D}_{N}+\mathbf{C}_{k^2}-k_y^2\mathbf{I}\right)\bm{\hat{\Psi}}
		= \mu^2 \bm{\hat{\Psi}}
\end{equation}
where $\mathbf{I}$ is the identity matrix. The above equation is a matrix eigenvalue problem, but the boundary conditions are not considered at this time.

For the waveguide in Eq.~\eqref{eq.20}, the modal equations (Eqs.~\eqref{eq.5} and \eqref{eq.28}) must be established in both the water column and the bottom sediment. In the flat segments, a single Chebyshev polynomial approximation (continuous with continuous derivatives) cannot be used, as the pressure perturbation $\tilde{p}$ is continuous at $z=h$ but its vertical derivative is not. Thus, we apply the domain decomposition strategy \cite{Min2005} to Eq.~\eqref{eq.5} and split the domain interval into two subintervals:
\begin{equation}
		\label{eq.29}
		\Psi(z)= \begin{cases}
			\Psi_w(z) = \Psi_w(t) \approx\sum_{i=0}^{N_w}\hat{\Psi}_{w,i}T_i(t_w),\quad t_w=-\frac{2z}{h}+1,&
			0\leq z\leq h \\
			\Psi_b(z) = \Psi_b(t) \approx\sum_{i=0}^{N_b}\hat{\Psi}_{b,i}T_i(t_b),\quad
			t_b=-\frac{2z}{H-h}+\frac{H+h}{H-h},&
			h\leq z\leq H
		\end{cases}
\end{equation}
where $N_w$ and $N_b$ are the spectral truncation orders in the water column and bottom sediment, respectively, and $\{\hat{\Psi}_{w,i}\}_{i=0}^{N_w}$ and $\{\hat{\Psi}_{b,i}\}_{i=0}^{N_b}$ are the spectral coefficients in these two layers. Similar to Eq.~\eqref{eq.28}, the modal equations in the water column and bottom sediment can be discretized into matrix-vector form:

\begin{subequations}
		\begin{align}
			\label{eq.30a}
			&\mathbf{A}
			\bm{\hat{\Psi}}_w
			= k_x^2
			\bm{\hat{\Psi}}_w,
			&\mathbf{A} 
			=\frac{4}{h^2}\mathbf{C}_{\rho_w}\mathbf{D}_{N_w}\mathbf{C}_{1/\rho_w}\mathbf{D}_{N_w}+\mathbf{C}_{k_w^2}-k_y^2\mathbf{I}\\
			\label{eq.30b}
			&\mathbf{B}
			\bm{\hat{\Psi}}_b
			= k_x^2
			\bm{\hat{\Psi}}_b,&\mathbf{B}= 
			\frac{4}{(H-h)^2}\mathbf{C}_{\rho_b}\mathbf{D}_{N_b}\mathbf{C}_{1/\rho_b}\mathbf{D}_{N_b}+\mathbf{C}_{k_b^2}-k_y^2\mathbf{I}
		\end{align}
\end{subequations}
where $\mathbf{A}$ and $\mathbf{B}$ are square matrices of order $(N_w+1)$ and $(N_b+1)$, respectively, and $\bm{\hat{\Psi}}_w$ and $\bm{\hat{\Psi}}_b$ are column vectors composed of $\{\hat{\Psi}_{w,i}\}_{i=0}^{N_w}$ and $\{\hat{\Psi}_{b,i}\}_{i=0}^{N_b}$, respectively. Since the interface conditions are related to both the water column and the bottom sediment, Eqs.~\eqref{eq.30a} and \eqref{eq.30b} can be solved simultaneously as follows:
\begin{equation}
		\label{eq.31}
		\left[\begin{array}{cc}
			\mathbf{A}&\mathbf{0}\\
			\mathbf{0}&\mathbf{B}\\
		\end{array}\right]
		\left[\begin{array}{c}
			\bm{\hat{\Psi}}_w\\
			\bm{\hat{\Psi}}_b\\
		\end{array}
		\right]=\mu^2\left[
		\begin{array}{c}
			\bm{\hat{\Psi}}_w\\
			\bm{\hat{\Psi}}_b\\
		\end{array}\right],\quad \mathbf{E}=\left[\begin{array}{cc}
		\mathbf{A}&\mathbf{0}\\
		\mathbf{0}&\mathbf{B}\\
	\end{array}\right]
\end{equation}
The boundary conditions and interface conditions in Eqs.~\eqref{eq.21}, \eqref{eq.22} and \eqref{eq.24} must also be expanded to the spectral space and explicitly added to Eq.~\eqref{eq.31}. For details regarding the treatment of the boundary conditions and interface conditions, please see Eq.~(36) in Ref.~\cite{Tuhw2021a}. Rearranging and combining the modified rows by an elementary row transformation, Eq.~\eqref{eq.31} can be rewritten into the form of the following block matrix:
\begin{equation}
		\label{eq.32}
		\left[
		\begin{array}{cc}
			\mathbf{L}_{11}&\mathbf{L}_{12}\\
			\mathbf{L}_{21}&\mathbf{L}_{22}\\
		\end{array}
		\right]\left[
		\begin{array}{c}
			\bm{\hat{\Psi}}_1\\
			\bm{\hat{\Psi}}_2
		\end{array}
		\right]=\mu^2\left[
		\begin{array}{c}
			\bm{\hat{\Psi}}_1\\
			\mathbf{0}
		\end{array}\right],\quad \mathbf{L}=\left[
	\begin{array}{cc}
	\mathbf{L}_{11}&\mathbf{L}_{12}\\
	\mathbf{L}_{21}&\mathbf{L}_{22}\\
\end{array}
\right]
\end{equation}
where $\mathbf{L}_{11}$ is a square matrix of order $(N_w+N_b-2)$ and $\mathbf{L}_{22}$ is a square matrix of order 4, $\mathbf{\hat{\Psi}}_1=[\hat{\psi}_{w,0},\hat{\psi}_{w,1},\cdots,\hat{\psi}_{w,N_w-2},\\ \hat{\psi}_{b,0},\hat{\psi}_{b,1},\cdots,\hat{\psi}_{b,N_b-2}]^\mathrm{T}$ and $\mathbf{\hat{\Psi}}_2=[\hat{\psi}_{w,N_w-1},\hat{\psi}_{w,N_w},\hat{\psi}_{b,N_b-1},\hat{\psi}_{b,N_b}]^\mathrm{T}$. Solving this linear algebraic system yields the wavenumbers and spectral coefficients of the eigenmodes $(\mu,\bm{\hat{\Psi}}_w^j,\bm{\hat{\Psi}}_b^j)$. This matrix eigenvalue problem can be solved by a mature numerical software library, such as the $\texttt{zgesv()}$ function in LAPACK \cite{LAPACK}.

For the acoustic half-space boundary condition in Eq.~\eqref{eq.23}, since $\gamma_\infty$ contains the eigenvalue to be solved, $\mu$, the elements of $\mathbf{L}$ on the left side of Eq.~\eqref{eq.32} also contain $\mu$; thus, Eq.~\eqref{eq.32} is no longer an algebraic eigenvalue problem and can be solved iteratively only by a root-finding algorithm \cite{Sabatini2019}. Since a prior estimate of $\mu$ is usually not available, many of the existing numerical programs following similar principles fail to converge to a specific root in some cases. To avoid the same issue, we apply the method proposed by \cite{Sabatini2019}: we use $k_{z,\infty}=\sqrt{k_\infty^{2}-\mu^{2}}$ to transform the modal equation and Eq.~\eqref{eq.23} as follows:

\begin{subequations}
		\label{eq.33}
		\begin{gather}
			\label{eq.33a}
			\rho(z) \frac{\mathrm{d}}{\mathrm{d} z}\left(\frac{1}{\rho(z)} \frac{\mathrm{d} \Psi}{\mathrm{d}z}\right)+\left[k^{2}(z)-k_y^2-k_{\infty}^{2}+k_{z,\infty}^{2}\right] \Psi=0 \\
			\label{eq.33b}		
			\frac{\mathrm{i}\rho_\infty}{\rho_b(H)}\frac{\mathrm{d} \Psi(z)}{\mathrm{d} z}\bigg\vert_{z=H}+k_{z,\infty} \Psi(H)=0
		\end{gather}
\end{subequations}
In the Chebyshev--Tau spectral method, Eq.~\eqref{eq.33a} can be discretized into:
\begin{equation}
		\label{eq.34}
		\left[\mathbf{U}+k_{z,\infty}^{2} \mathbf{I}\right] \bm{\hat{\Psi}}=\mathbf{0},\quad \mathbf{U}=\mathbf{E}-k_{\infty}^{2}\mathbf{I}
\end{equation}
After adding the boundary condition in Eq.~\eqref{eq.33b} related to $k_{z,\infty}$, Eq.~\eqref{eq.34} finally expresses the following polynomial eigenvalue problem:
\begin{equation}
		\label{eq.35}
		\left[\mathbf{U}+k_{z,\infty} \mathbf{V}+k_{z,\infty}^{2} \mathbf{W}\right] \bm{\hat{\Psi}}=\mathbf{0}
\end{equation}
$\mathbf{U}$ in Eq.~\eqref{eq.35} is not identical to that in Eq.~\eqref{eq.34}, as it has been modified by boundary conditions and interface conditions; nevertheless, we denote it as $\mathbf{U}$. This polynomial eigenvalue problem can be efficiently solved by transforming into a general matrix eigenvalue problem, although the matrices are twice as large:
\begin{equation}
		\label{eq:36}
		\begin{gathered}
			\tilde{\mathbf{U}} \tilde{\bm{\Psi}}=k_{z,\infty}^{2} \tilde{\mathbf{V}} \tilde{\bm{\Psi}}, \\
			\tilde{\mathbf{U}}=\left[\begin{array}{cc}
				-\mathbf{V} & -\mathbf{U} \\
				\mathbf{I} & 0
			\end{array}\right], \quad \tilde{\mathbf{V}}=\left[\begin{array}{cc}
				\mathbf{W} & 0 \\
				0 & \mathbf{I}
			\end{array}\right], \quad \tilde{\bm{\Psi}}=\left[\begin{array}{c}
				k_{z,\infty} \bm{\hat{\Psi}} \\
				\bm{\hat{\Psi}}
			\end{array}\right]
	\end{gathered}
\end{equation}
Here, $\mathbf{U}$, $\mathbf{V}$, and $\mathbf{W}$ are all very sparse. Similarly, this generalized matrix eigenvalue problem can be solved by the $\texttt{zggev()}$ function in LAPACK \cite{LAPACK}.

Using Eq.~\eqref{eq.26}, the eigenvectors $\bm{\hat{\Psi}}_w$ and $\bm{\hat{\Psi}}_b$ can be easily transformed into $\bm{\Psi}_w(z),z\in[0,h]$ and $\bm{\Psi}_b(z),z\in[h,H]$. The vectors $\bm{\Psi}_w$ and $\bm{\Psi}_b$ are stacked into a single column vector to form $\bm{\Psi}$. Then, either Eq.~\eqref{eq.6} or \eqref{eq.7} is used to normalize $\bm{\Psi}$, and finally, a set of eigenpairs $\{\mu,\Psi(z)\}$ is obtained.

\subsection{Numerical sound field synthesis}

\begin{figure}[htbp]
		\centering
\begin{tikzpicture}[node distance=2cm,scale = 0.7,samples=600,domain=-6:6]
			\draw[very thick, ->](2,-5)--(14,-5) node[right]{Re($k_y$)};
			\draw[very thick, ->](8,-8.5)--(8,-1.5) node[above]{{Im($k_y$)}};
			\filldraw [red] (8.7,-3.15) circle [radius=2.5pt];   		
			\filldraw [red] (9.4,-4.05) circle [radius=2.5pt];
			\filldraw [red] (10.2,-4.5) circle [radius=2.5pt];
			\filldraw [red] (11,-4.75) circle [radius=2.5pt];
			\filldraw [red] (12,-5) circle [radius=2.5pt];
			\filldraw [red] (12.9,-5) circle [radius=2.5pt];			
			\draw[color=blue,very thick,smooth] plot(\x+8,{-1.5*((exp(\x))-(exp(-\x)))/((exp(\x))+(exp(-\x)))-5})
			node[below left] {$k_{y}(q)=q-\mathrm{i} \varepsilon \tanh (\delta q)$};
			\draw[color=cyan,very thick](9,-5)--(9,-6.15);
			\draw[color=cyan,very thick](13.5,-5)--(13.5,-6.5);
			\draw[color=cyan,dashed](9,-6.15)--(9,-7.6)node[below]{$k_{\min}$};
			\draw[color=cyan,dashed](13.5,-6.5)--(13.5,-7.6)node[below]{$k_{\max}$};
\end{tikzpicture}
\caption{Complex integration contour to avoid singularities on the real-$k_y$-axis \cite{Luowy2016a}; red dots indicate the eigenvalues of the two-dimensional problem.}
\label{Figure3}
\end{figure}
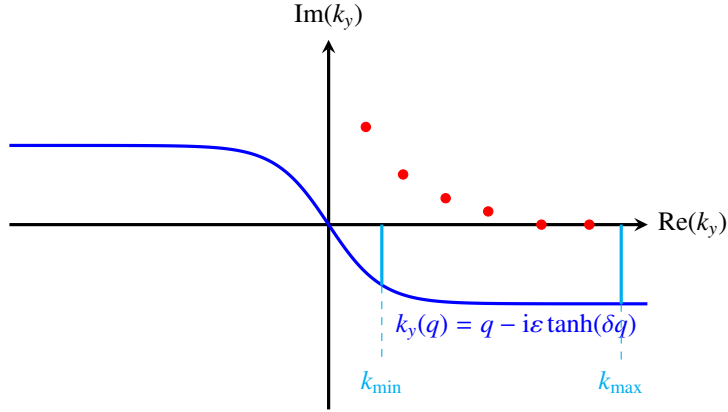

According to the above analysis, after calculating $\tilde{p}(x,k_y,z)$, the spatial sound field can be obtained by the inverse Fourier transform in Eq.~\eqref{eq.2b}. However, $\tilde{p}(x,k_y,z)$ has singularities along the real $k_y$ axis (please see Sec.~\uppercase\expandafter{\romannumeral 1}C of Ref. \cite{Fawcett1990}); therefore, we refer to the approach in Ref.~\cite{Fawcett1990,Luowy2016a} and convert the integral of Eq.~\eqref{eq.2b} into a contour integral on the complex plane. The contour, as shown in Fig.~\ref{Figure3}, takes the following form:
\begin{equation}
		\label{eq.37}
		k_{y}(q)=q-\mathrm{i} \varepsilon \tanh (\delta q), \quad-\infty<q<\infty
\end{equation}

The integral in Eq.~\eqref{eq.2b} is transformed into the following form:
\begin{equation}
		\label{eq.38}
		p(x, y, z)=\frac{1}{\pi} \int_{0}^{\infty} \tilde{p}\left(x, k_{y}(q), z\right) \cos \left(y k_{y}(q)\right)\left[1-\mathrm{i} \varepsilon \delta \operatorname{sech}^{2}(\delta q)\right] \mathrm{d} q
\end{equation}
Since the interval is infinite, integrating over this interval is impossible in an actual numerical evaluation. Luo et al. demonstrated that sufficiently accurate results can be obtained by setting the integration interval of $q$ in the above formula as $[0, 1.5k_0]$, $k_0=2\pi f/\min\{c_w(x,z), c_b(x,z)\}$. Here, $q$ takes equidistant, discrete values in the interval; thus, the above formula can be transformed into a general summation problem \cite{Luowy2015}.

\section{Algorithm and Complexity}
\subsection{Algorithm}
Based on the above derivation, the proposed algorithm can be summarized as follows:

\textbf{Input:} Data from the marine environment and the program parameters.

\textbf{Output:} The quasi-three-dimensional sound pressure field.
\begin{enumerate}
\item
Set the parameters.

The parameters include the frequency $f$, the truncation interval $q$ of the Fourier integral (Eq.~\eqref{eq.38}), the number of sampling points $N_q$ in the $k_y$ wavenumber domain, the location of the source $(x_\mathrm{s},z_\mathrm{s})$, the total depth of the ocean $H$, the topography of the seabed, the number of acoustic profiles, and the specific information pertaining to each group of acoustic profiles. In addition, the parameters should include the spectral truncation orders ($N_w$ and $N_b$), the horizontal and vertical resolutions ($\Delta x$ and $\Delta z$), and the number of modes $M$ to be coupled (the choice of $M$ is usually in the interval $\left[\lfloor \frac{2fH}{\max\{c_w(x,z),c_b(x,z)\}} \rfloor, \lfloor \frac{2fH}{\min\{c_w(x,z),c_b(x,z)\}}+\frac{1}{2} \rfloor\right]$ based on the cutoff frequency estimate for the normal modes). If the lower boundary is the upper interface of an acoustic half-space, the speed $c_\infty$, density $\rho_\infty$ and attenuation $\alpha_\infty$ of the half-space must also be specified.

\item
Segment the marine environment based on the seabed topography, frequency of the source and acoustic profiles.

We assume that the $x$-dependence is divided into $J$ segments. The larger $J$ is, the higher the computational accuracy is, but this generates expensive computational overhead. Jensen confirmed that a strict segmentation criterion is $\Delta x \le \lambda/4$, where $\lambda=\min\{c_w(x,z),c_b(x,z)\}/f$ \cite{Jensen2011,Jensen1998}.

\item
Discretize the $q$ wavenumber domain into $N_q$ equidistant points and perform the calculation from the fourth to the seventh step for each $k_y(q)$.

\item
Apply the Chebyshev--Tau spectral method to solve for the eigenpairs $\{(\mu_m^j,\Psi_m^j)\}_{m=1}^{M}$ of the $J$ $x$-independent segments (see Sec.~\ref{section3.2}).

\item
Calculate the coupling submatrices $\{\mathbf{R}_{1}^{j}\}_{j=1}^{J-1}$, $\{\mathbf{R}_{2}^{j}\}_{j=1}^{J-1}$, $\{\mathbf{R}_{3}^{j}\}_{j=1}^{J-1}$, and $\{\mathbf{R}_{4}^{j}\}_{j=1}^{J-1}$.

\item
Calculate $\mathbf{s}$, construct the global matrix according to either Eq.~\eqref{eq.17} or Eq.~\eqref{eq.19}, and solve the system of linear equations to obtain the coupling coefficients $(\{\mathbf{a}^j\}_{j=1}^J,\{\mathbf{b}^j\}_{j=1}^J)$ of the $J$ segments.

\item
Calculate the $k_y$--domain sound field $\tilde{p}(x,k_y(q),z)$.

The sound pressure subfields of the segments are calculated according to Eq.~\eqref{eq.4}. The subfields of the $J$ segments are then spliced to form the sound pressure field of the entire $xoz$--plane. The fourth to the seventh steps are naturally parallel because the solution of $\tilde{p}(x,k_y(q),z)$ is independent for each $k_y(q)$ (see Sec.~\ref{section2.1}).

\item
Calculate the sound field $\tilde{p}(x,y,z)$.

According to Eq.~\eqref{eq.38}, the $N_q$ discrete $\tilde{p}(x,k_y(q),z)$ fields are accumulated in the following manner to obtain the spatial sound field $\tilde{p}(x,y,z)$:
\begin{equation}
		\label{eq.39}
		p(x, y, z)=\frac{3k}{2\pi(N_q-1)} \sum_{i=1}^{N_q} \tilde{p}\left(x, k_{y,i}, z\right) \cos \left(y k_{y,i}\right)\times\left\{1- \frac{\mathrm{i}}{4\pi\lg \mathrm{e}}\operatorname{sech}^{2}\left[\frac{(N_q-1)k_{y,i}}{9}\right]\right\}			
\end{equation}
The above formula is equivalent to taking $\frac{3k}{2(N_q-1)}$ for $\mathrm{d}q$, $\frac{3\mathrm{d}q}{2\pi\lg \mathrm{e}}$ for $\varepsilon$ and $\frac{1}{6\mathrm{d}q}$ for $\delta$ in Eq.~\eqref{eq.38} \cite{Jensen2011,Tuhw2022d,Luowy2016a}.
\end{enumerate}

\subsection{Computational complexity}
The computational cost of the above algorithm can be approximately divided into two parts: one is to solve Eq.~\eqref{eq.3}, and the other is to inverse integral transform Eq.~\eqref{eq.2b}, corresponding to the fourth to eighth steps of the above algorithm. Next, we analyze the algorithm step by step.

In step 4, when using the Chebyshev--Tau spectral method to solve the local modes, it is necessary to solve the eigenvalues of the $(N_w+N_b+2)$-order dense matrix. If the lower boundary is an acoustic half-space, the generalized matrix eigenvalues of the $2(N_w+N_b+2)$-order matrix are solved. Let $N=N_w+N_b+2$; then, the computational complexity is $O(N^3)$. $N_w$ and $N_b$ should be adapted to the complexity of the sound speed profiles. The complicated marine environment requires a larger spectral truncation order; generally speaking, $N \ge 2M$. For each $k_y$, the $J$-segment local modes need to be solved: $J=\lceil\frac{4fx_{\max}}{\min\{c_w(x,z),c_b(x,z)\}}\rceil$. Thus, the total computational complexity of the fourth step is $O(N_q\times J\times N^3)$. In step 5, the process of forming the global matrix, the matrices $\widetilde{\mathbf{C}}^{j}$ and $\widehat{\mathbf{C}}^{j}$ need to be computed. The integrals are calculated by the trapezoidal rule, and the calculation amount is related to the number of vertical discrete points, i.e., $O\left(J \times \frac{H}{\Delta z}\right)$. In addition, $M$-order matrix inverse and multiplication is performed, and the computational complexity is $O(J \times M^3)$. Thus, the total computational complexity of the fifth step is $O\left(N_q \times J \times \frac{H}{\Delta z}\right)+O\left(N_q \times J \times M^3\right)$. In step 6, the global matrix is a sparse band matrix of order $(2J-2) \times M$; the upper bandwidth is $(2M-1)$, and the lower bandwidth is $(3M-1)$. The direct solution method of the banded linear equation system typically uses LU decomposition first and then solves the problem step by step. The total computational cost is $O\left[(12M^3-8M^2-2M)\times(2J-2)\right]$. The seventh step is essentially $J$ matrix multiplication operations, and the calculation amount is $O(N_q\times J\times \frac{H}{\Delta z}\times\frac{x_{\max}}{\Delta x})$. The inverse integral transform in Eq.~\eqref{eq.39} is essentially a matrix multiplication with a computational complexity of $O\left(N_q\times\frac{H}{\Delta z}\times\frac{x_{\max}}{\Delta x}\right)$. Therefore, the total computational complexity of this algorithm is approximately:
\[
	O(N_q\times J\times N^3)+O(N_q\times J\times M^3)+ O\left(N_q\times J \times \frac{H}{\Delta z}\times\frac{x_{\max}}{\Delta x}\right)
\]

\section{Numerical Simulation}
Here, we present a program named SPEC3D that we developed based on the above algorithm, and we verify the accuracy of the algorithm through several numerical experiments. The transmission loss (TL) field is often used in actual displays to compare and analyze sound fields \cite{Jensen2011}. Therefore, to present the sound field results, the TL of the acoustic pressure is defined as $\text{TL}=-20\log_{10}(\vert p\vert/\vert p_0\vert)$ in dB, where $p_0=\exp(\mathrm{i}k_\mathrm{s})/4\pi$ is the acoustic pressure 1 m from the source and $k_\mathrm{s}$ is the wavenumber of the medium at the location of the source.

For convenience, in SPEC3D, the spectral truncation orders of all the examples are set to $N_w=N_b=20$; in an actual simulation, however, the user can arbitrarily specify the spectral truncation orders in the SPEC3D input file. In the following eight numerical experiments, the speed of sound in seawater is set to 1500.0 m/s, and the density of seawater is set to 1.0 g/cm$^3$; the speed of sound in the sediment is set to 1700.0 m/s, the density in sediment is set to 1.5 g/cm$^3$, and the attenuation coefficient in the sediment is set to 0.5 dB$/\lambda$; the sound speed in the acoustic half-space is set to 2000.0 m/s, the density is set to 2.0 g/cm$^3$, and the attenuation coefficient is set to 1.0 dB$/\lambda$.

\subsection{Analytical example: a three-dimensional ideal fluid waveguide}
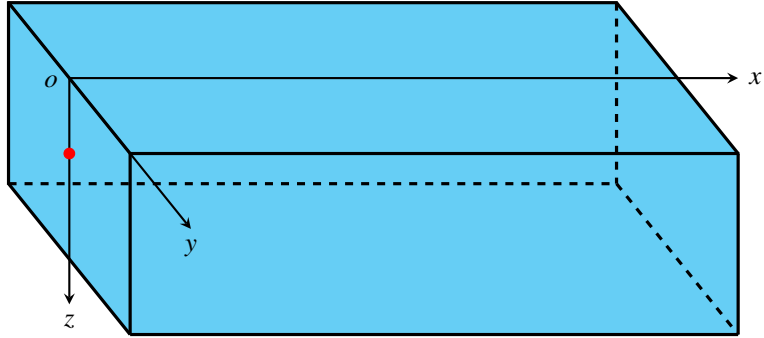
\begin{figure}[htbp]
\centering
\begin{tikzpicture}[node distance=2cm,scale = 0.8,domain=-5:5]
			\fill[cyan,opacity=0.6] (2,0)--(12,0)--(14,-2.5)--(4,-2.5)--(4,-5.5)--(2,-3)--cycle;
			\fill[cyan,opacity=0.6] (4,-5.5)--(14,-5.5)--(14,-2.5)--(4,-2.5)-- cycle; 		
			\draw[thick, ->](3,-1.25)--(14,-1.25) node[right]{$x$};
			\draw[thick, ->](3,-1.25)--(3,-5) node[below]{$z$};	    		
			\draw[very thick](1.98,0)--(12.02,0);
			\draw[very thick](4,-2.5)--(14.02,-2.5);
			\draw[very thick](2,0)--(4,-2.5);
			\draw[thick, ->](2,0)--(5,-3.75) node[below]{$y$};	
			\draw[very thick](12,0)--(14,-2.5);
			\draw[very thick](2,0.02)--(2,-3.02);
			\draw[very thick](4,-2.5)--(4,-5.5);
			\draw[very thick](2,-3)--(4,-5.5);		
			\draw[dashed, very thick](12,0)--(12,-3.0);
			\draw[very thick](14,-2.5)--(14,-5.5);
			\draw[dashed, very thick](12,-3.0)--(14,-5.5);
			\draw[very thick] (4,-5.5)--(14,-5.5);
			\draw[dashed, very thick](2,-3)--(12,-3);		
			\filldraw [red] (3,-2.5) circle [radius=2.5pt];
			\node at (2.7,-1.3){$o$};				
\end{tikzpicture}
\caption{Schematic diagram of the three-dimensional ideal fluid waveguide.}
\label{Ideal}
\end{figure}

\begin{figure}[htbp]
\centering\includegraphics[height=8cm]{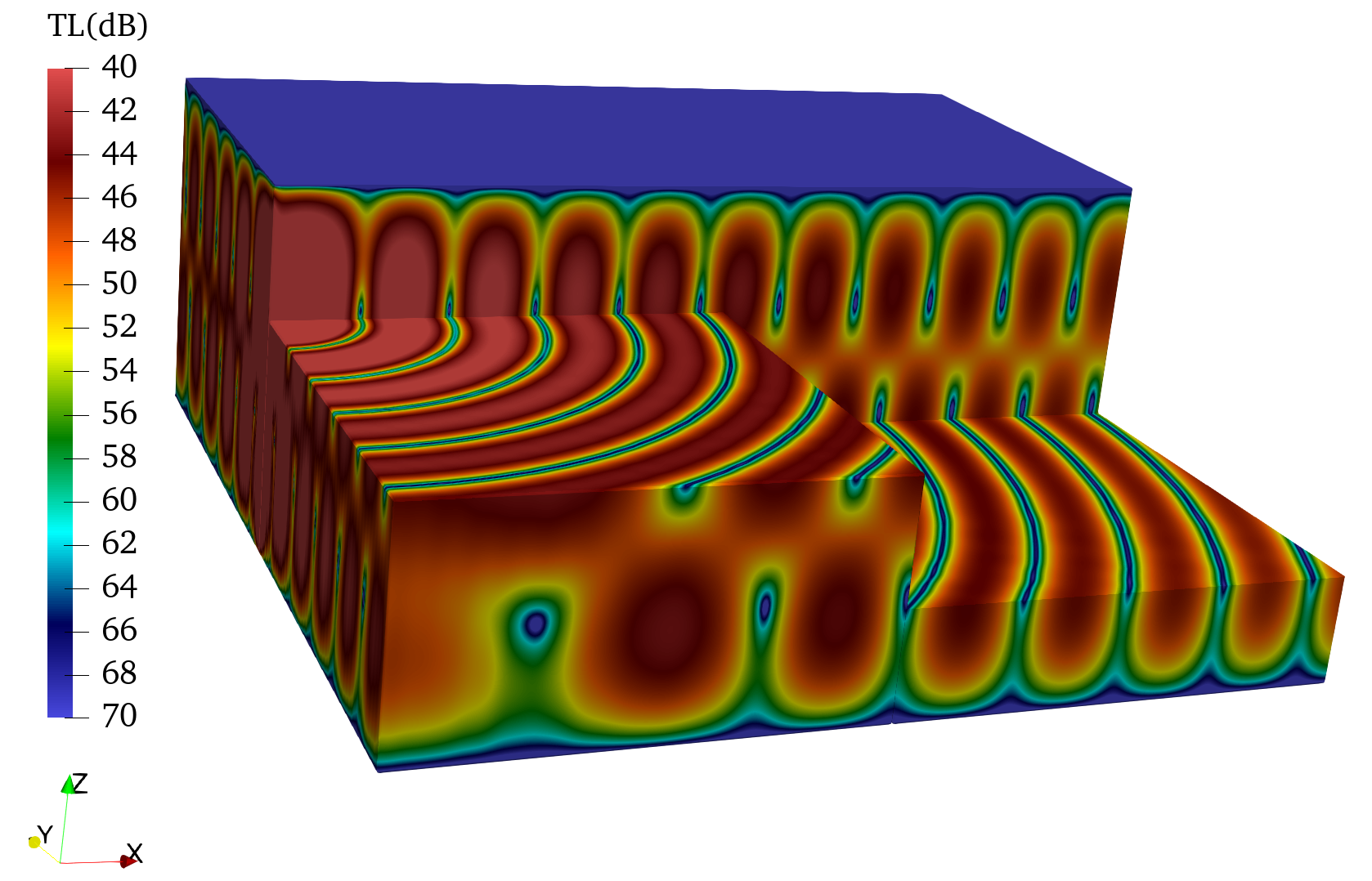}
\caption{Sound field of the three-dimensional ideal fluid waveguide calculated by SPEC3D.}
\label{Figure5}
\end{figure}

Here, we use an ideal fluid waveguide, which is a highly simplified case in underwater acoustics, to validate SPEC3D. The structure of the three-dimensional ideal fluid waveguide is shown in Fig.~\ref{Ideal}. The frequency of the sound source is $f=20$ Hz, and the source is located at $z_\mathrm{s}=36$ m. The number of coupled modes is $M=2$. The number of discrete points in the $k_y$--domain is $N_q=512$. Due to the horizontal independence of the marine environment, there is no coupling in the $x$-direction, so segmentation is not needed, but for the smooth operation of SPEC3D, $J$ is taken as 2. The horizontal independence also makes the sound field cylindrically symmetrical around the axis where the sound source is located. First, consider the case when the lower boundary is also the pressure-release boundary. Fig.~\ref{Figure5} presents the sound field calculated by SPEC3D, featuring prominent cylindrical symmetry.

\begin{figure}[htbp]
\centering
\subfigure[]{\includegraphics[width=6.5cm]{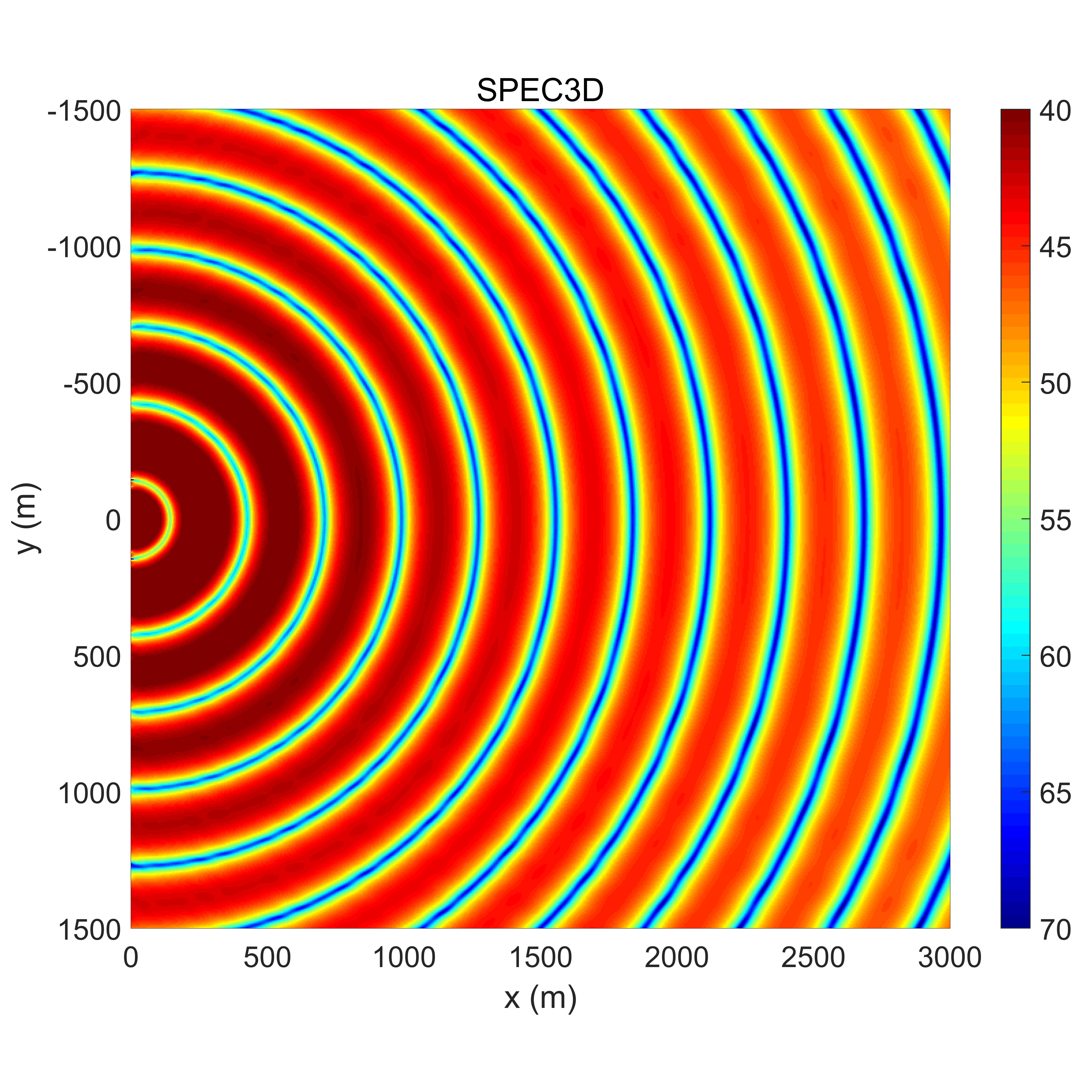}\label{Figure6a}}
\subfigure[]{\includegraphics[width=6.5cm]{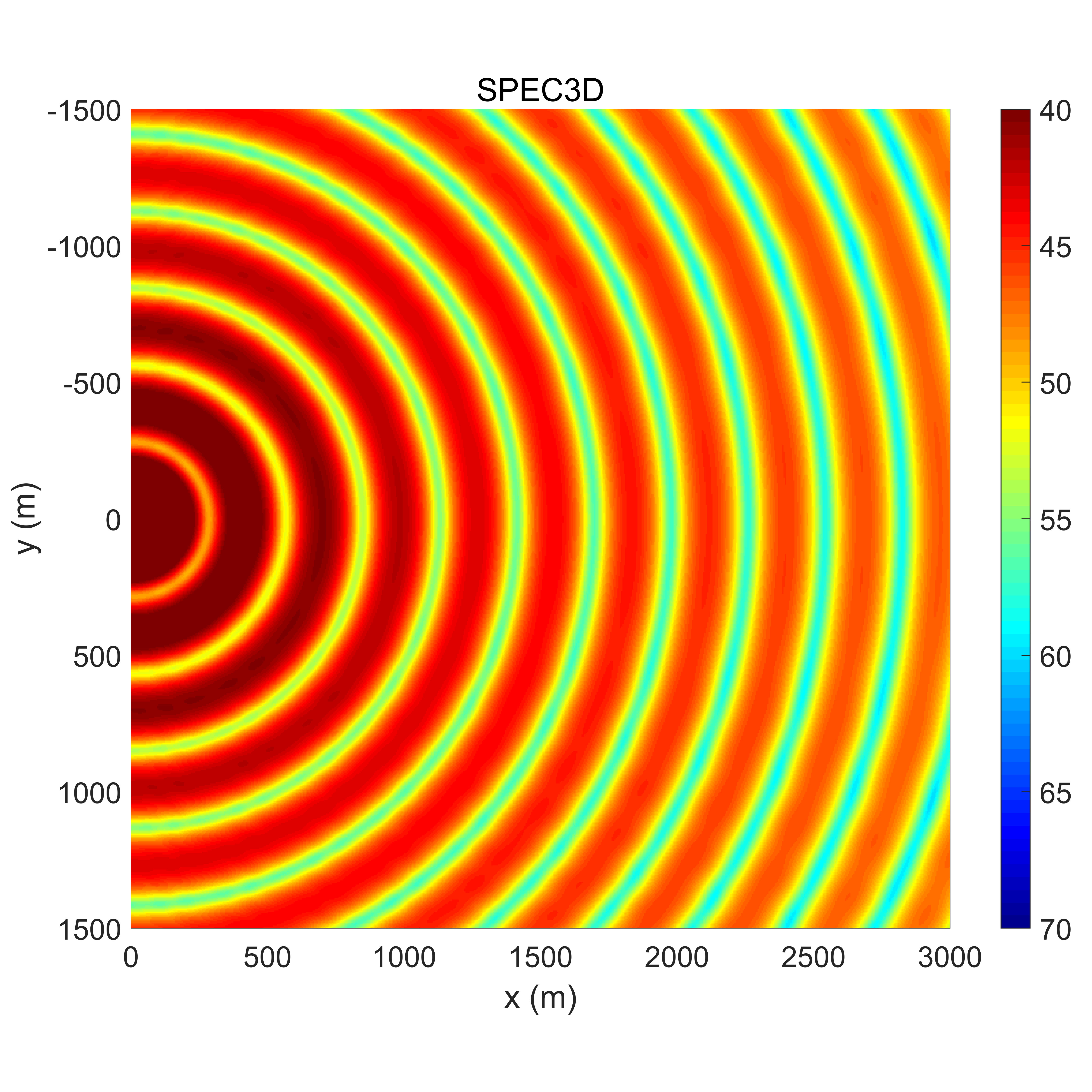}\label{Figure6b}}
\subfigure[]{\includegraphics[width=6.5cm]{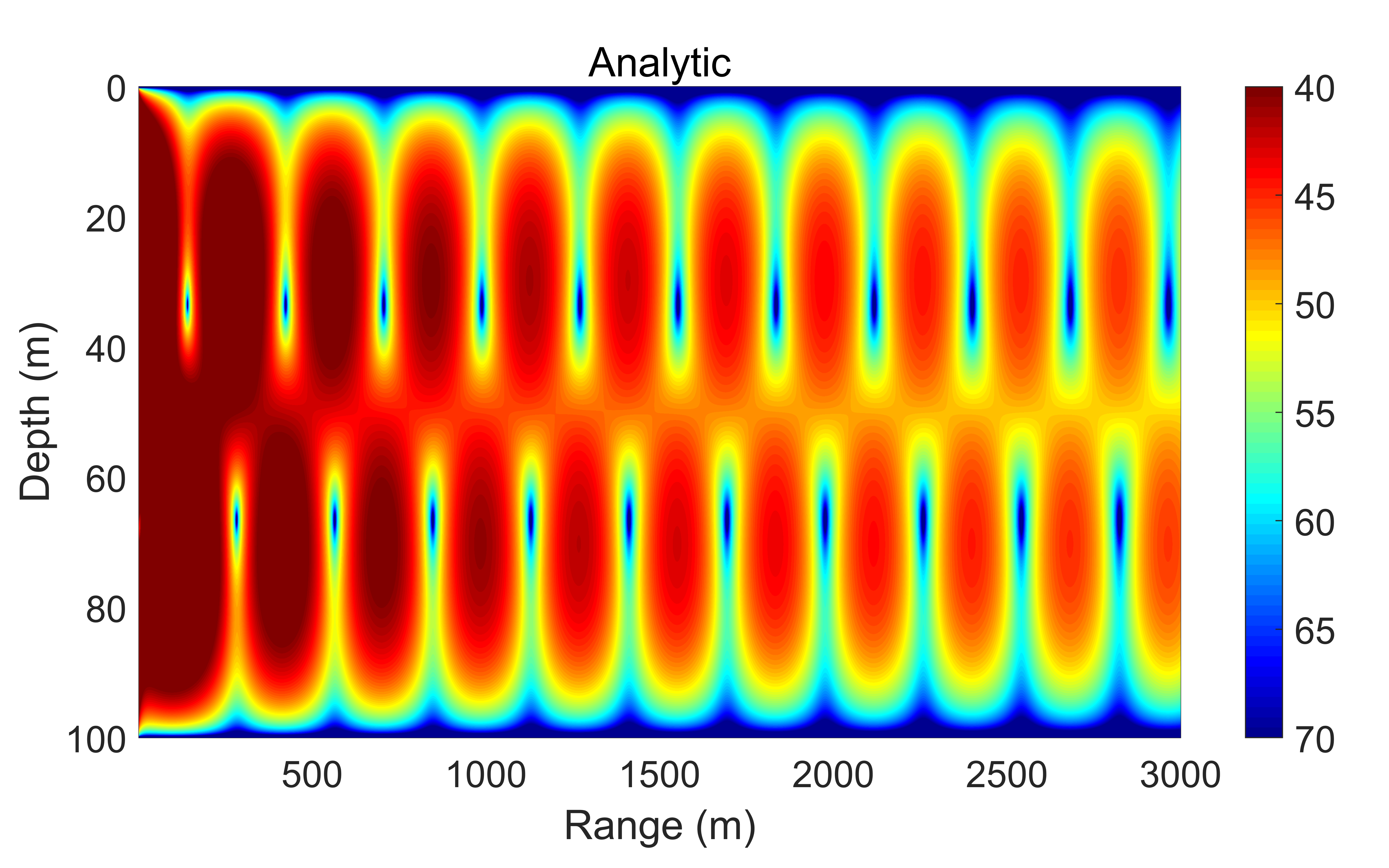}\label{Figure6c}}
\subfigure[]{\includegraphics[width=6.5cm]{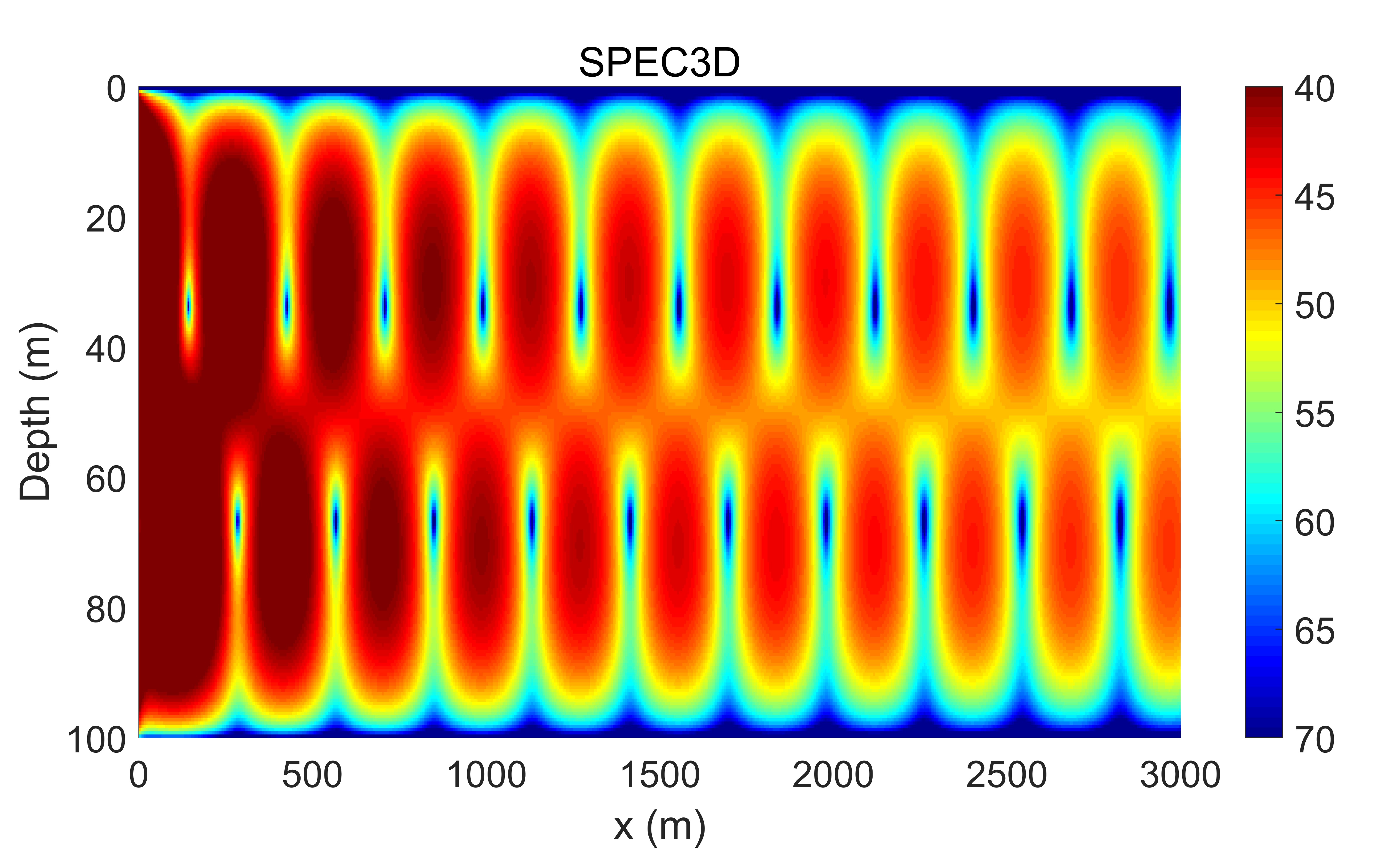}\label{Figure6d}}
\subfigure[]{\includegraphics[width=13cm]{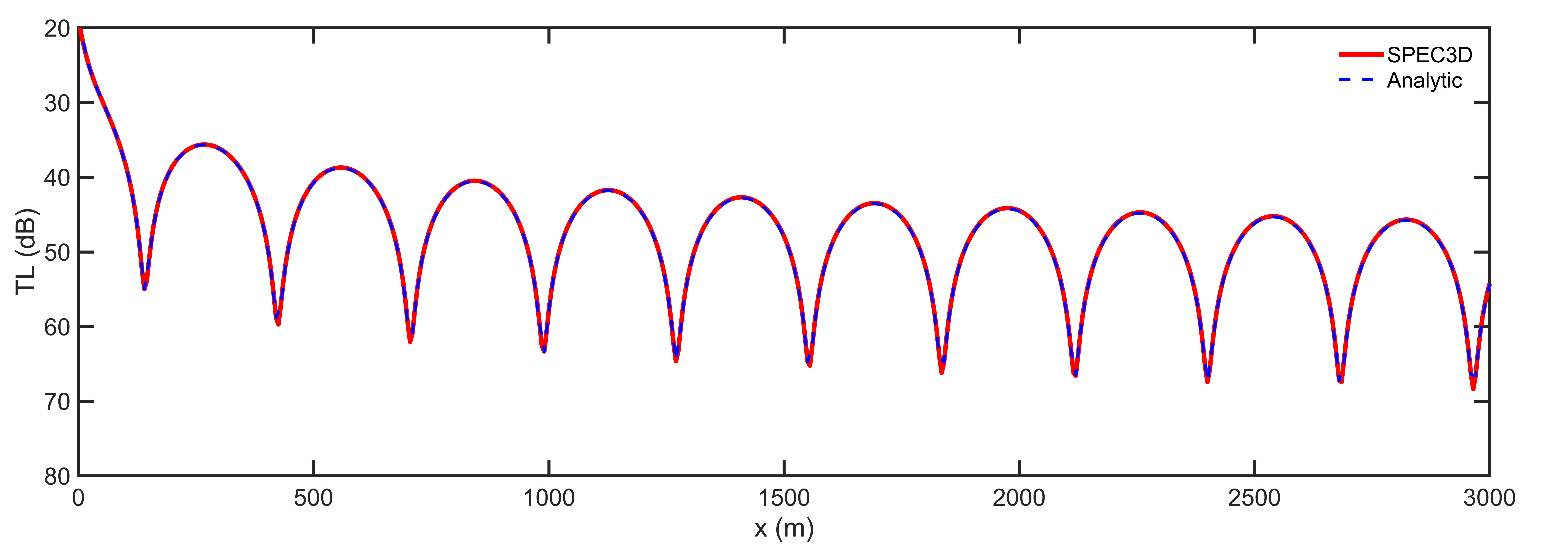}\label{Figure6e}}
\caption{Sound fields of the three-dimensional ideal fluid waveguide with a free bottom calculated by SPEC3D at depths of 36 m (a) and 80 m (b); analytical solution (c) and sound field calculated by SPEC3D (d) on the $y$--plane at $y=0$ m; TL along the $x$-direction at the depth of the sound source (e).}
\label{Figure6}
\end{figure}

\begin{figure}[htbp]
\centering
\subfigure[]{\includegraphics[width=6.5cm]{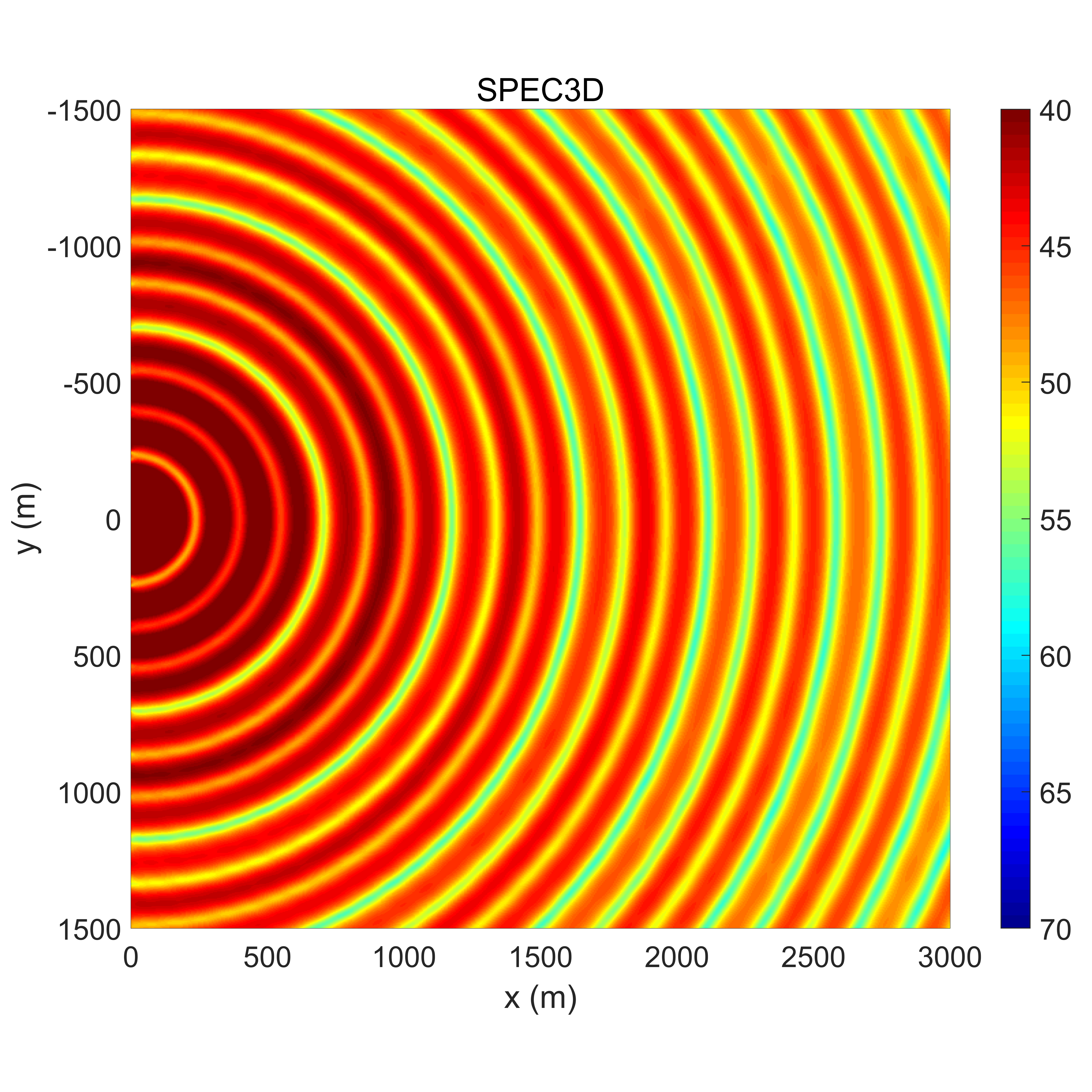}}
\subfigure[]{\includegraphics[width=6.5cm]{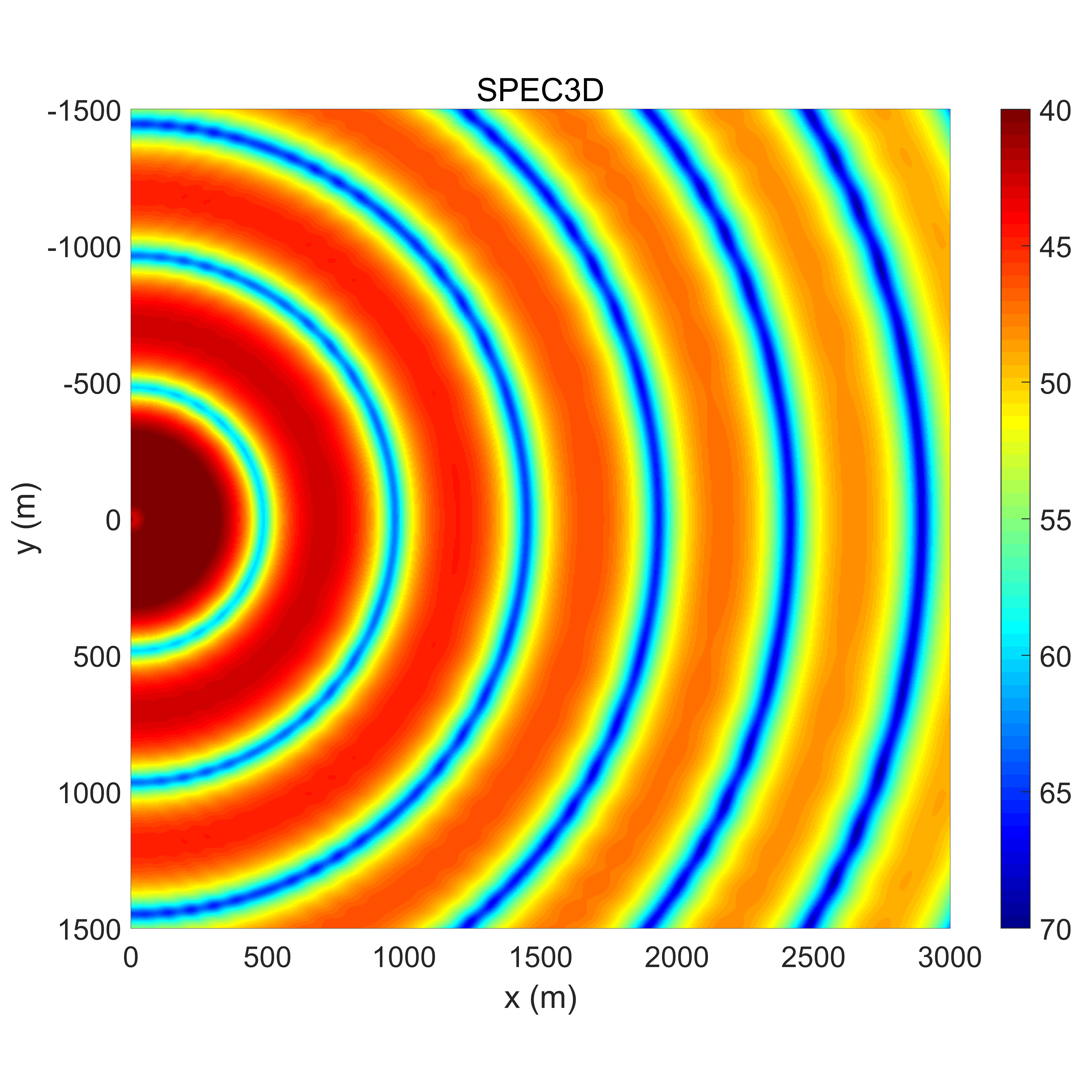}}
\subfigure[]{\includegraphics[width=6.5cm]{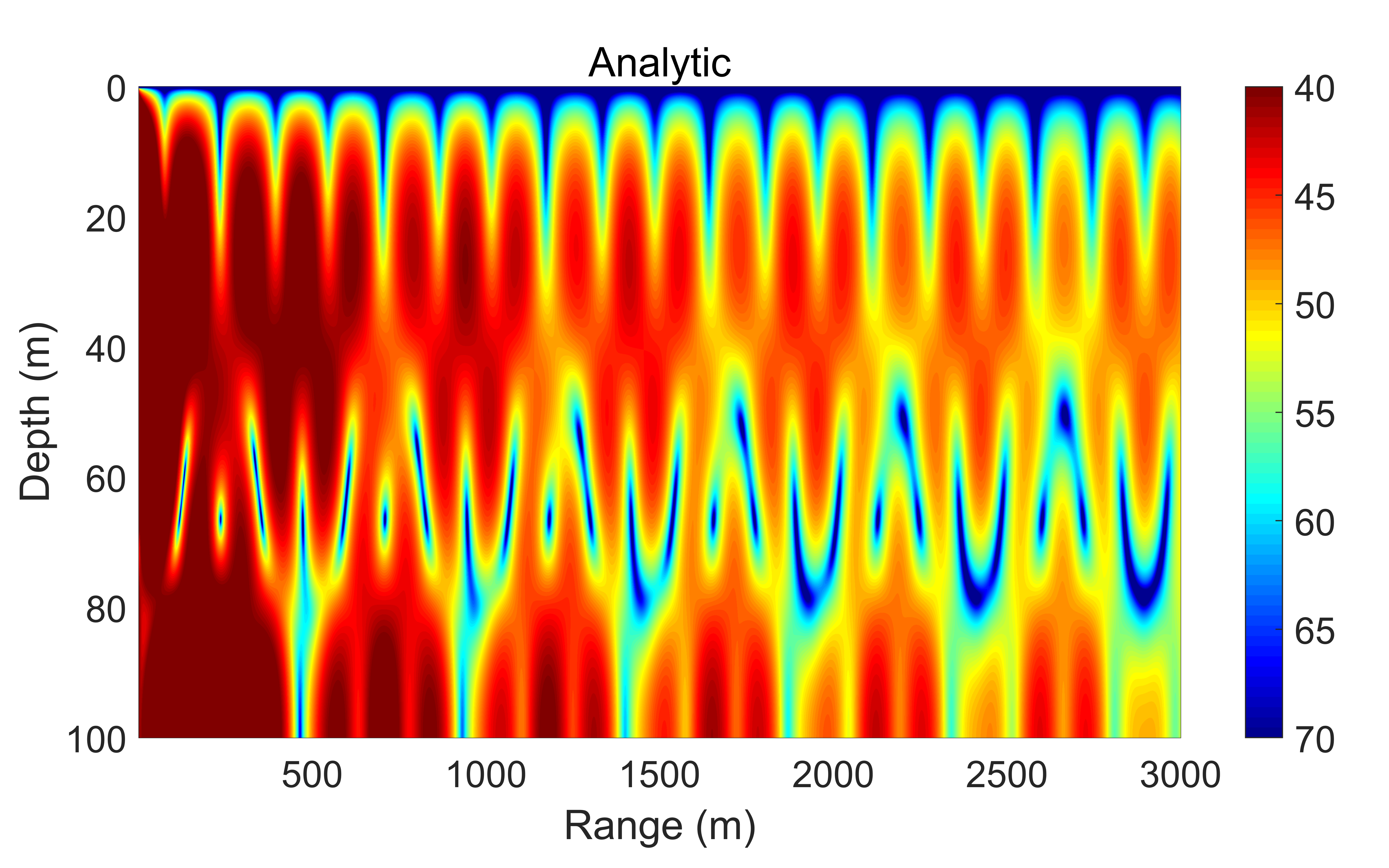}}
\subfigure[]{\includegraphics[width=6.5cm]{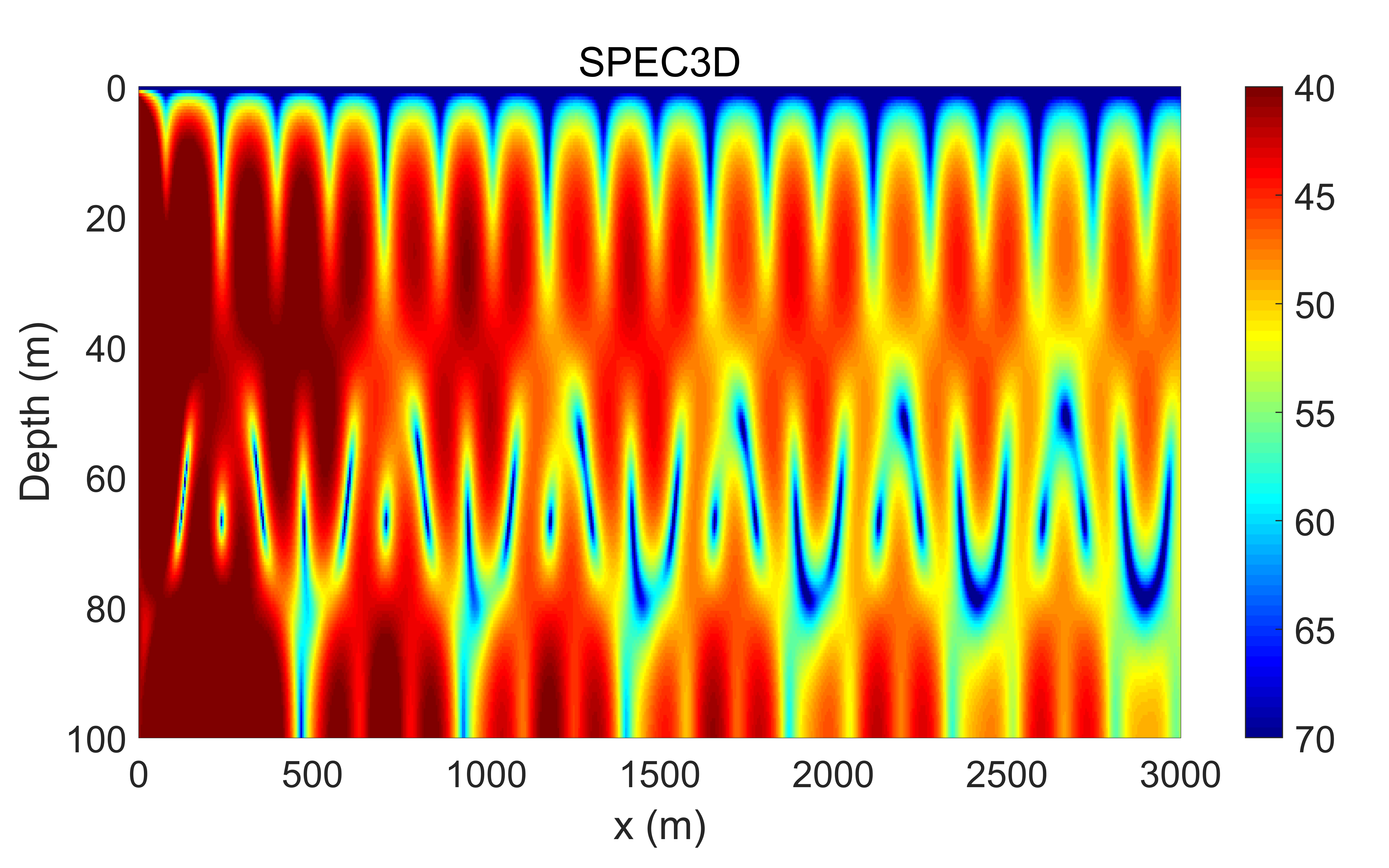}}
\subfigure[]{\includegraphics[width=13cm]{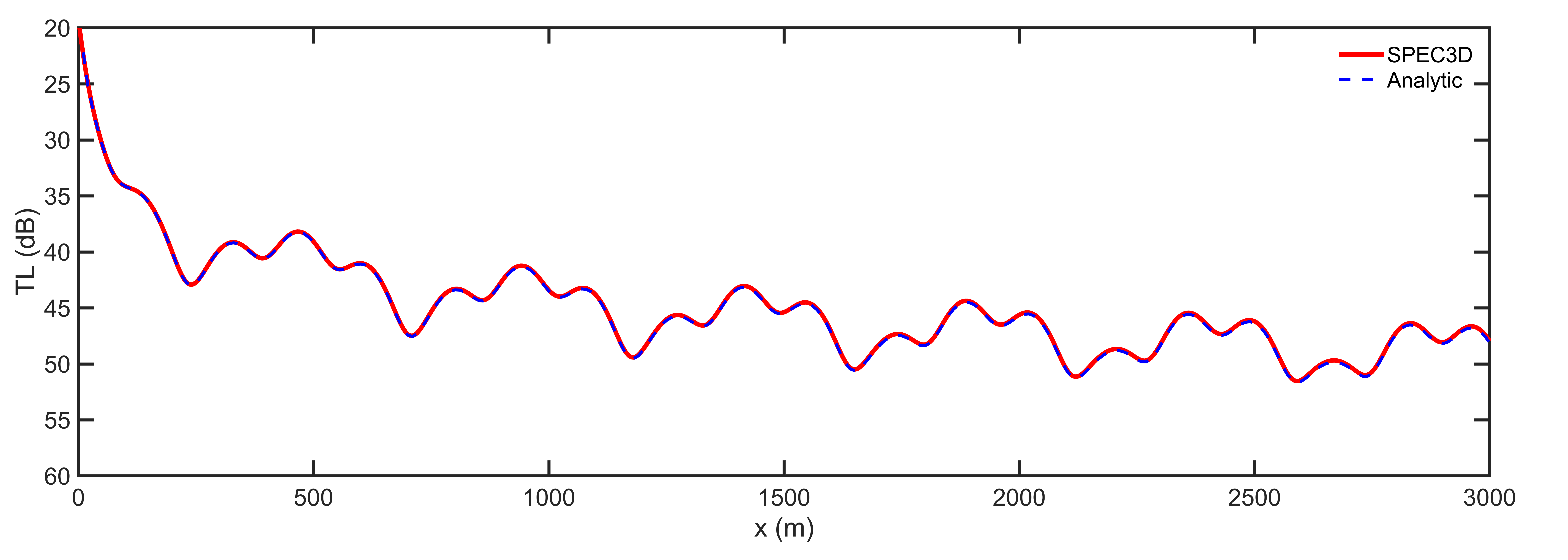}}
\caption{Sound fields of the three-dimensional ideal fluid waveguide with rigid bottom calculated by SPEC3D at depths of 25 m (a) and 80 m (b); analytical solution (c) and sound field calculated by SPEC3D (d) on the $y$--plane at $y=0$ m; TL along the $x$-direction at the depth of the sound source (e).}
\label{Figure7}
\end{figure}

Figs.~\ref{Figure6a} and \ref{Figure6b} depict slices of the sound field calculated by SPEC3D at two depths. According to the above analysis, the rings formed by the sound field on each slice should be strictly concentric, but the sound fields are not standard rings, so there are some errors. There is an analytical solution for an ideal fluid waveguide with a pressure-release boundary. The analytical solution for each plane of symmetry is:
\begin{equation}
	\label{eq.40}
		\begin{gathered}
			p(x,z)=\frac{2\pi \mathrm{i}}{H}\sum_{n=1}^{\infty}\sin{(k_{z,n}z_\mathrm{s})}\sin{(k_{z,n}z)}\mathcal{H}_0^{(1)}(k_{x,n}x)\\
			k_{z,n}=\frac{n\pi}{H},\quad k_{x,n}=\sqrt{k_0^2-k_{z,n}^2}, \quad n=1,2,3,\cdots
		\end{gathered}
\end{equation}
where $\mathcal{H}_0^{(1)}(\cdot)$ is the Hankel function. Figs.~\ref{Figure6c} and \ref{Figure6d} show the analytical solution and the sound field calculated by SPEC3D, respectively, indicating that the two sound fields are reasonably consistent. Fig.~\ref{Figure6e} plots the TL curve along the $x$-direction at the depth of the sound source. The consistency between the SPEC3D solution and the analytical solution verifies the accuracy and reliability of SPEC3D.

When the lower boundary is perfectly rigid, the analytical solution of the ideal fluid waveguide is the same as Eq.~\eqref{eq.40}, except that the vertical wavenumber becomes:
\begin{equation}
		k_{z,n}=\left(n-\frac{1}{2}\right)\frac{\pi}{H},\quad k_{x,n}=\sqrt{k_0^2-k_{z,n}^2},\quad n=1,2,3\dots
\end{equation}
The number of coupled modes is $M=3$. Similarly, the results of Fig.~\ref{Figure7} can lead to the same conclusion as Fig.~\ref{Figure6}.

\subsection{Analytical example: an ideal wedge-shaped waveguide}

\begin{figure}[htbp]
\centering
\begin{tikzpicture}[node distance=2cm,scale = 0.8]
			\fill[cyan,opacity=0.6] (2,0)--(12,0)--(14,-2.5)--(4,-8.5)--(2,-6)--cycle;
			\draw[thick, ->](3,-1.25)--(14,-1.25) node[right]{$x$};
			\draw[thick, ->](3,-1.25)--(3,-8) node[below]{$z$};	    		
			\draw[very thick](1.98,0)--(12.02,0);
			\draw[very thick](4,-2.5)--(14.02,-2.5);
			\draw[very thick](2,0)--(4,-2.5);
			\draw[thick, ->](2,0)--(4.5,-3.125) node[below]{$y$};	
			\draw[very thick](12,0)--(14,-2.5);
			\draw[very thick](2,0.02)--(2,-6.02);
			\draw[very thick](4,-2.5)--(4,-8.5);
			
			\draw[dashed, very thick](2,-6)--(12,0);		
			\draw[very thick](4,-8.5)--(14,-2.5);
			
			\draw[very thick](2,-6)--(4,-8.5);	
			\draw[dashed, very thick](4,-8.5)--(8,-8.5);			
			\filldraw [red] (8,-2.7) circle [radius=2.5pt];
			\node at (2.7,-1.3){$0$};
			\draw[very thick](5,-8.5) arc (0:30:1);
			\node at (6.2,-8.2){$\theta_0$=2.86°};
		\end{tikzpicture}
\caption{Schematic of the ideal wedge-shaped waveguide.}
\label{Wedge}
\end{figure}
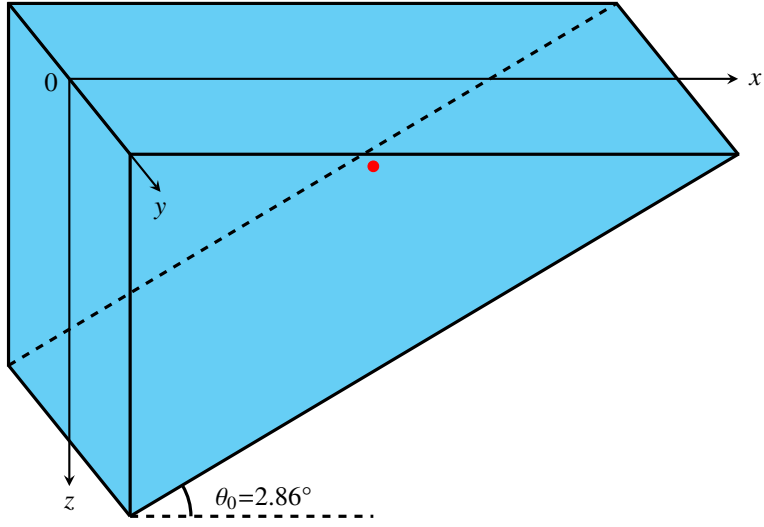

Consider the ideal wedge-shaped waveguide shown in Fig.~\ref{Wedge}, which is a primary benchmark problem for range-dependent waveguides. Both the sea surface and the seabed are pressure-release boundaries, and the source is located at $(x_\mathrm{s}, 0, z_\mathrm{s})$. Buckingham \cite{Buckingham1984,Buckingham1987,Doolittle1988} developed an analytical solution to this benchmark problem, which was originally proposed and discussed at two consecutive conferences of the Acoustic Society of America (ASA). Here, we briefly introduce this problem. The velocity potential in the water body is:

\begin{subequations}
		\label{eq.42}
		\begin{gather}
			\Phi=\frac{1}{\theta_{0}} \sum_{n=1}^{\infty} I_{\nu_n}\left(r, r_\mathrm{s}, z\right) \sin (\nu_n \theta) \sin \left(\nu_n \theta_\mathrm{s}\right) \\
			\label{eq.42b}
			I_{\nu_n}\left(r, r_\mathrm{s}, z\right)=\mathrm{i} \int_{0}^{\infty} k_r \frac{\exp (\mathrm{i} \eta|z|)}{\eta} J_{\nu_n}(k_r r) J_{\nu_n}\left(k_r r_\mathrm{s}\right) \mathrm{d} k_r\\
			\eta=\sqrt{k^2-k_r^2},\quad \nu_n = \frac{n\pi}{\theta_0}, \quad n=1,2,\cdots
		\end{gather}		
\end{subequations}
where $r$ and $r_\mathrm{s}$ are the distances from the receiver and source to the apex of the wedge, respectively; $\theta$ and $\theta_\mathrm{s}$ are the angles measured from the apex to the depths of the receiver and source, respectively; $\theta_0$ is the wedge angle; $k$ is the wavenumber of the seawater; and $J_{\nu_n}(\cdot)$ is the Bessel function of order $\nu_n$. The TL field is calculated by the following formula:

\begin{subequations}
		\label{eq.43}
		\begin{gather}
			\mathrm{TL}=-20 \log _{10}\bigg\vert\frac{\Phi\left(r, r_\mathrm{s}, \theta, \theta_\mathrm{s}\right)}{\Phi_{0}(1)}\bigg\vert \\
			\Phi_{0}(r)=\frac{\exp(\mathrm{i}k r)}{4\pi r}
		\end{gather}		
\end{subequations}
Note that directly calculating the integral in Eq.~\eqref{eq.42b} may trigger a numerical overflow, which can be approximated by taking the Debye asymptotic expansion \cite{Milton1972}. In this paper, we use the Gauss--Kronrod quadrature.

\begin{figure}[htbp]
\centering\includegraphics[height=7cm]{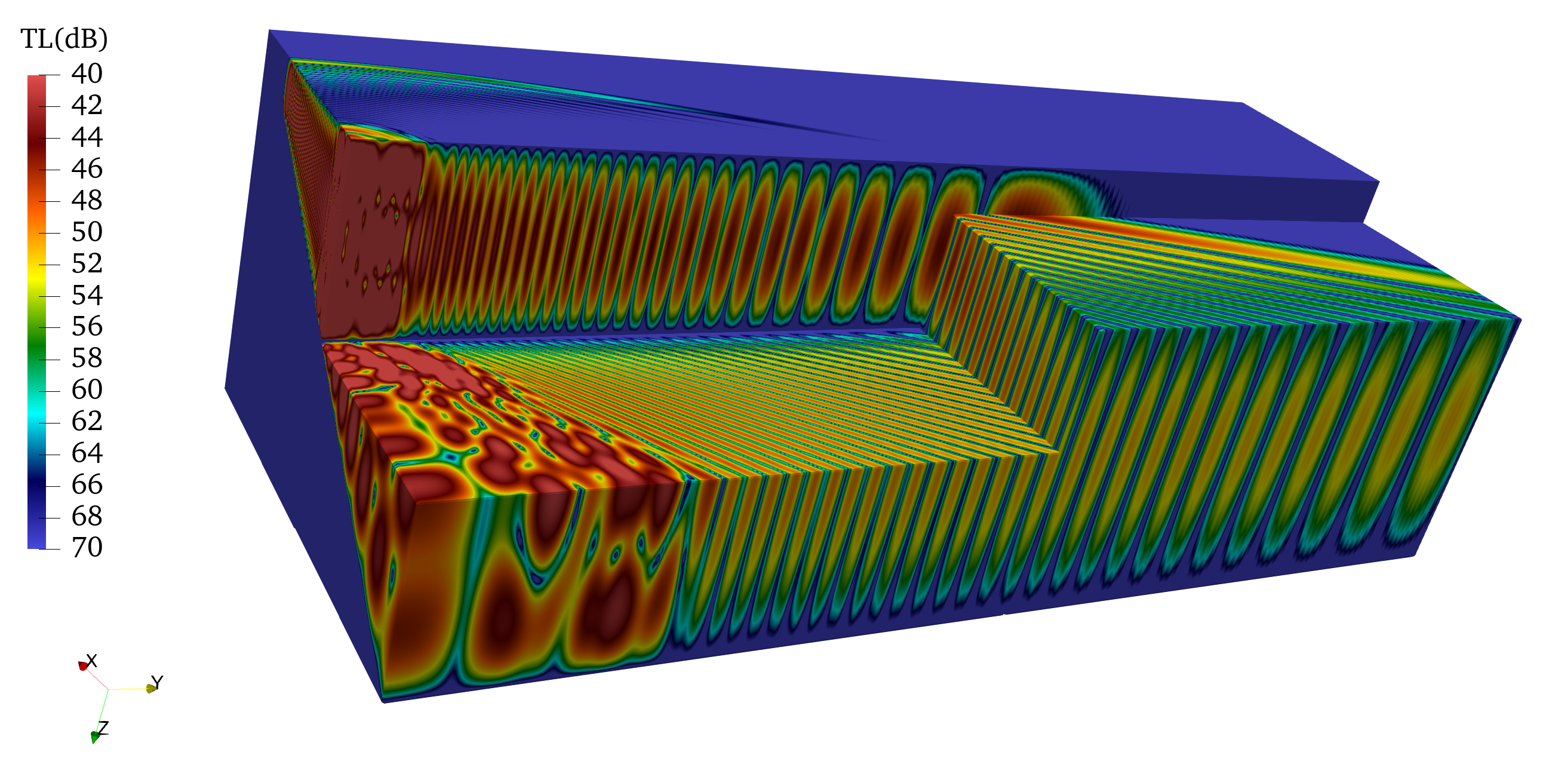}
\caption{Sound field of the three-dimensional wedge-shaped waveguide with a free bottom calculated by SPEC3D.}
\label{Figure9}
\end{figure}

In this case, the frequency of the sound source is $f=25$ Hz, and the source is located at $x_\mathrm{s}=2000$ m and $z_\mathrm{s}=50$ m. The wedge is $x=4000$ m long in the horizontal direction and $z=200$ m deep in the vertical direction. The number of coupled modes is $M=6$, $J=267$. The number of discrete points in the $k_y$--domain is $N_q=2048$. Fig.~\ref{Figure9} shows the sound field of the three-dimensional wedge-shaped waveguide calculated by SPEC3D; the sound field exhibits prominent three-dimensional characteristics.

\begin{figure}[htbp]
\centering
\subfigure[]{\includegraphics[width=6.5cm]{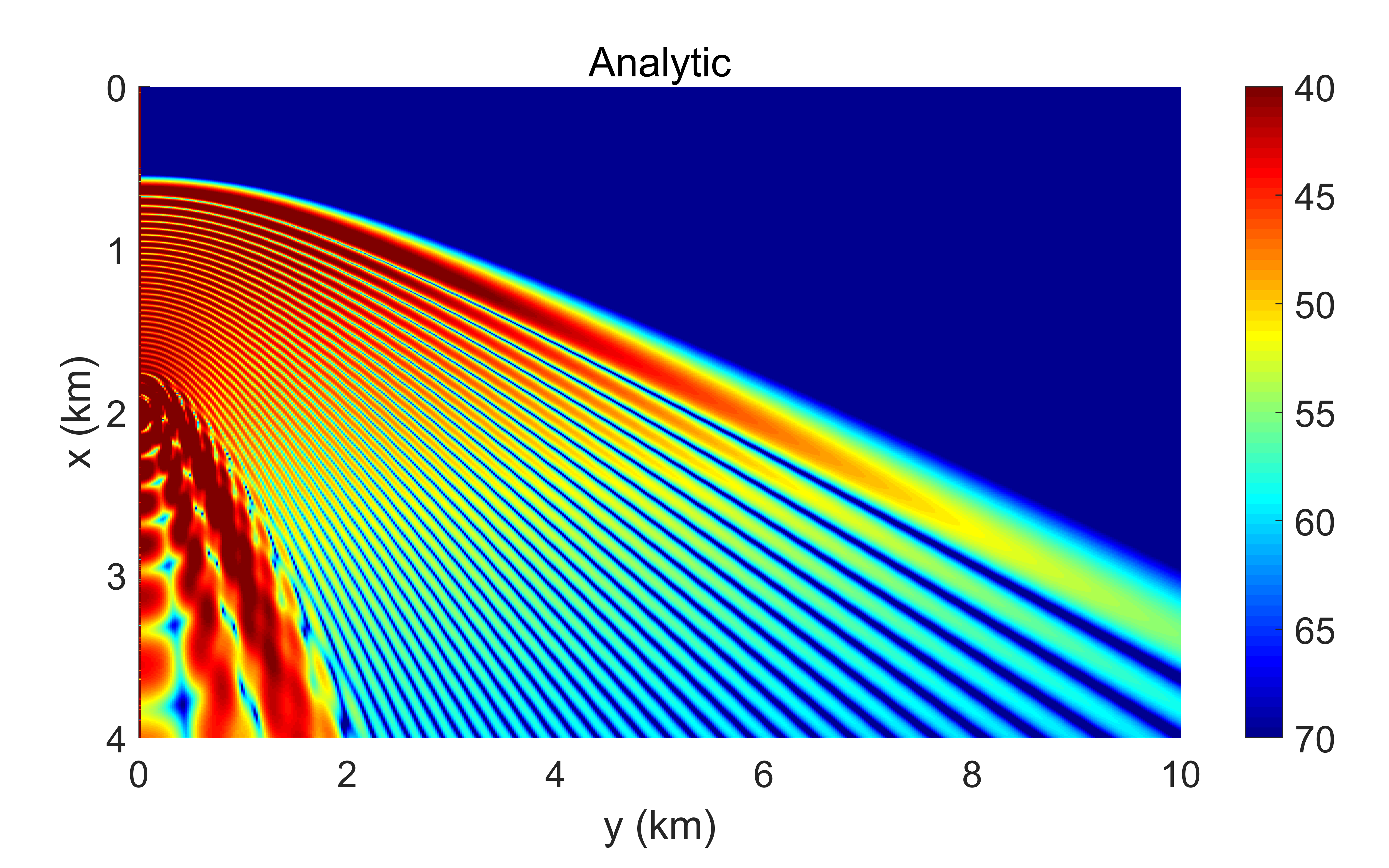}}
\subfigure[]{\includegraphics[width=6.5cm]{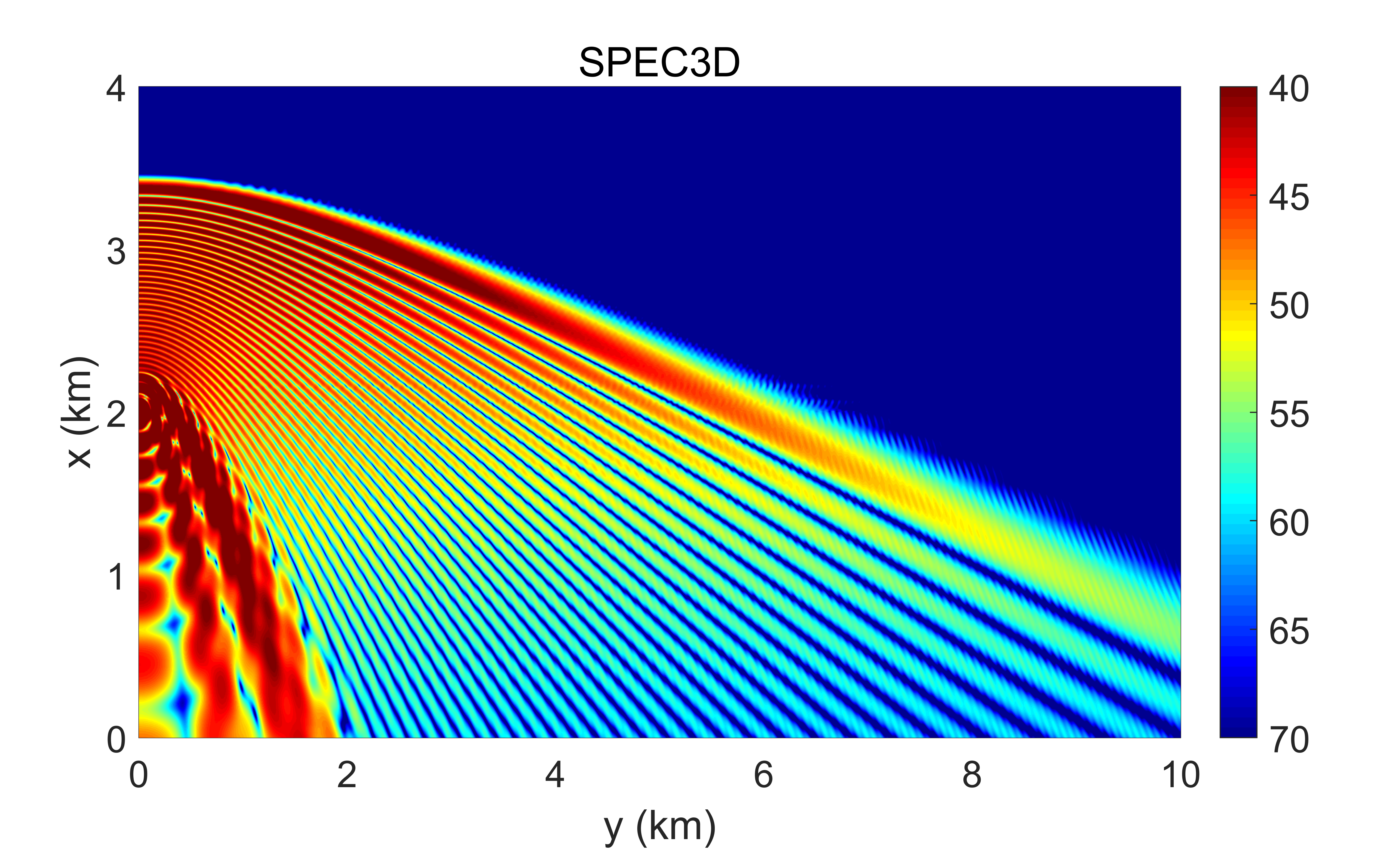}}
\subfigure[]{\includegraphics[width=6.5cm]{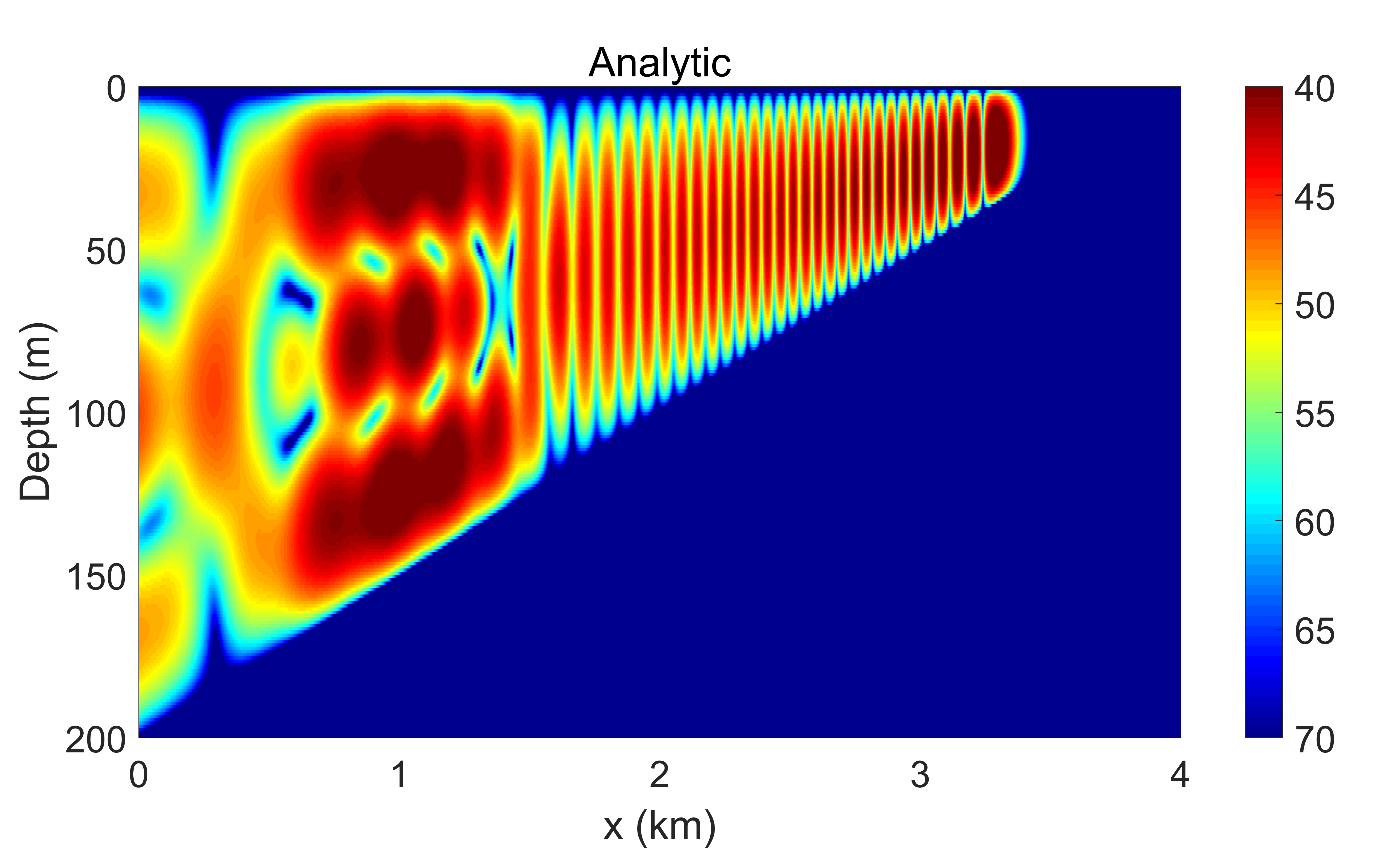}}
\subfigure[]{\includegraphics[width=6.5cm]{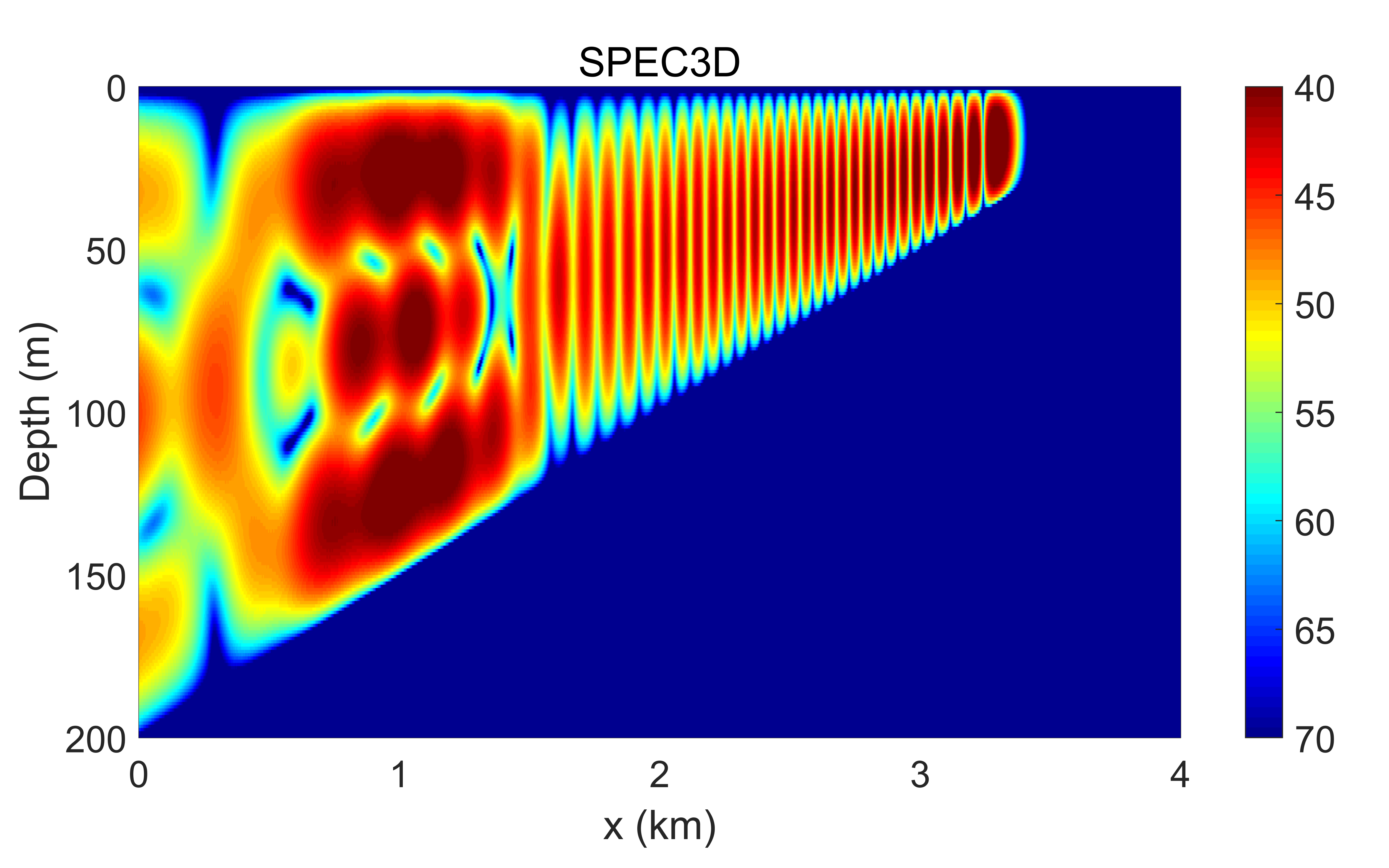}}
\subfigure[]{\includegraphics[width=6.5cm]{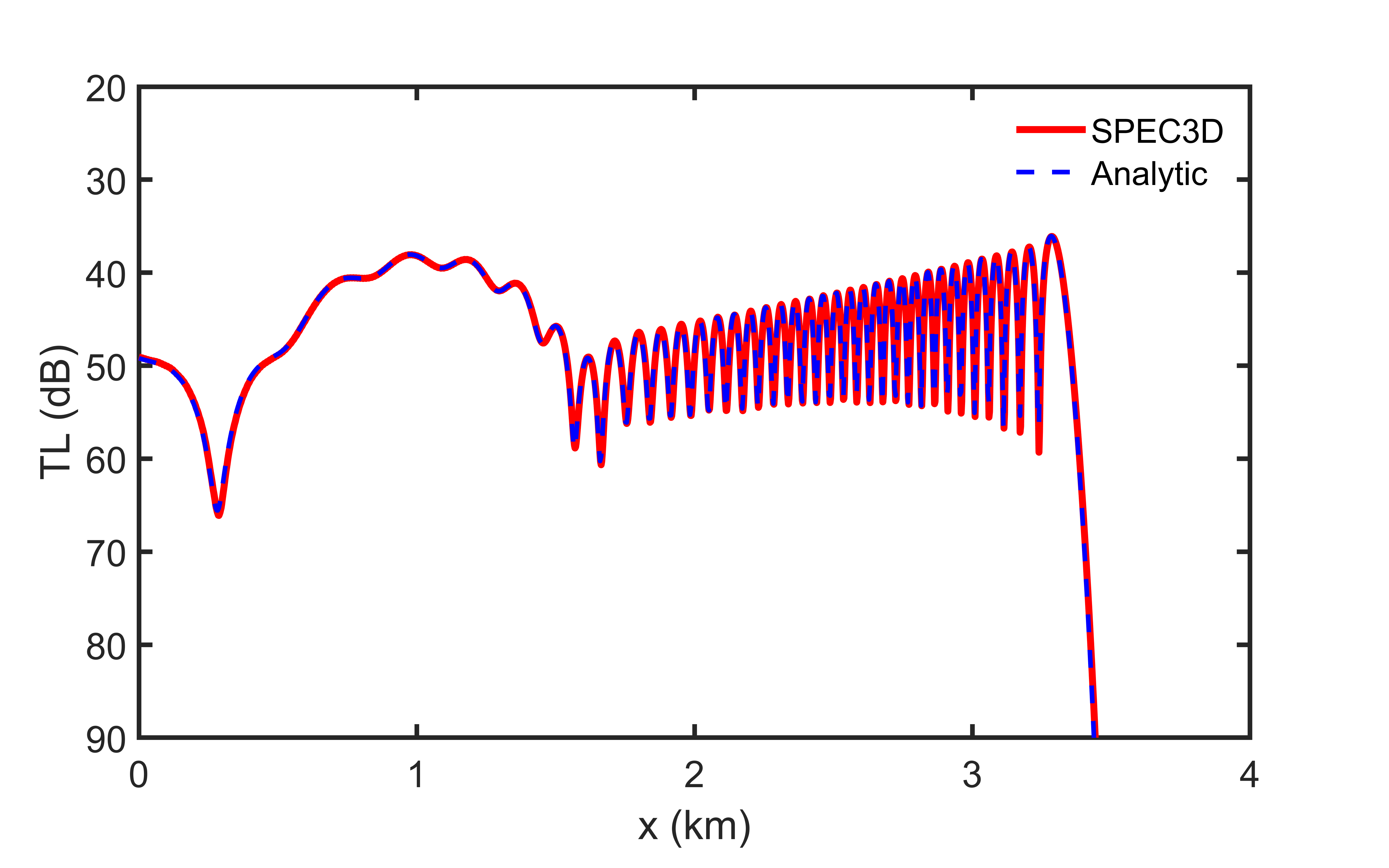}}
\subfigure[]{\includegraphics[width=6.5cm]{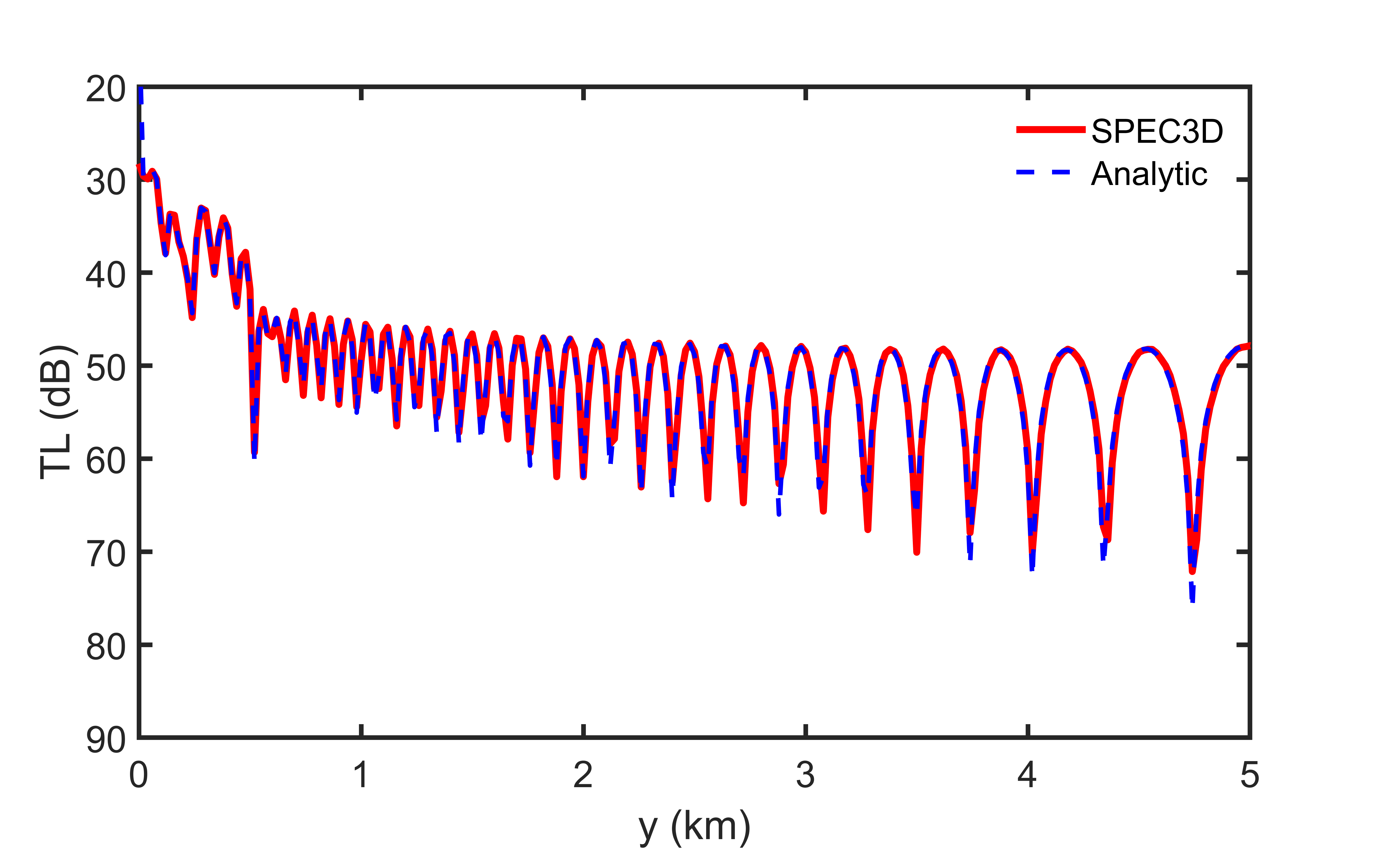}}
\caption{Sound fields of the three-dimensional wedge-shaped waveguide with a free bottom calculated by the analytical solution (a) and SPEC3D (b) at a depth of $z=25$ m; sound fields calculated by the analytical solution (c) and SPEC3D (d) on the $y$--plane at $y=1000$ m; TLs along the $x$-direction (e) and $y$-direction (f) at a depth of $z=25$ m.}
\label{Figure10}
\end{figure}

Next, we compare slices through the sound field with slices from the analytical solution in detail. As observed from the slices through the sound field in Fig.~\ref{Figure10}, the sound field calculated by SPEC3D effectively matches the analytical sound field. Thus, despite the small (typically less than 1 dB) error at long distances, SPEC3D can successfully reproduce the sound field of the waveguide.

\begin{figure}[htbp]
\centering
\subfigure[]{\includegraphics[width=6.5cm]{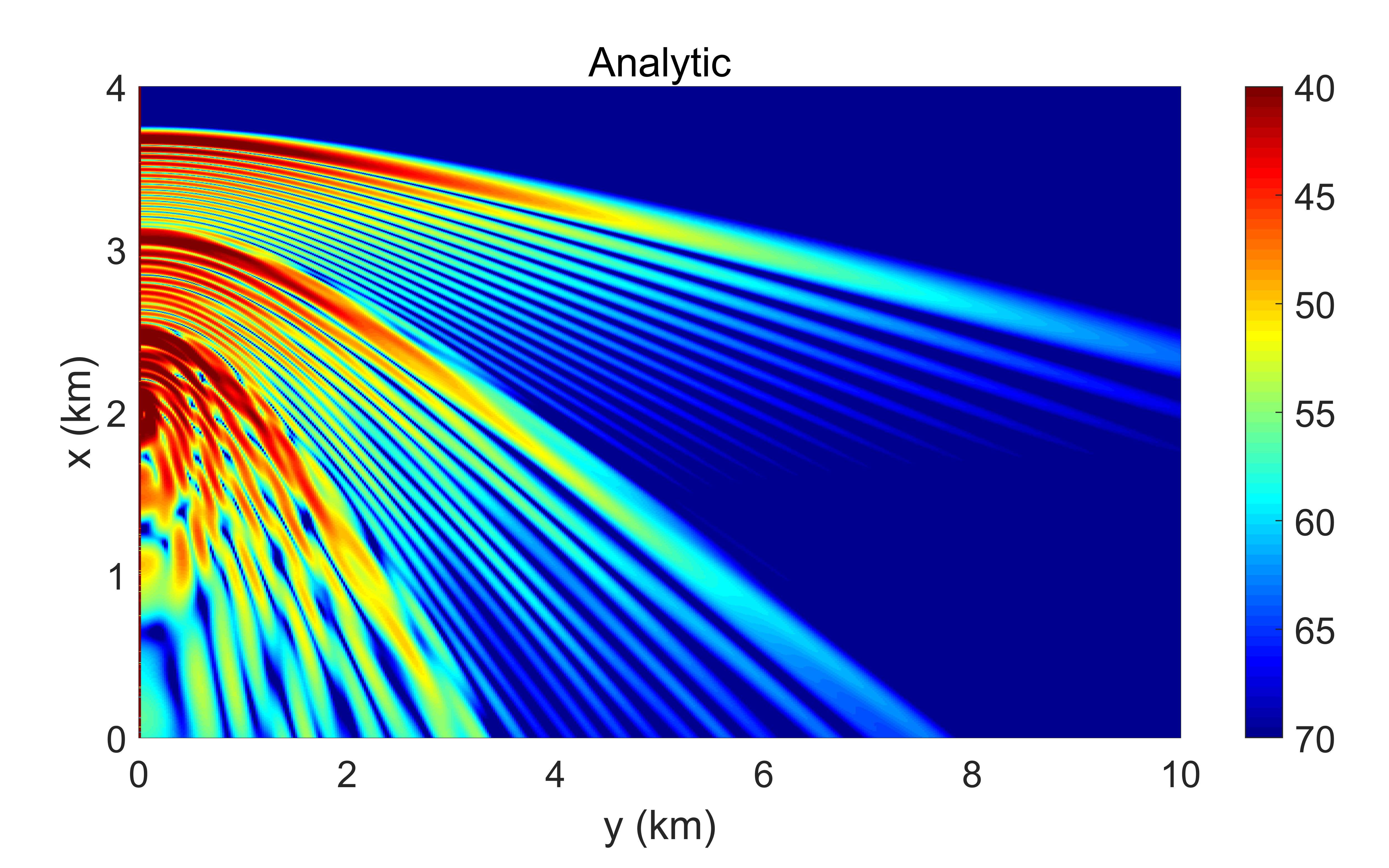}}
\subfigure[]{\includegraphics[width=6.5cm]{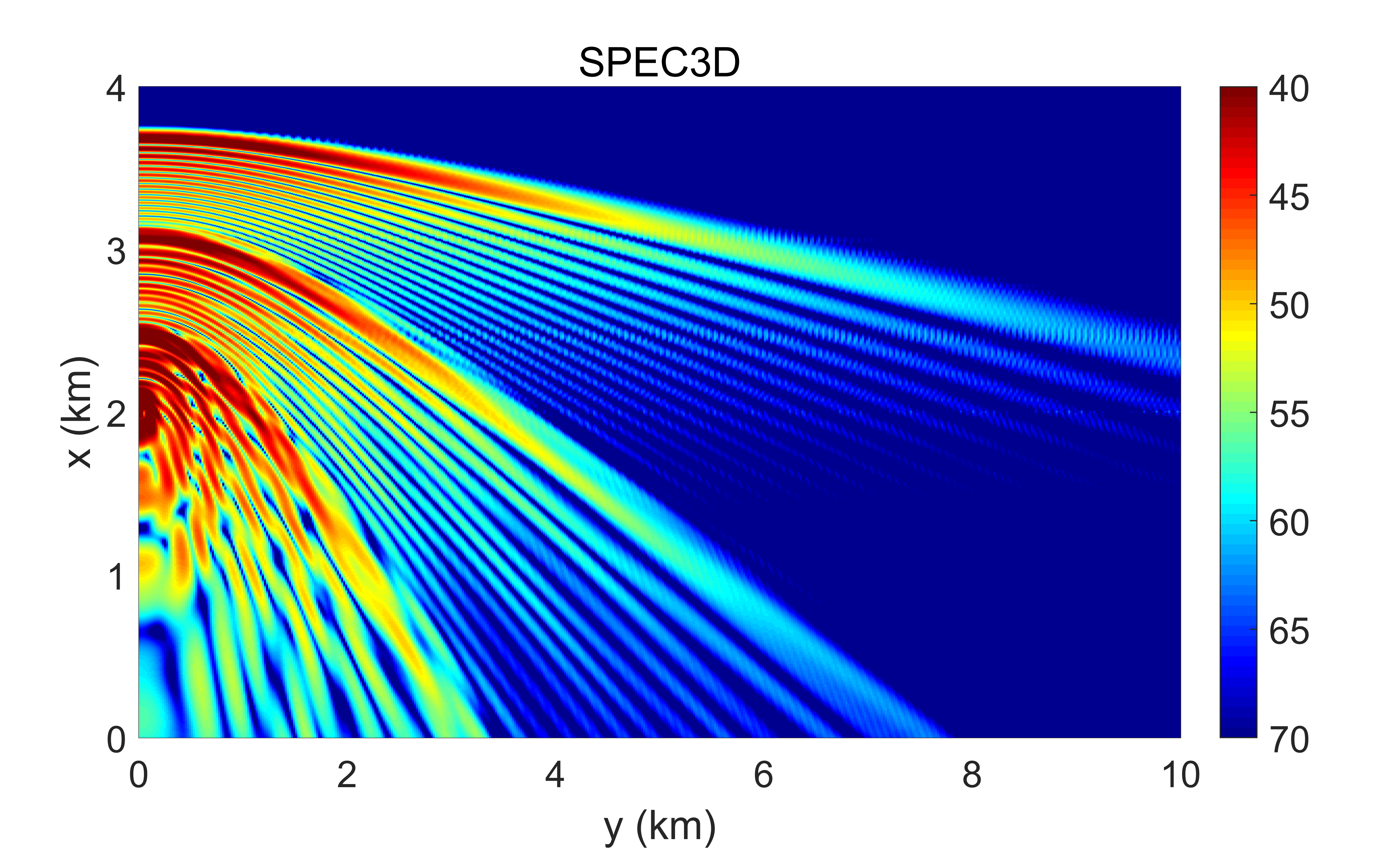}}
\subfigure[]{\includegraphics[width=6.5cm]{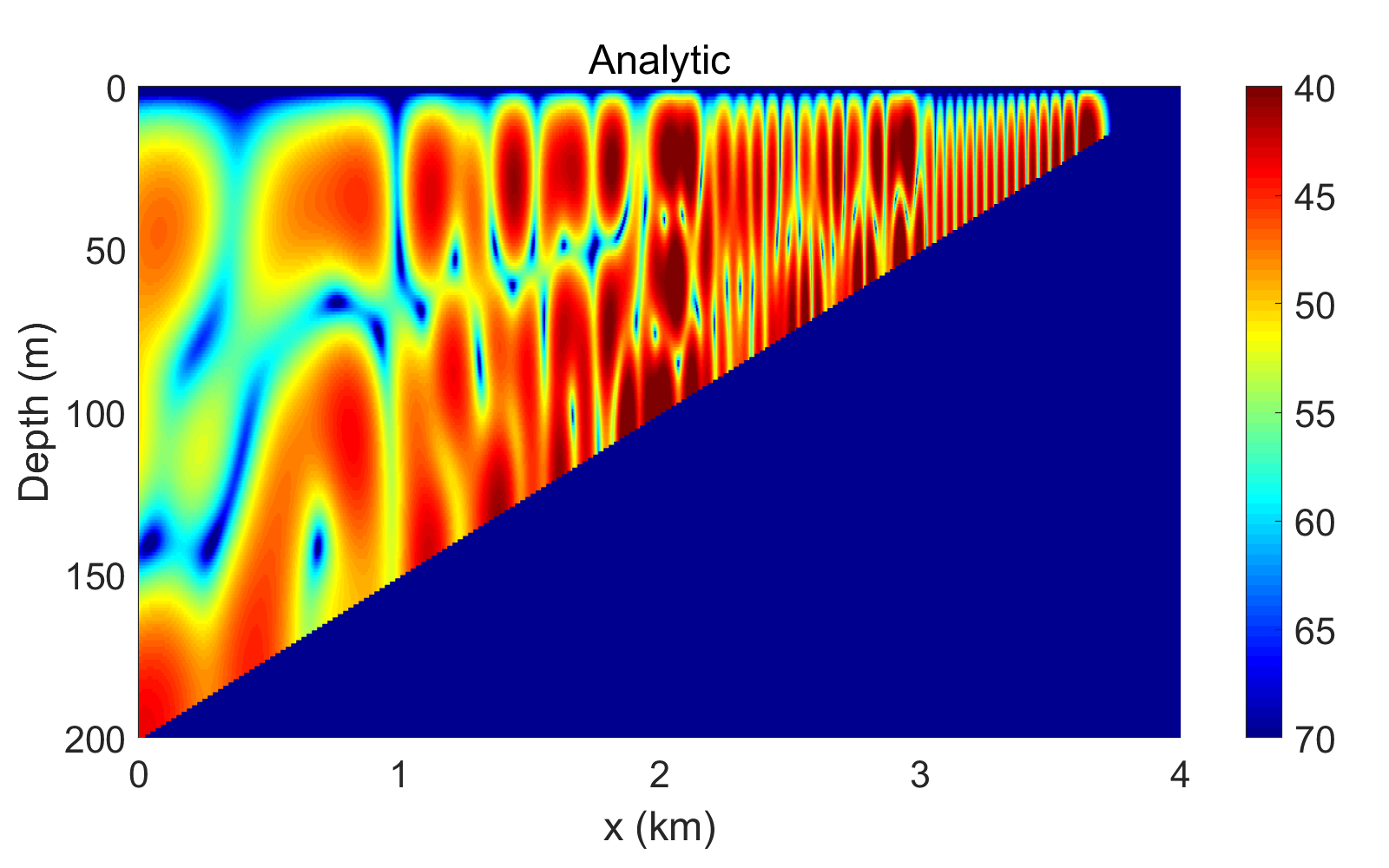}}
\subfigure[]{\includegraphics[width=6.5cm]{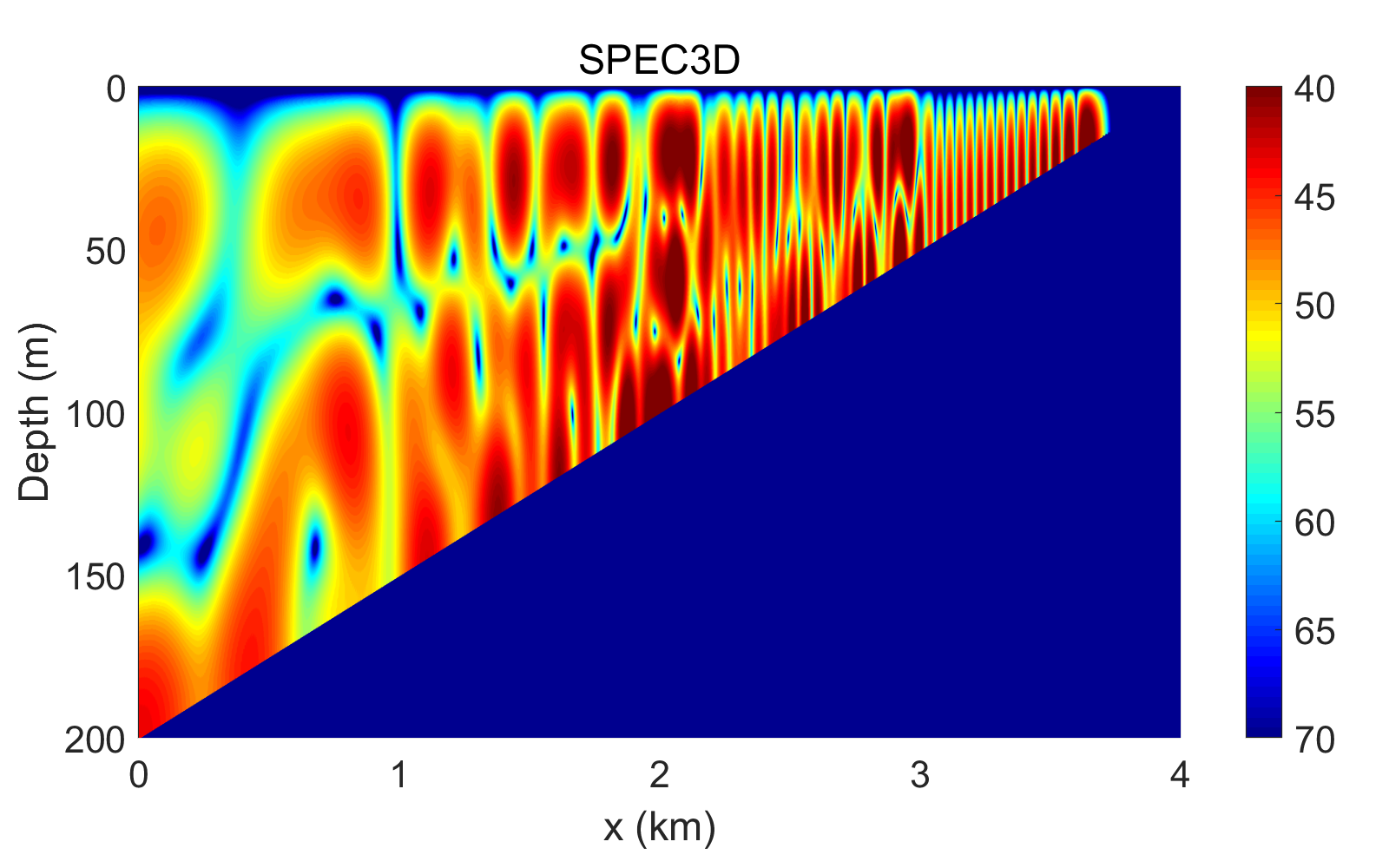}}
\subfigure[]{\includegraphics[width=6.5cm]{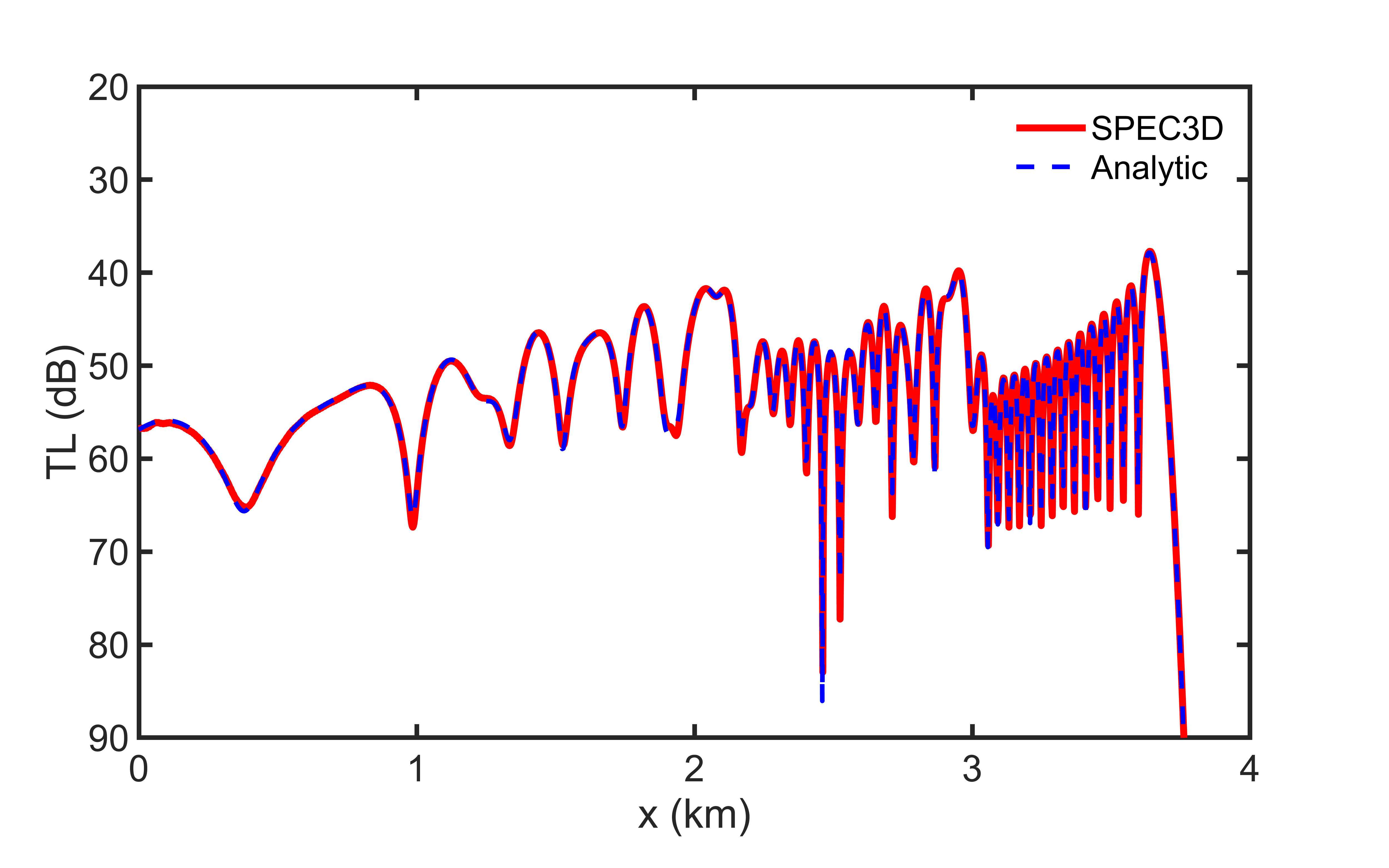}}
\subfigure[]{\includegraphics[width=6.5cm]{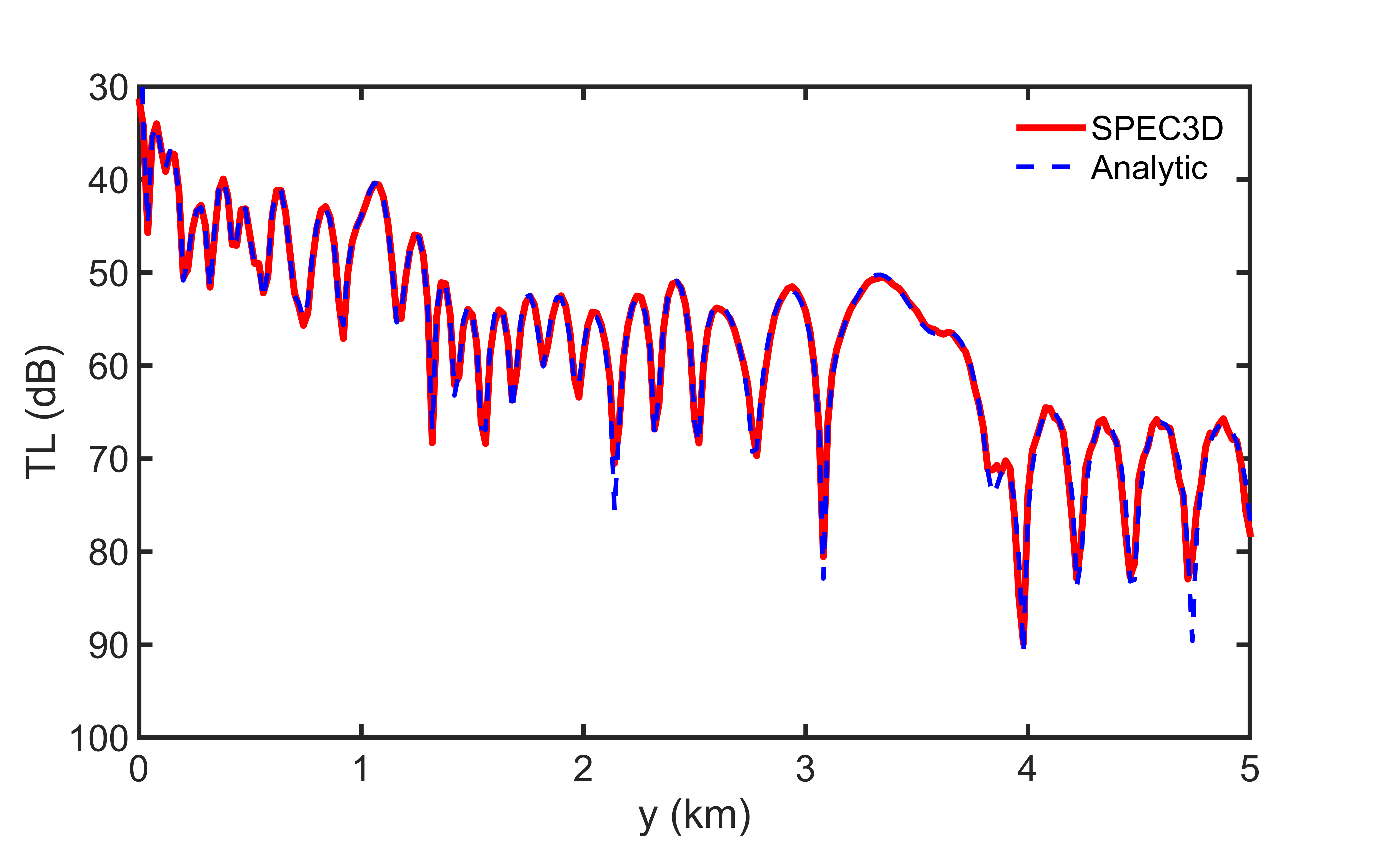}}
\caption{Sound fields of the three-dimensional wedge-shaped waveguide with a rigid bottom calculated by the analytical solution (a) and SPEC3D (b) at a depth of $z=10$ m; sound fields calculated by the analytical solution (c) and SPEC3D (d) on the $y$--plane at $y=1000$ m; TLs along the $x$-direction (e) and $y$-direction (f) at a depth of $z=10$ m.}
\label{Figure11}
\end{figure}

The analytical solution for the three-dimensional wedge-shaped waveguide with a rigid bottom has the same form as Eq.~\eqref{eq.42}, but:
\begin{equation}
	\nu_n = \left(n-\frac{1}{2}\right)\frac{\pi}{\theta_0}, \quad n=1,2,\cdots
\end{equation}
Fig.~\ref{Figure11} displays a comparison between the results of SPEC3D and the analytical solution. The number of coupled modes is $M=7$, revealing clear agreement, which further confirms the accuracy and reliability of SPEC3D.

\begin{figure}[htbp]
	\centering
	\includegraphics[height=7cm]{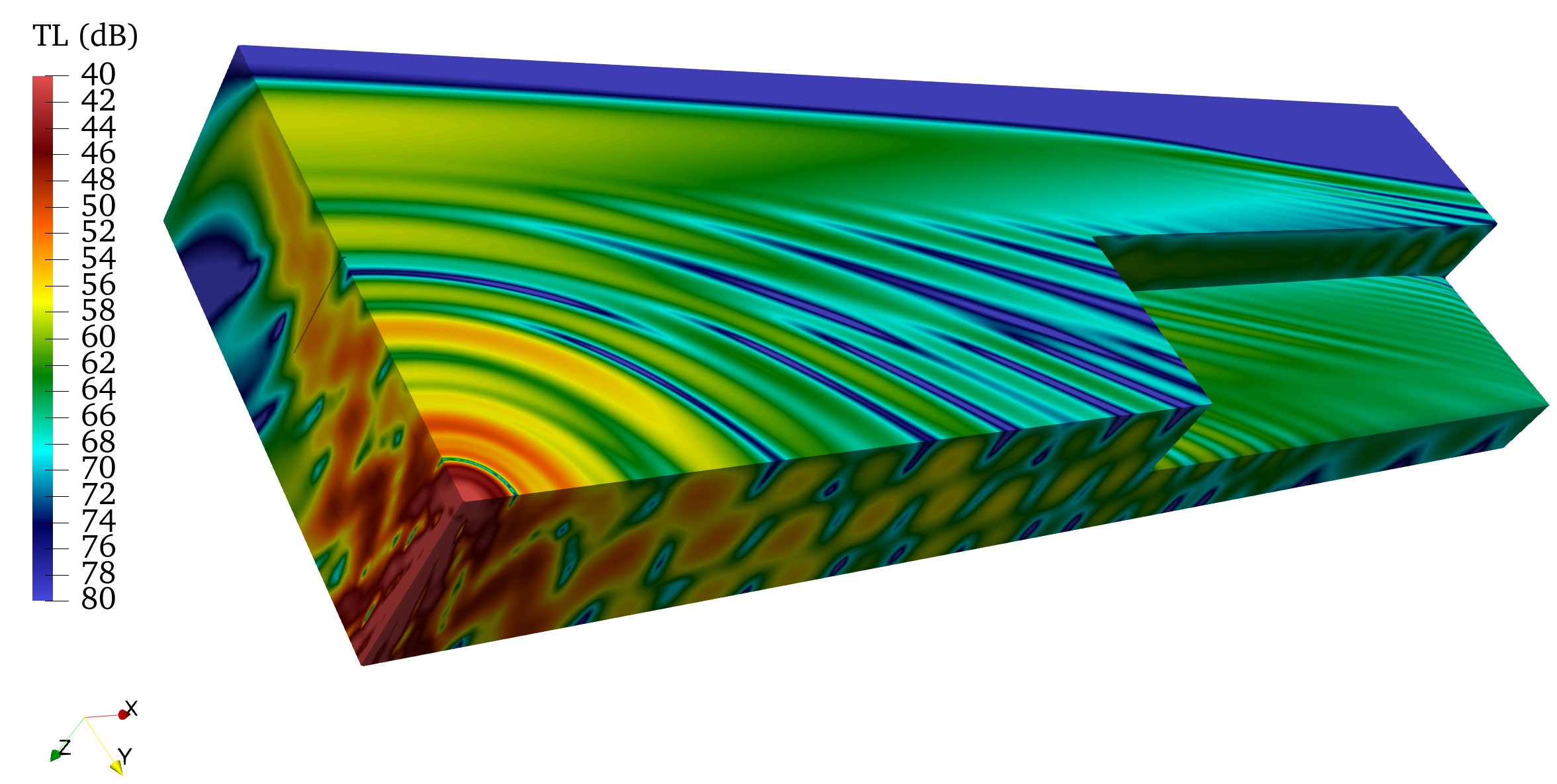}
\caption{Sound field of the three-dimensional wedge-shaped waveguide with a penetrable bottom calculated by SPEC3D.}
	\label{penetrable}
\end{figure}
\begin{figure}[htbp]
	\centering
\subfigure[]{\includegraphics[width=6.5cm]{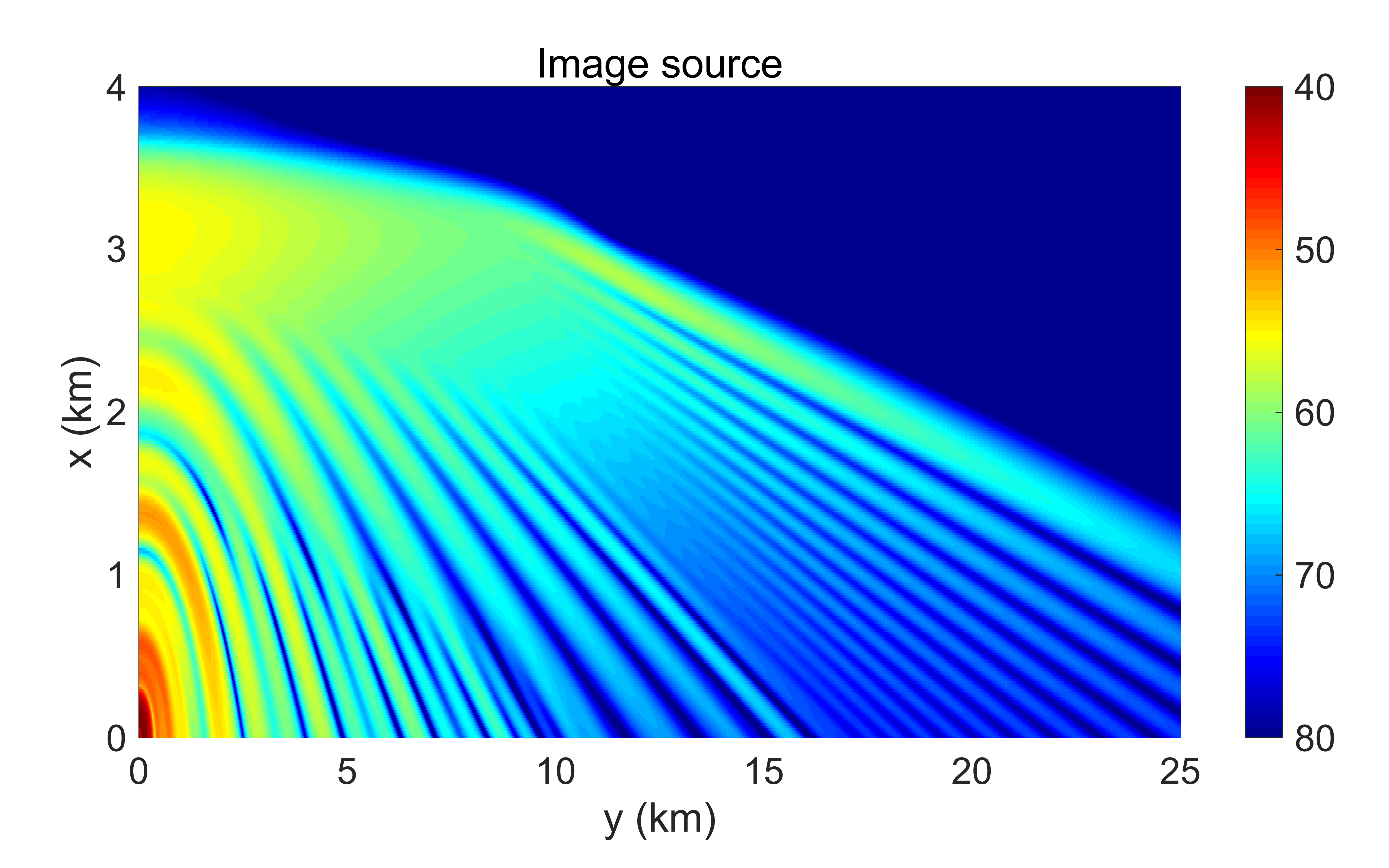}}
\subfigure[]{\includegraphics[width=6.5cm]{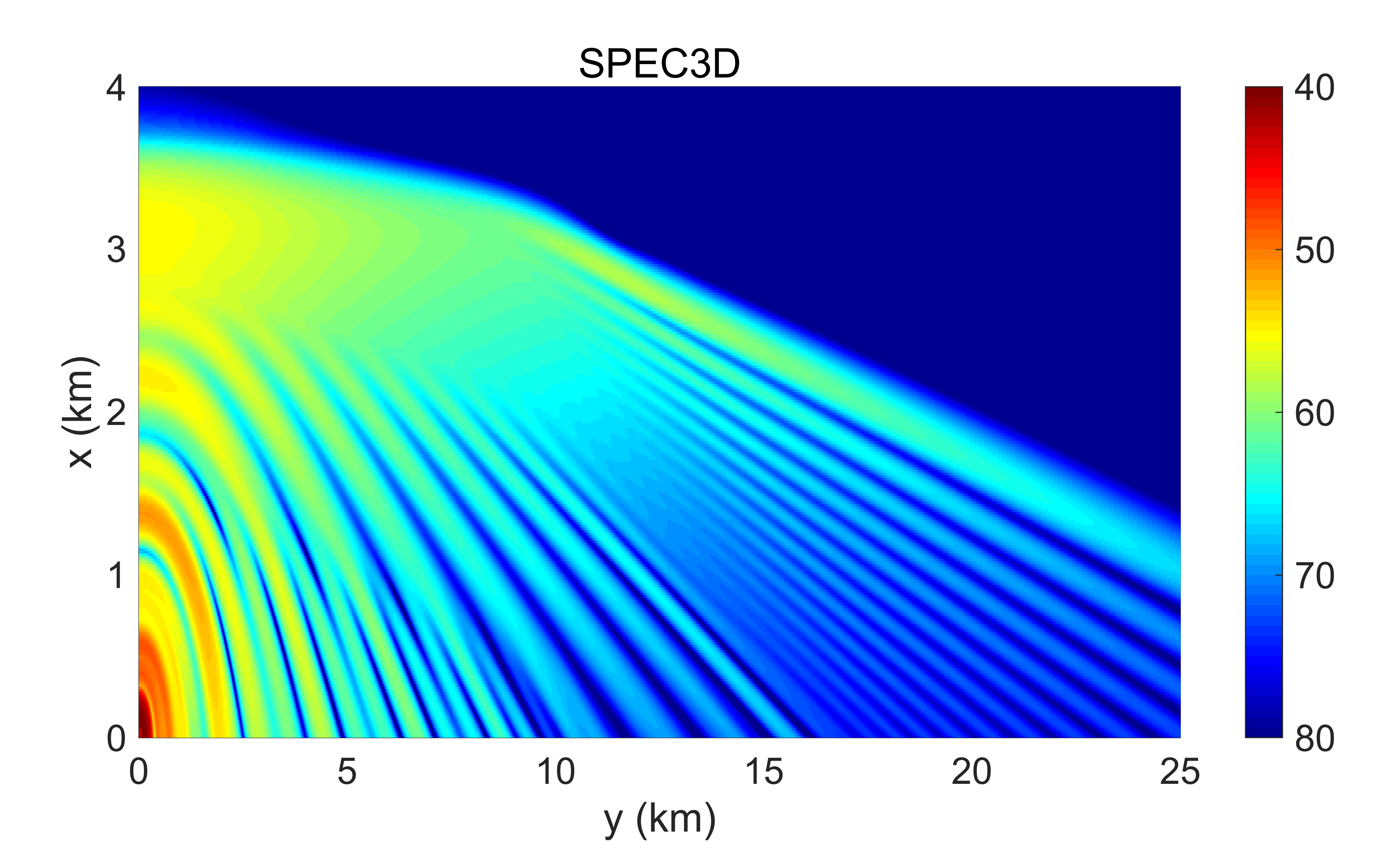}}
\subfigure[]{\includegraphics[width=6.5cm]{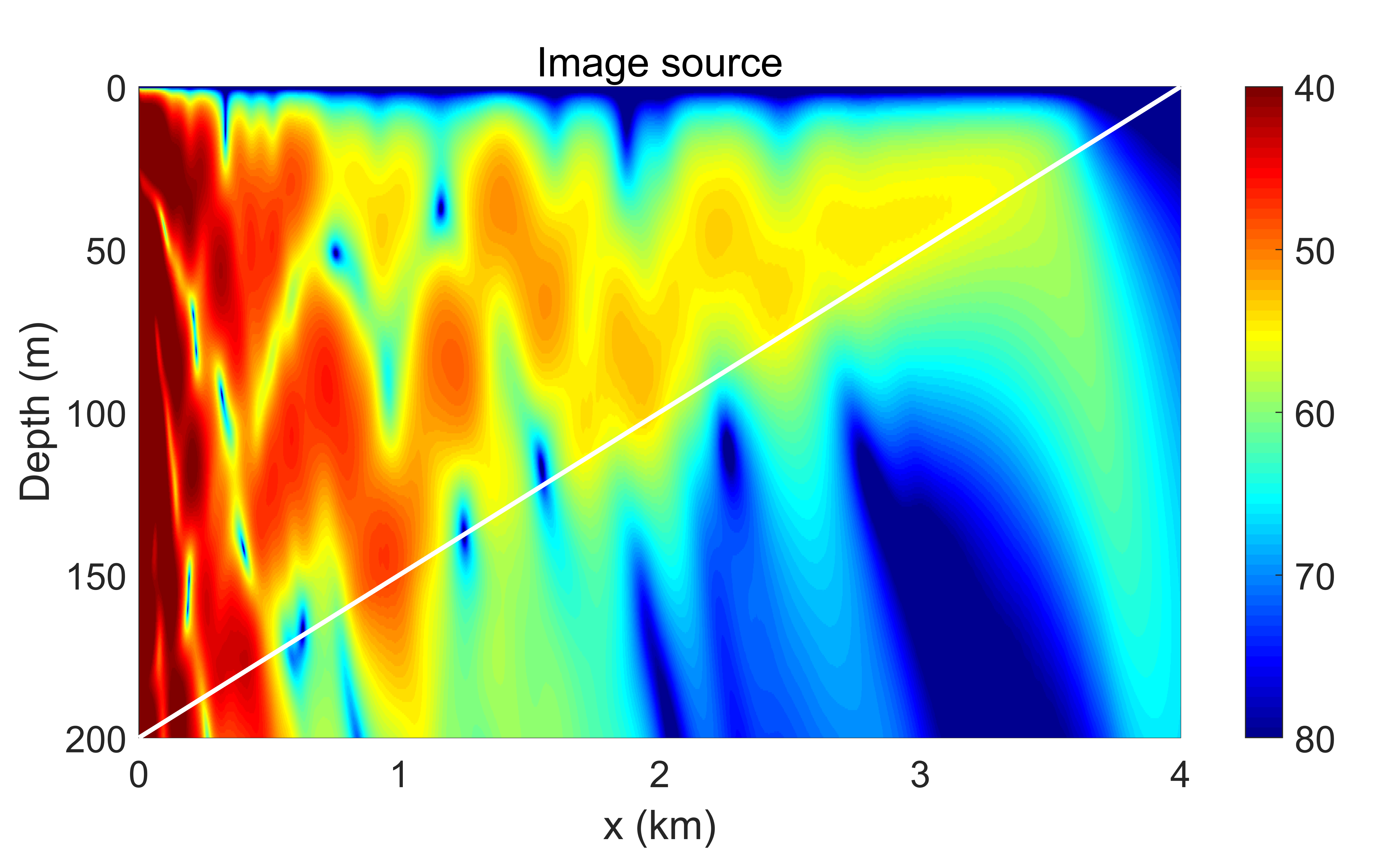}}
\subfigure[]{\includegraphics[width=6.5cm]{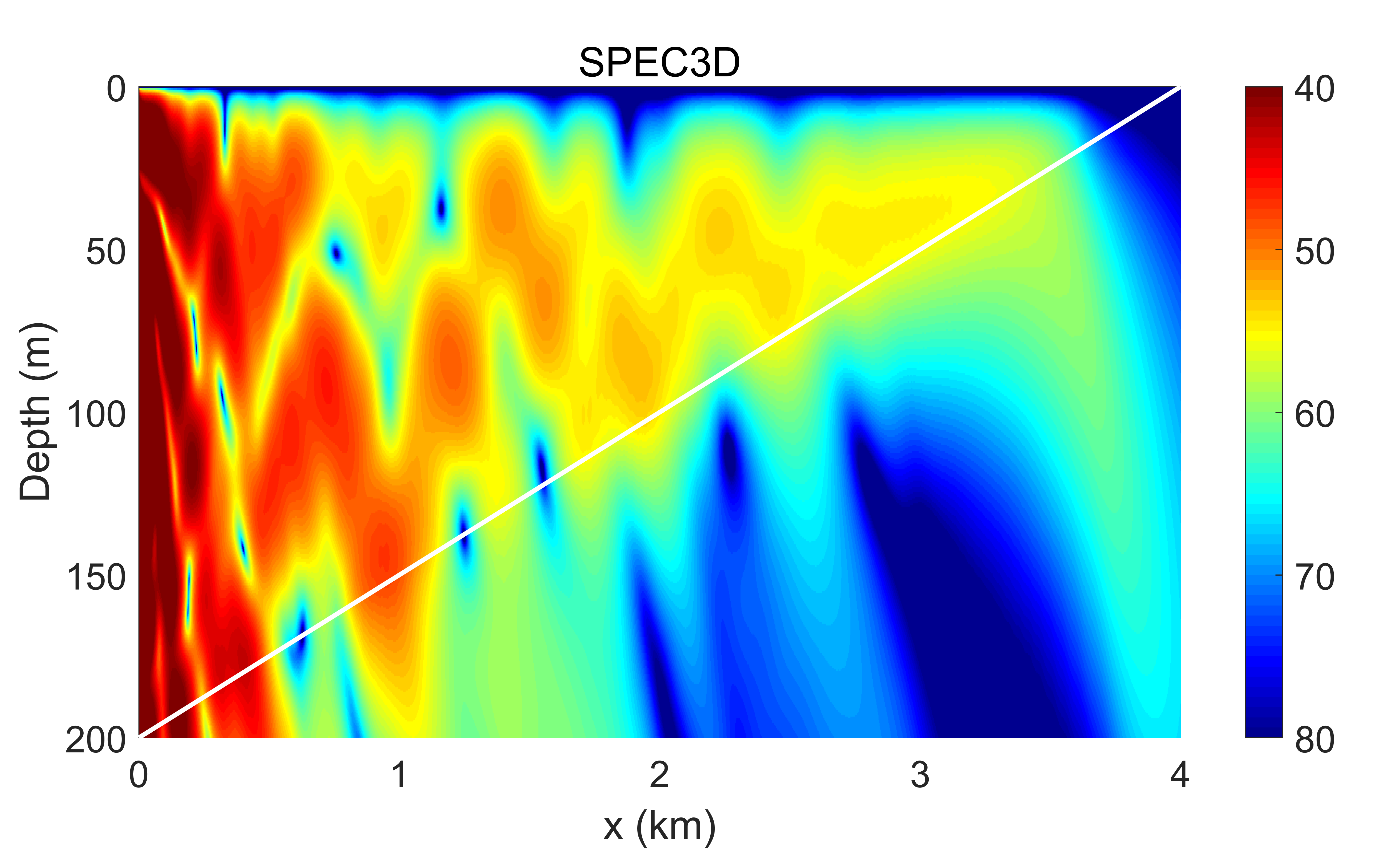}}
\subfigure[]{\includegraphics[width=6.5cm]{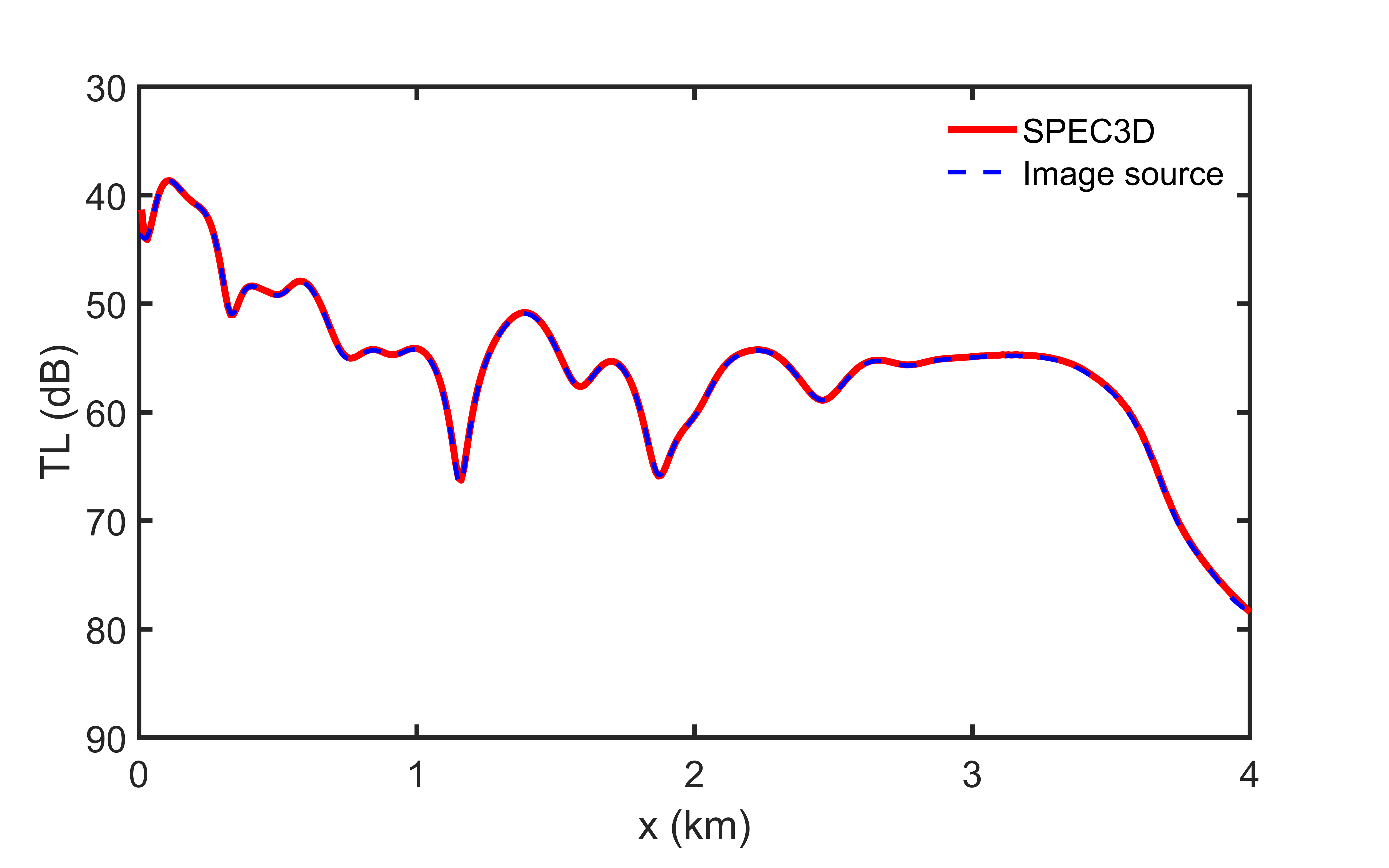}}
\subfigure[]{\includegraphics[width=6.5cm]{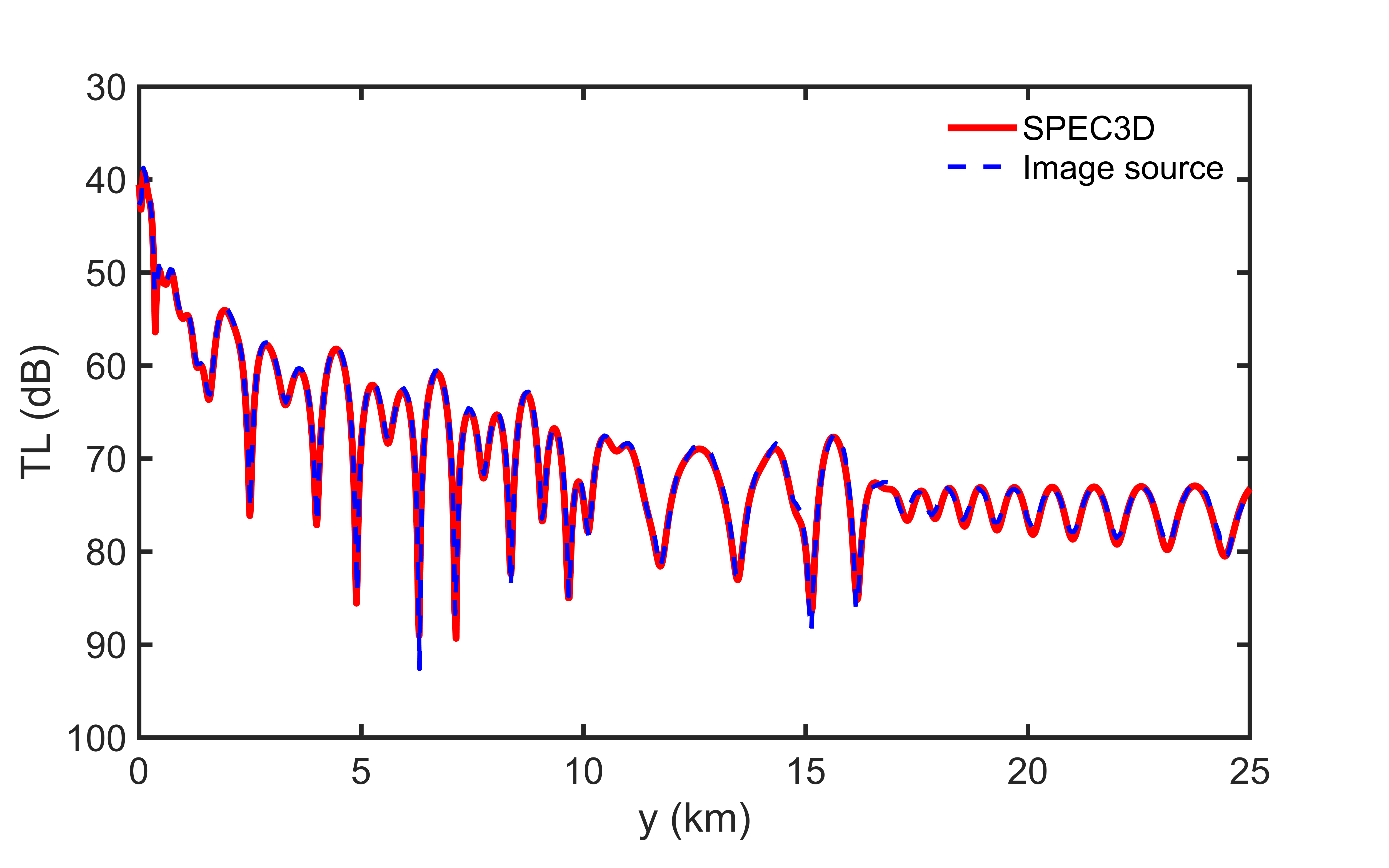}}
	\caption{Sound fields of the three-dimensional wedge-shaped waveguide with a penetrable bottom calculated by the image source method (a) and SPEC3D (b) at a depth of $z=30$ m; sound fields calculated by the image source (c) and SPEC3D (d) on the $y$--plane at $y=0$ m; TLs along the $x$-direction (e) and $y$-direction (f) at a depth of $z=30$ m.}
	\label{Figure13}
\end{figure}

In addition to the ideal bottom, the penetrable bottom is a classic configuration of a wedge-shaped waveguide. Fig.~\ref{penetrable} shows the sound field simulated by SPEC3D. The sound source is located at $x_\mathrm{s}=0$, $z_\mathrm{s}=100$ m, and the other configurations are similar to the above example. Since there is a homogeneous half-space under the penetrable bottom, SPEC3D is truncated from $H=400$ m in the simulation, and a total of $M=13$ modes are involved in the coupling. For a wedge-shaped waveguide with a penetrable bottom, there is only an exact solution constructed by the image source method. Deane and Buckingham proposed the image source solution to calculate the three-dimensional acoustic field in wedge-shaped sea water \cite{Deane1993}. Yang et al. further improved the above three-dimensional image source solution, enabling it to calculate not only the sound field in seawater but also the sound field in the sediment \cite{Yangcm2015b}. Fig.~\ref{Figure13} shows a comparison between slices of the sound field calculated by SPEC3D and the image source solution. Both Figs.~\ref{penetrable} and \ref{Figure13} fully demonstrate the capability of SPEC3D in simulating this example.

\subsection{Acoustic half-space}
Here, we consider an example involving an acoustic half-space, as shown in Fig.~\ref{Halfspace}, where the seafloor topography is also $x$-independent. In this case, the frequency of the sound source is $f=50$ Hz, and the source is located at $x_\mathrm{s}=0$ m, $z_\mathrm{s}=36$ m, $h=50$ m, and $H=100$ m. The number of coupled modes is $M=6$, and the number of discrete points in the $k_y$--domain is $N_q=512$. Due to the horizontal independence of the marine environment, $J$ is taken as 2. Fig.~\ref{Figure15} approximately shows the three-dimensional sound field structure of the waveguide.
\begin{figure}[htbp]
\centering
\begin{tikzpicture}[node distance=2cm,scale=0.7,domain=-5:5]
			\fill[cyan,opacity=0.6] (2,0)--(12,0)--(14,-2.5)--(4,-2.5)--(4,-5.5)--(2,-3)--cycle;
			\fill[cyan,opacity=0.6] (4,-5.5)--(14,-5.5)--(14,-2.5)--(4,-2.5)-- cycle; 		
			\fill[orange,opacity=0.6] (2,-3)--(4,-5.5)--(4,-8.5)--(2,-6)--cycle;	
			\fill[orange,opacity=0.6]  (4,-5.5)--(14,-5.5)--(14,-8.5)--(4,-8.5)--cycle; 
			\fill[brown] (2,-6)--(4,-8.5)--(14,-8.5)--(14,-9.5)--(4,-9.5)--(2,-7)--cycle;
			\draw[thick, ->](3,-1.25)--(14,-1.25) node[right]{$x$};
			\draw[thick, ->](3,-1.25)--(3,-9.5) node[below]{$z$};	    		
			\draw[very thick](1.98,0)--(12.02,0);
			\draw[very thick](4,-2.5)--(14.02,-2.5);
			\draw[very thick](2,0)--(4,-2.5);
			\draw[thick, ->](2,0)--(5,-3.75) node[below]{$y$};	
			\draw[very thick](12,0)--(14,-2.5);
			\draw[very thick](2,0.02)--(2,-7);
			\draw[very thick](4,-2.5)--(4,-9.5);
			\draw[very thick](2,-3)--(4,-5.5);
			
			\draw[dashed, very thick](12,0)--(12,-6.0);
			\draw[very thick](14,-2.5)--(14,-9.5);
			\draw[dashed, very thick](12,-3.0)--(14,-5.5);
			\draw[very thick] (4,-5.5)--(14,-5.5);
			\draw[dashed, very thick](2,-3)--(12,-3);
			\draw[dashed, very thick](12,-6.0)--(14,-8.5);
			\draw[dashed, very thick](2,-6.0)--(12,-6.0);
			
			\draw[very thick](2,-6)--(4,-8.5);	
			\draw[very thick](4,-8.5)--(14,-8.5);			
			\filldraw [red] (3,-3.5) circle [radius=2.5pt];
			\node at (2.7,-1.3){$o$};				
		\end{tikzpicture}
\caption{Schematic diagram of the three-dimensional horizontally independent waveguide with an acoustic half-space.}
\label{Halfspace}
\end{figure}
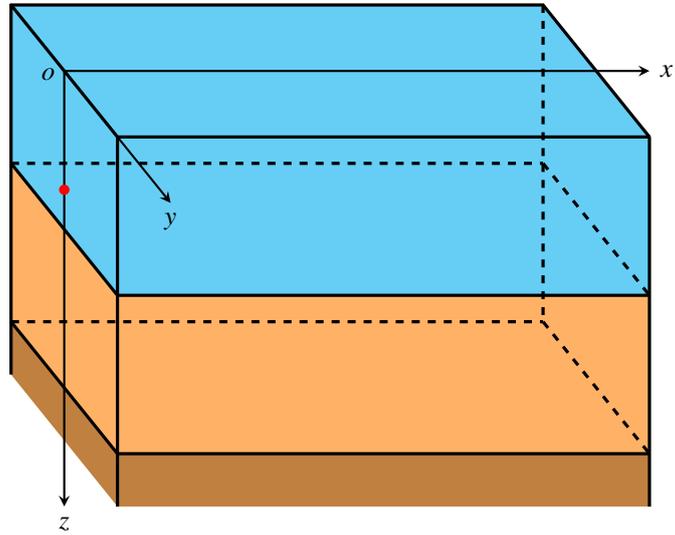
\begin{figure}[htbp]
\centering\includegraphics[height=8cm]{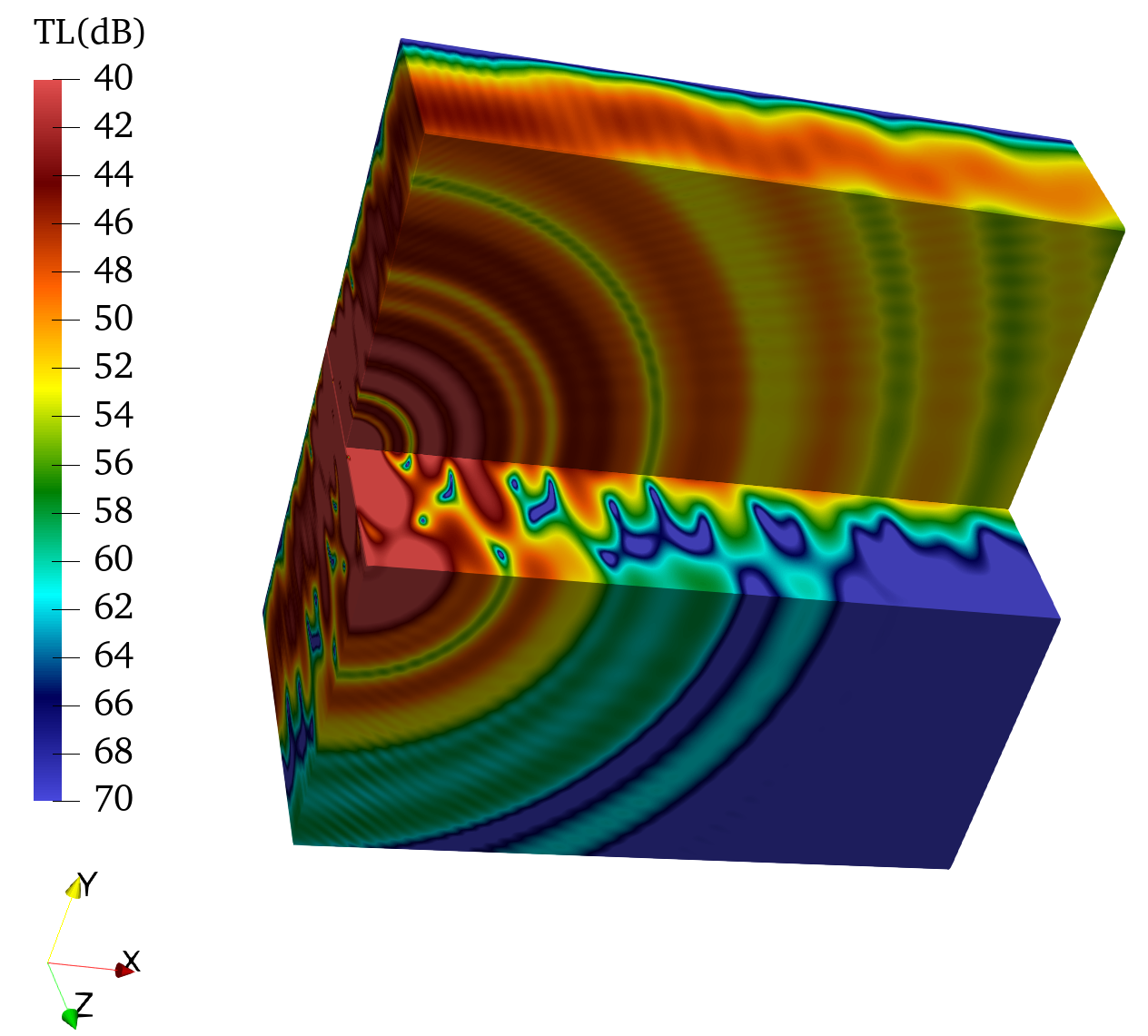}
\caption{Sound field of the three-dimensional horizontally independent waveguide with an acoustic half-space calculated by SPEC3D.}
\label{Figure15}
\end{figure}
\begin{figure}[htbp]
	\centering
\subfigure[]{\includegraphics[width=6.5cm]{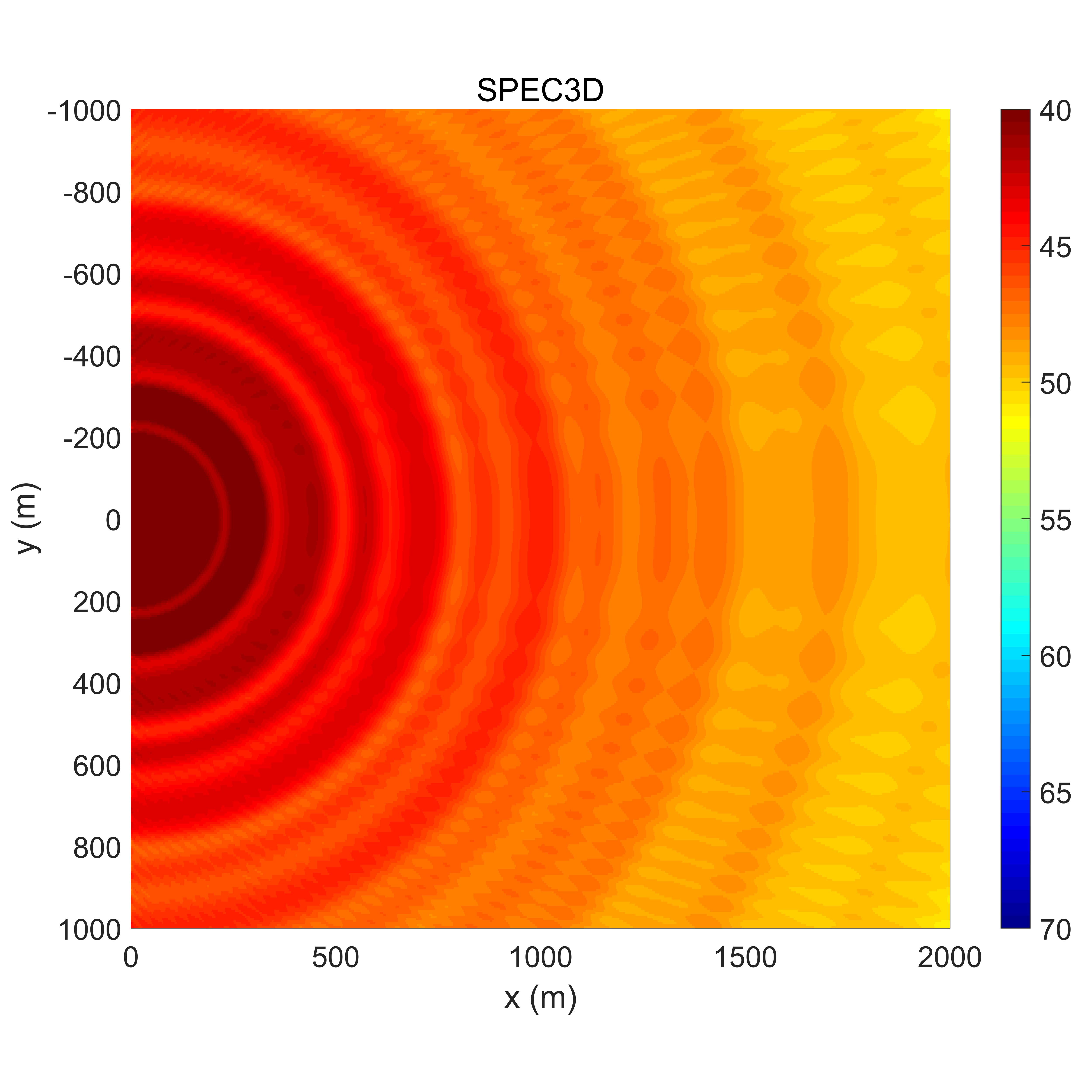}\label{Figure16a}}
\subfigure[]{\includegraphics[width=6.5cm]{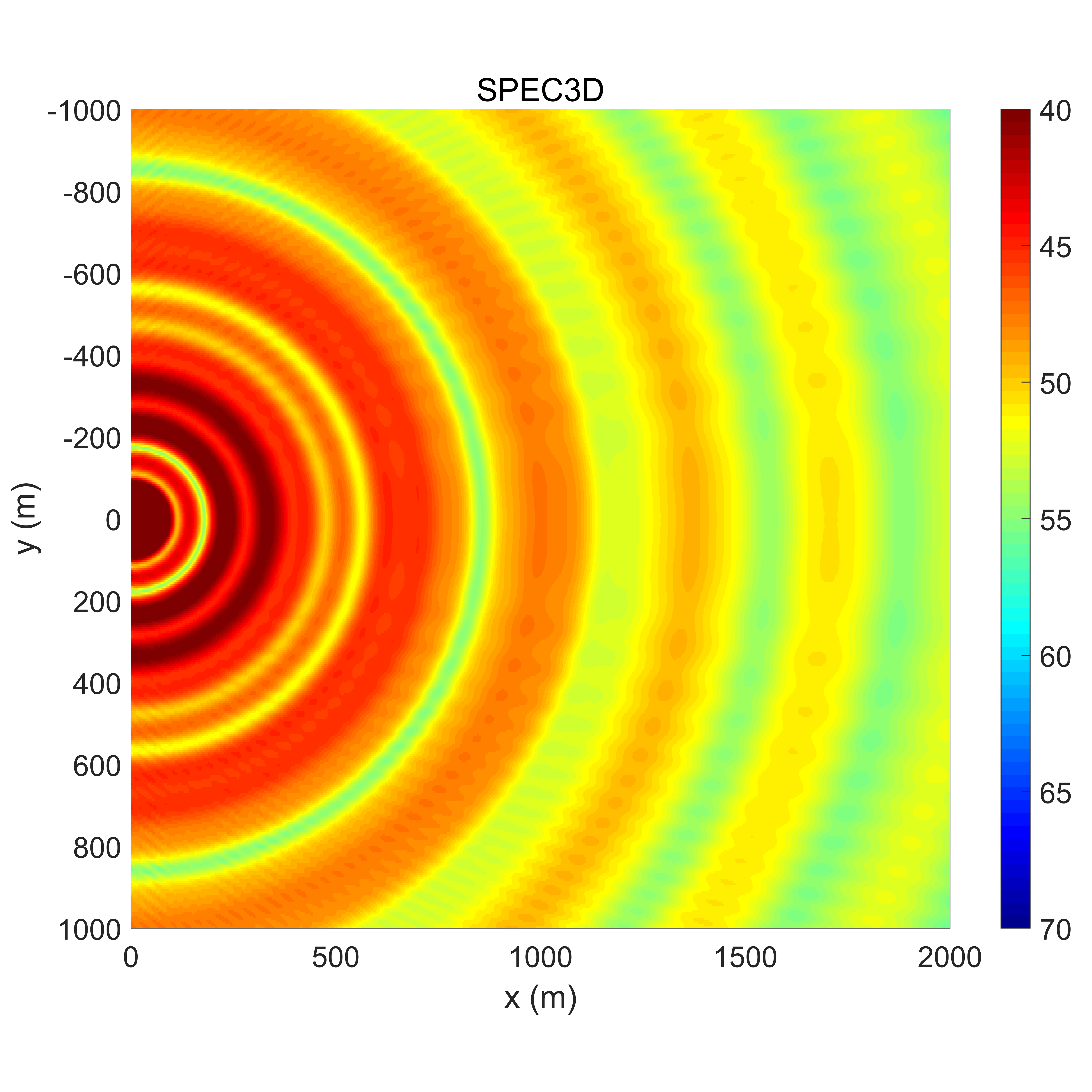}\label{Figure16b}}
\subfigure[]{\includegraphics[width=6.5cm]{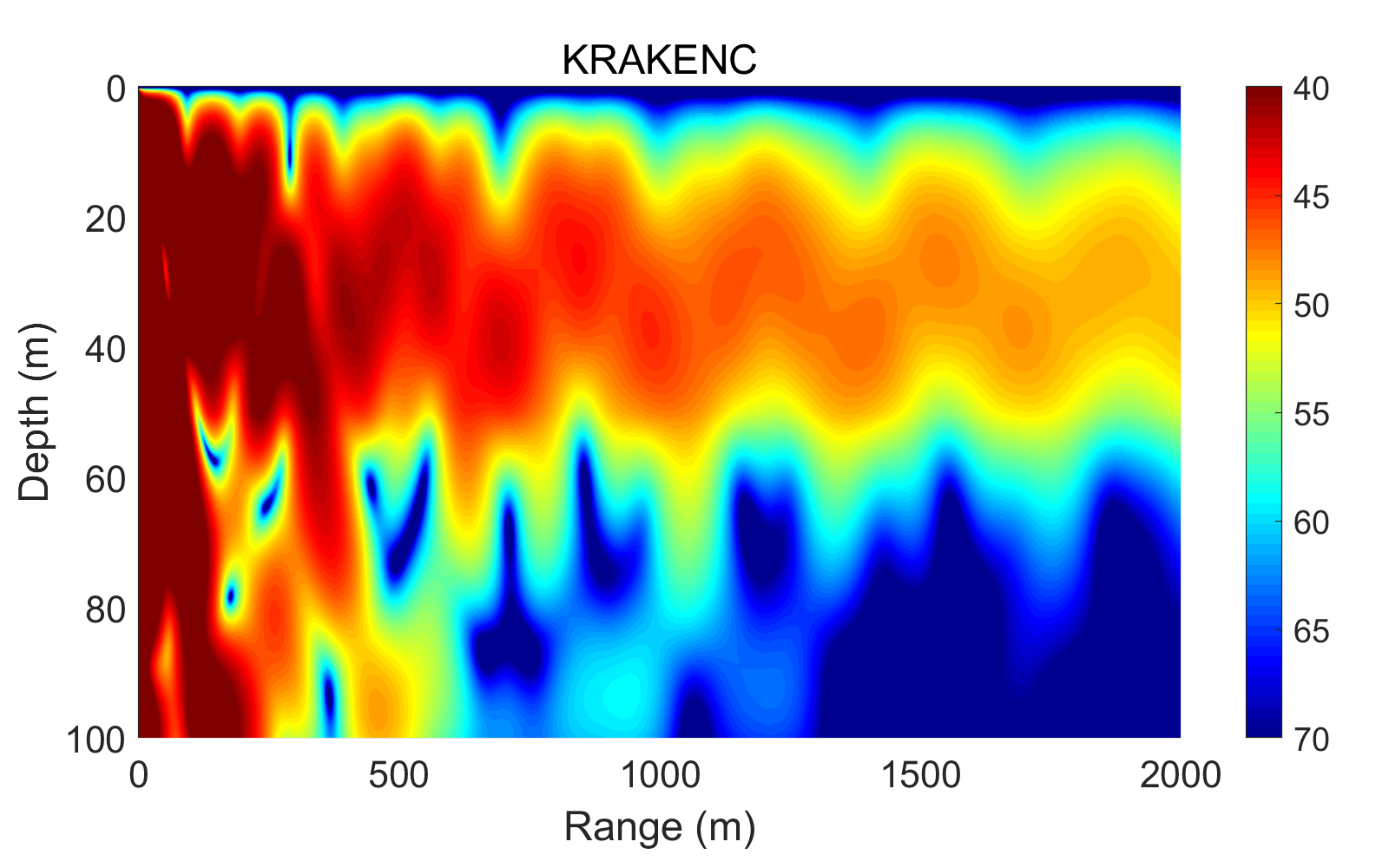}\label{Figure16c}}
\subfigure[]{\includegraphics[width=6.5cm]{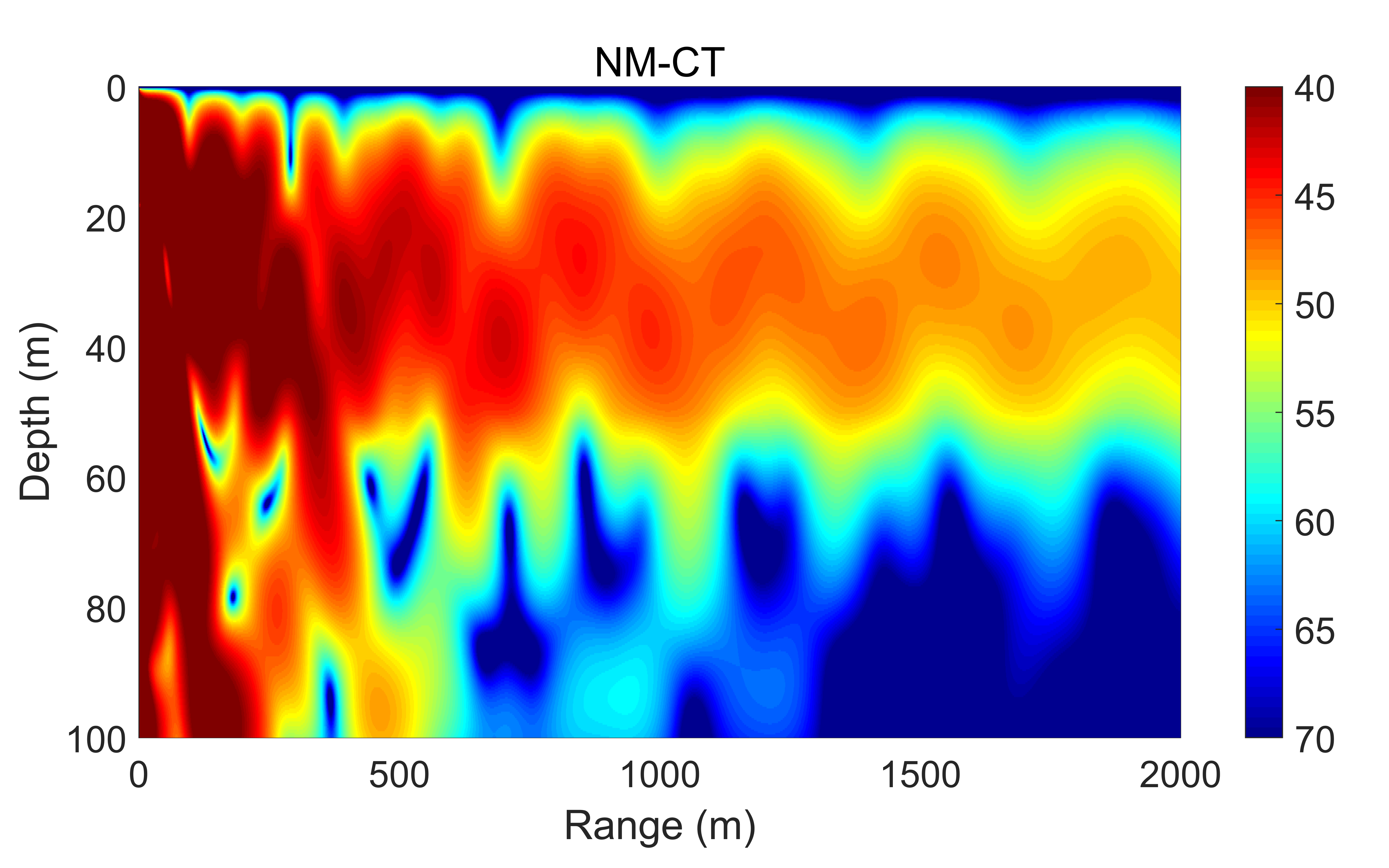}\label{Figure16d}}
\subfigure[]{\includegraphics[width=6.5cm]{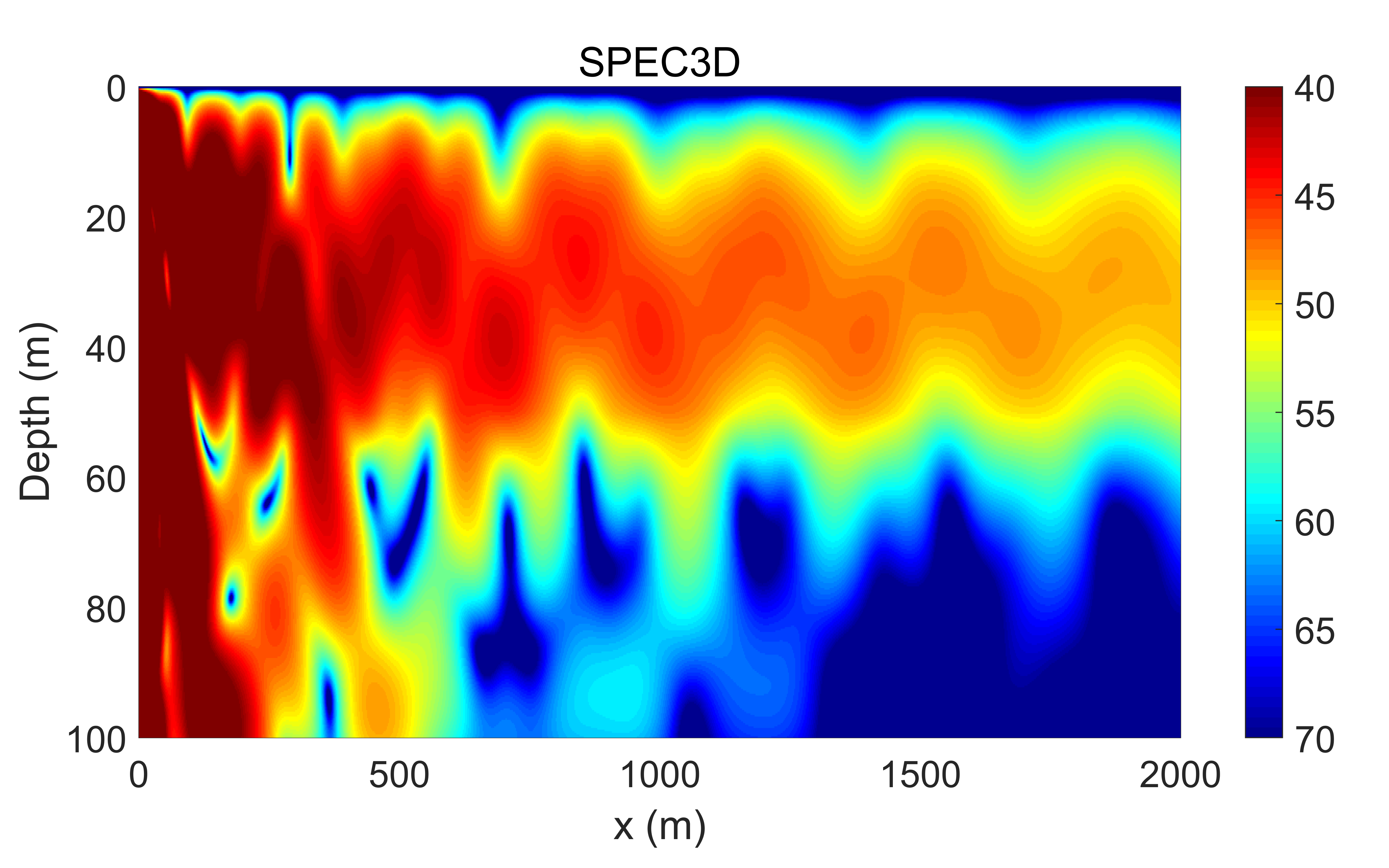}\label{Figure16e}}
\subfigure[]{\includegraphics[width=6.5cm]{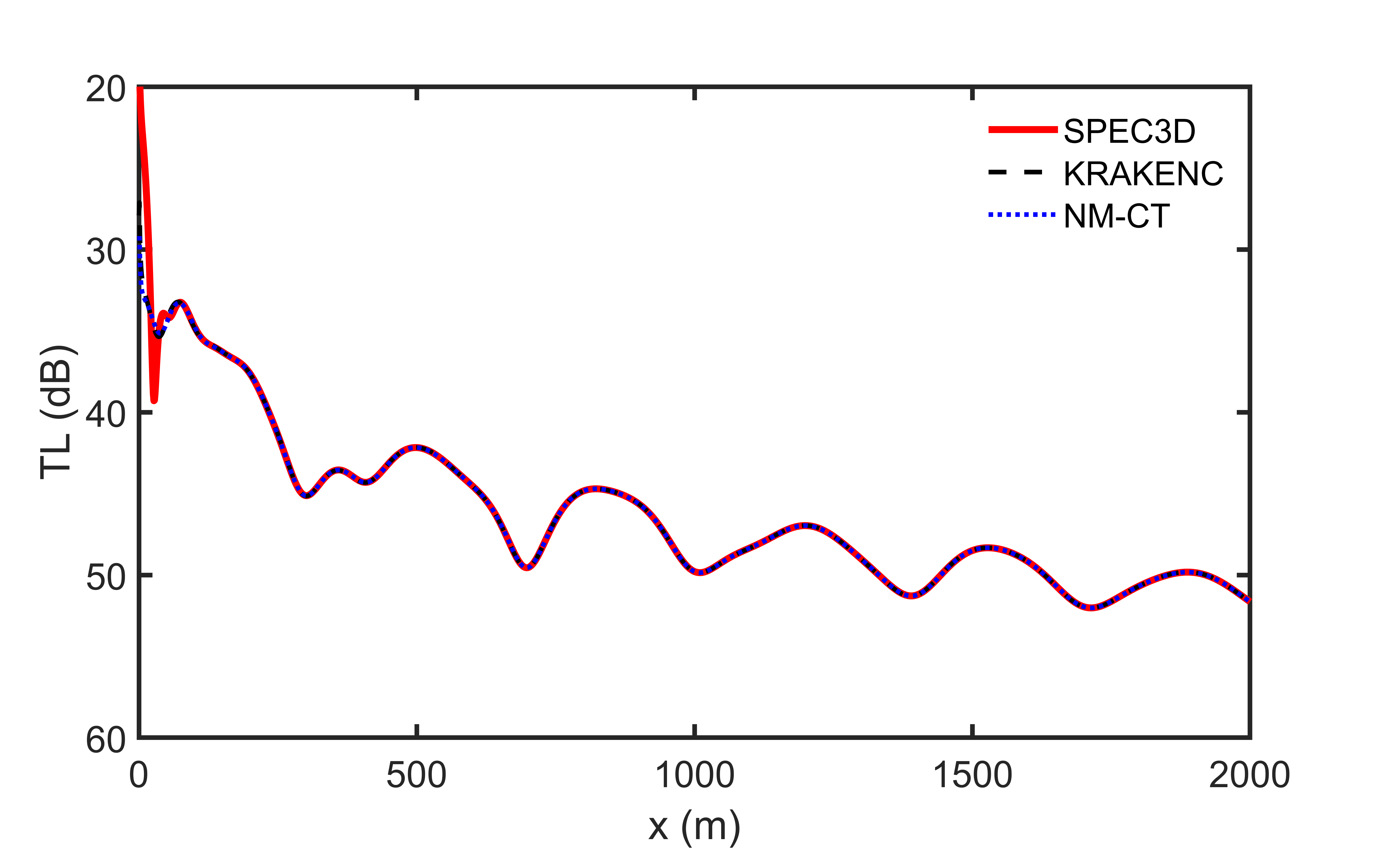}\label{Figure16f}}
\caption{Sound fields of the three-dimensional horizontally independent waveguide with an acoustic half-space calculated by SPEC3D at depths of $z=36$ m (a) and $z=50$ m (b); sound fields calculated by KRAKENC (c), NM-CT (d) and SPEC3D (e) on the $y$--plane at $y=0$ m; TLs on the $y=0$ m-plane along the $x$-direction at a depth of $z=20$ m (f).}
\end{figure}

Figs.~\ref{Figure16a} and \ref{Figure16b} present cross-sections of the sound field calculated by SPEC3D at depths of 36 m and 50 m, respectively. The sound field still clearly exhibits the symmetry of horizontally independent terrain, but a certain degree of error remains at different azimuth angles. Figs.~\ref{Figure16c} and \ref{Figure16d} display the two-dimensional sound fields calculated by KRAKENC \cite{Kraken2001} and NM-CT \cite{Tuhw2021a,NM-CT}, respectively, and Fig.~\ref{Figure16e} plots a slice of the sound field computed by SPEC3D at $y=0$ m; the simulation results of all three programs exhibit good agreement. Fig.~\ref{Figure16f} shows the line graph of TL in Figs.~\ref{Figure16c} to \ref{Figure16e}. The results of KRAKENC, NM-CT and SPEC3D are almost identical on the whole, although there are slight differences in the near field.

\subsection{Ridge-shaped waveguide}
Consider a ridge-shaped waveguide, as shown in Fig.~\ref{Ridge}. In the $xoz$--plane, the expression for the seabed topography is given by the following equation:
\begin{equation}
	h(x)= \begin{cases}50-25 \cos \left[ \frac{\pi(x-500)}{200}\right], & 300<x<700, \\ 75, & \text { elsewhere, }\end{cases}
\end{equation}
The frequency of the sound source is $f=25$ Hz, and the source is located at $x_\mathrm{s}=0$ m and $z_\mathrm{s}=25$ m. The ocean depth $H$ is 100 m. Fig.~\ref{Figure18} shows the three-dimensional sound field structure of the waveguide calculated by SPEC3D. Due to the flat terrain in both the near field and the far field, the sound field still exhibits a certain symmetry. The uplift of the ridge at 300--700 m induces prominent changes in the sound field within this region.

\begin{figure}[htbp]
	\centering
\begin{tikzpicture}[node distance=2cm,scale=0.8,samples=500,domain=7:10.5]
		
		\fill[orange,opacity=0.6](2,-3)--plot[domain=7:10.5,smooth](\x-2,{-1.9-1.1*cos(103*(\x))})--(12,-3)--(14,-5.5)--plot[domain=10.5:7,smooth](\x,{-4.4-1.1*cos(103*(\x))})--(4,-5.5)--cycle;
		\fill[cyan,opacity=0.6]plot[domain=8.57:9.98,smooth](\x,{-4.4-1.1*cos(103*(\x))})--cycle;\fill[orange,opacity=0.6]plot[domain=8.57:9.98,smooth](\x,{-4.4-1.1*cos(103*(\x))})--cycle;	
		\fill[cyan,opacity=0.6] (2,0)--(12,0)--(14,-2.5)--(4,-2.5)--(4,-5.5)--(2,-3)--cycle;
		\fill[cyan,opacity=0.6]  (4,-2.5)--(4,-5.5)--(7,-5.5)--plot(\x,{-4.4-1.1*cos(103*(\x))})--(10.5,-5.5)--(14,-5.5)--(14,-2.5)-- cycle;
		\fill[orange,opacity=0.6] (2,-3)--(4,-5.5)--(4,-6.5)--(2,-4)--cycle;
		\fill[orange,opacity=0.6]  (4,-5.5)--(7,-5.5)--plot(\x,{-4.4-1.1*cos(103*(\x))})--(10.5,-5.5)--(14,-5.5)--(14,-6.5)--(4,-6.5)--cycle;
		\fill[orange,opacity=0.6]plot[domain=9.1:9.98,smooth](\x-2,{-1.9-1.1*cos(103*(\x))}) --plot[domain=8.58:9.1,smooth](\x,{-4.4-1.1*cos(103*(\x))})--(8.7,-3)--(7.1,-1)--cycle;
		\fill[cyan,opacity=0.6]plot[domain=9.1:9.98,smooth](\x-2,{-1.9-1.1*cos(103*(\x))}) --plot[domain=8.58:9.1,smooth](\x,{-4.4-1.1*cos(103*(\x))})--(8.7,-3)--(7.1,-1)--cycle;
		\fill[brown] (2,-4)--(4,-6.5)--(14,-6.5)--(14,-7.5)--(4,-7.5)--(2,-5)--cycle;
		\draw[dashed](9.1,-3.5)--(7.1,-1);
		\draw[dashed](8.58,-3.35)--(9.98,-5.1);
		\draw[thick, ->](3,-1.25)--(14,-1.25) node[right]{$x$};
		\draw[thick, ->](3,-1.25)--(3,-7) node[below]{$z$};	    		
		\draw[very thick](1.98,0)--(12.02,0);
		\draw[very thick](4,-2.5)--(14.02,-2.5);
		\draw[very thick](2,0)--(4,-2.5);
		\draw[thick, ->](2,0)--(4.6,-3.25) node[below]{$y$};	
		\draw[very thick](12,0)--(14,-2.5);
		\draw[very thick](2,0.02)--(2,-5);
		\draw[very thick](4,-2.5)--(4,-7.5);
		\draw[dashed, very thick](12,0)--(12,-4);
		\draw[very thick](14,-2.5)--(14,-7.5);
		\draw[dashed, very thick](12,-3)--(14,-5.5);
		\draw[dashed, very thick](2,-4)--(12,-4);
		\draw[dashed, very thick](12,-4)--(14,-6.5);
		
		\draw[very thick](2,-4)--(4,-6.5);	
		\draw[very thick](4,-6.5)--(14,-6.5);			
		\filldraw [red] (3,-2.7) circle [radius=2.5pt];
		\node at (2.7,-1.3){$o$};
		
		\draw[very thick](2,-3)--(4,-5.5);
		\draw[very thick](4,-5.5)--(7,-5.5);
		\draw[very thick, smooth] plot(\x,{-4.4-1.1*cos(103*(\x))});
		\draw[very thick](10.48,-5.5)--(14,-5.5);
		\draw[dashed, very thick](2,-3)--(5,-3);
		\draw[dashed,very thick, smooth] plot(\x-2,{-1.9-1.1*cos(103*(\x))});
		\draw[dashed, very thick](8.48,-3)--(12,-3);		
	\end{tikzpicture}
\caption{Schematic diagram of the three-dimensional ridge-shaped waveguide.}
	\label{Ridge}
\end{figure}
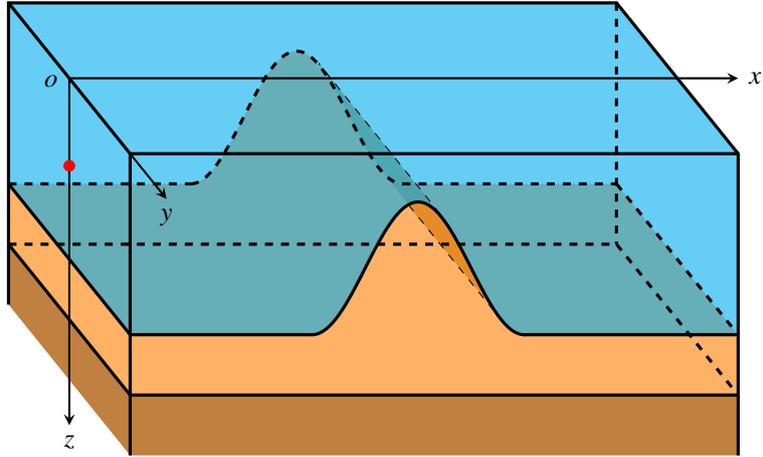
\begin{figure}[htbp]
\centering\includegraphics[height=8cm]{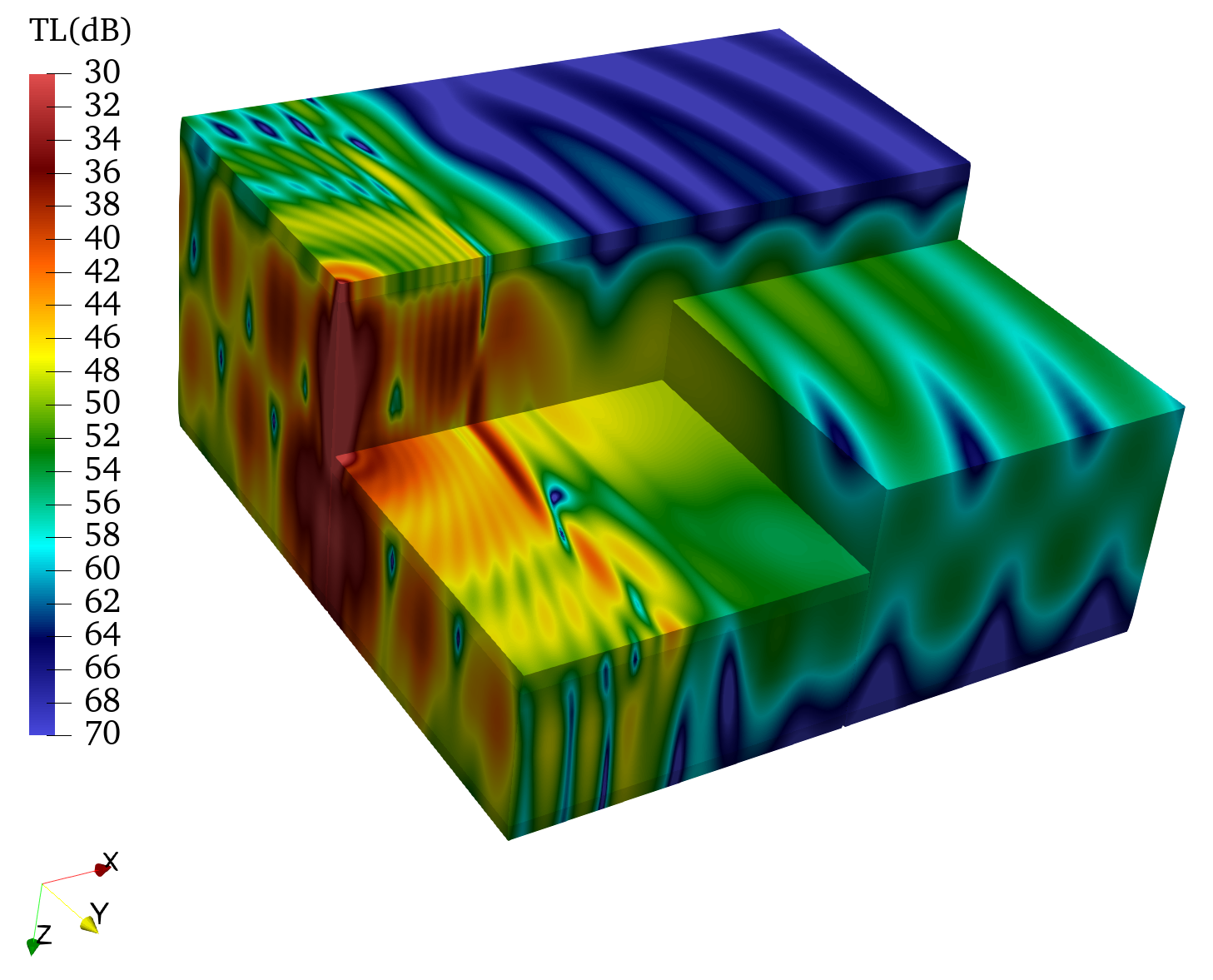}
\caption{Sound field of the three-dimensional ridge-shaped waveguide calculated by SPEC3D.}
\label{Figure18}
\end{figure}
\begin{figure}[htbp]
	\centering
	\label{Figure19}
\subfigure[]{\includegraphics[width=6.5cm]{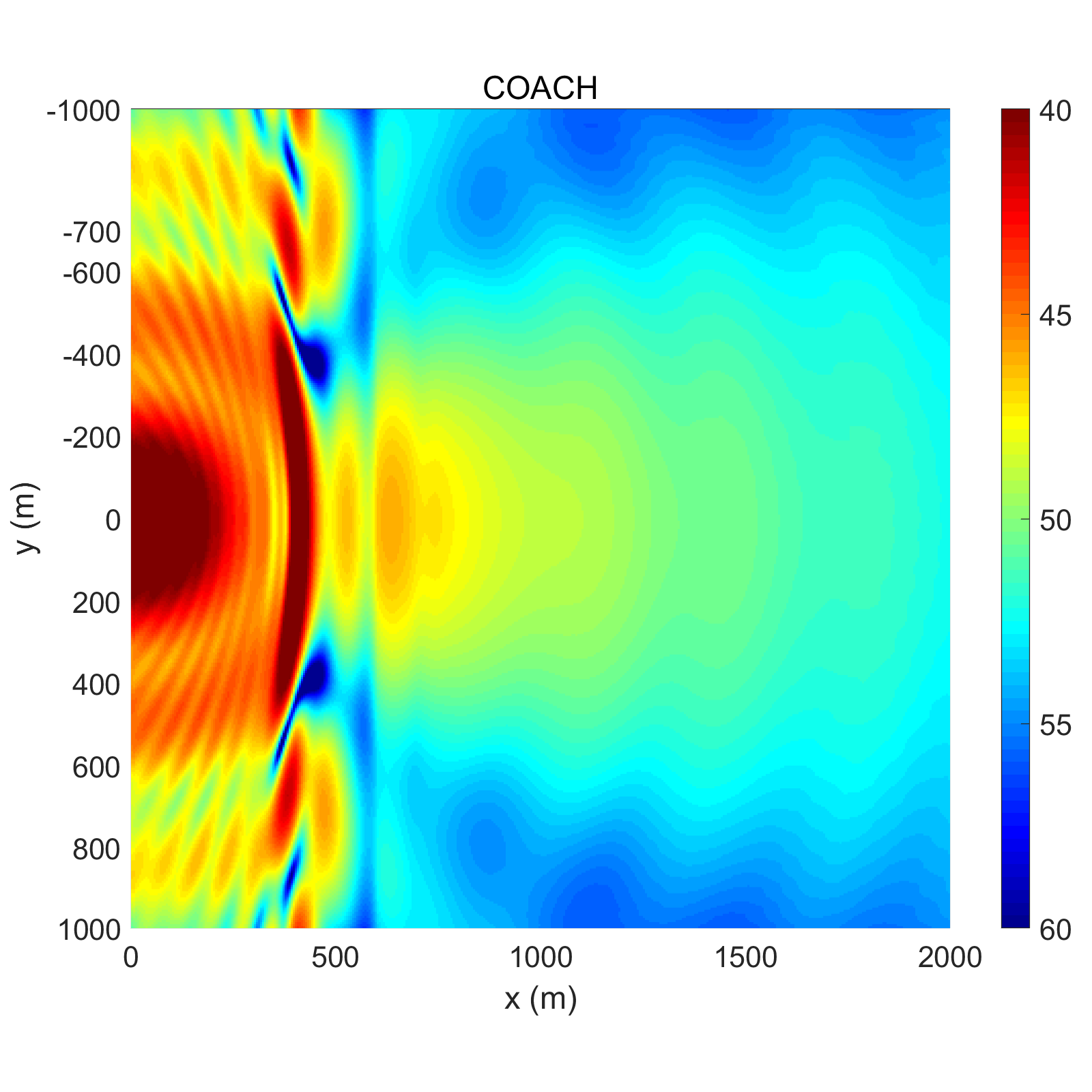}\label{Figure19a}}
\subfigure[]{\includegraphics[width=6.5cm]{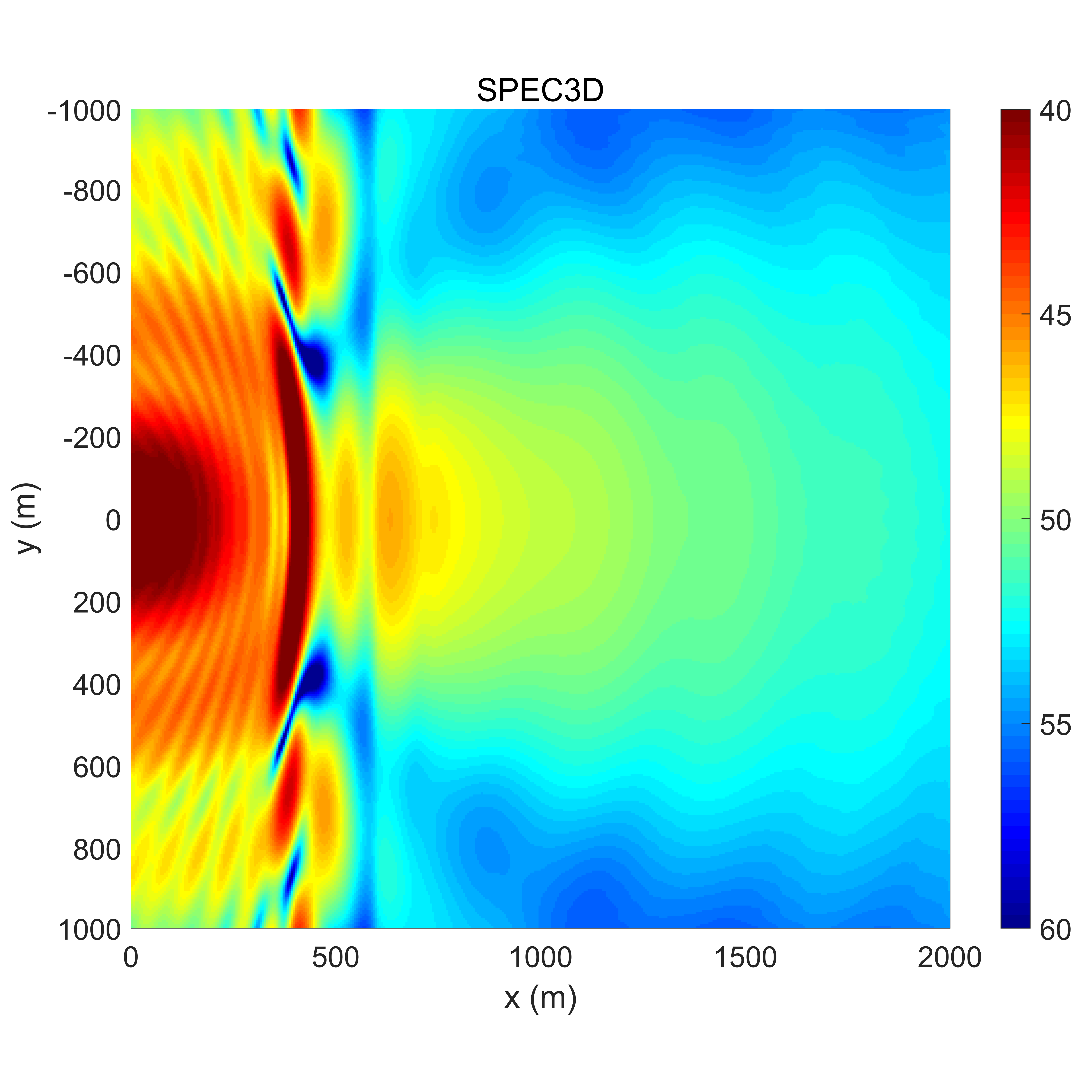}\label{Figure19b}}
\subfigure[]{\includegraphics[width=13cm]{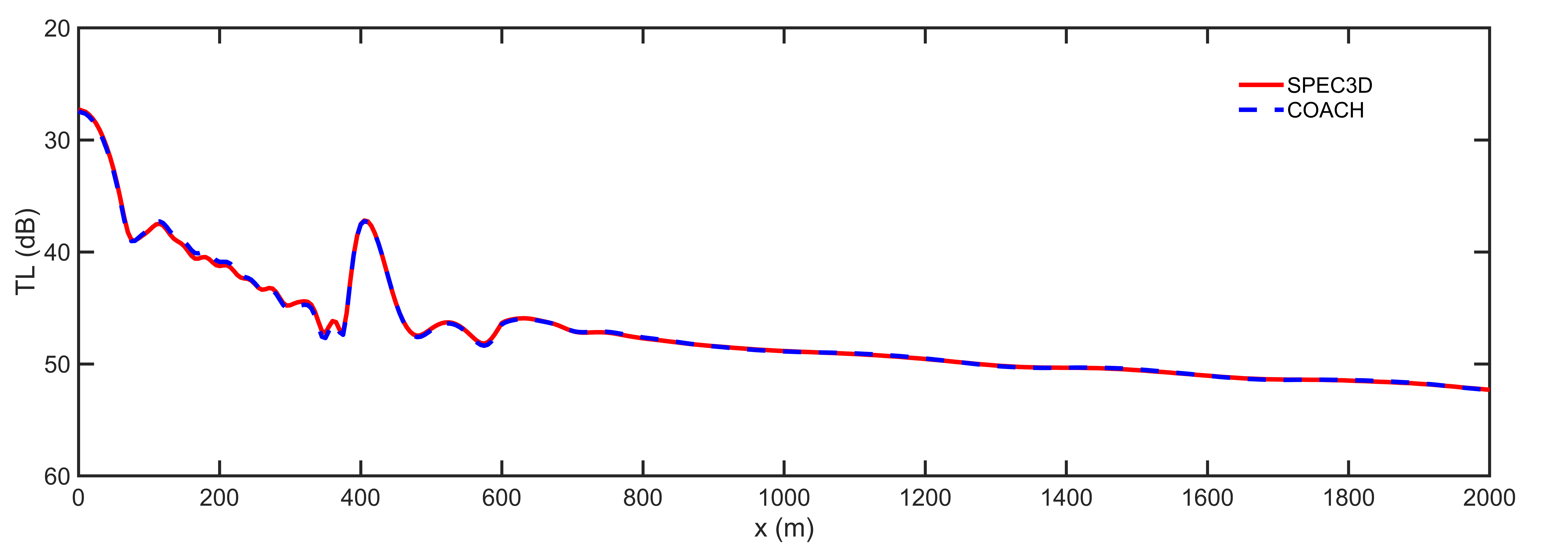}\label{Figure19c}}
\caption{Sound fields of the three-dimensional ridge-shaped waveguide calculated by COACH (a) and SPEC3D (b) at $z=50$ m; TLs along the $x$-direction at a depth of $z=50$ m (c).}
\end{figure}

For the cases where there is no analytical solution, we use the results of the benchmark program `COACH' developed by Liu et al. as a reference \cite{Liuw2021}. COACH is a three-dimensional finite difference program with fourth-order accuracy developed for the ocean acoustic Helmholtz equation, which can be used to address arbitrary bathymetry and provide more accurate benchmark solutions for other three-dimensional underwater acoustic approximation models. The derivatives in the acoustic Helmholtz equation are numerically discretized based on regular grids, and a perfectly matched layer is introduced to absorb unphysical reflections from the boundaries where Sommerfeld radiation conditions are deployed. Here, running COACH is equivalent to directly numerically solving Eq.~\eqref{eq.1}.

Figs.~\ref{Figure19a} and \ref{Figure19b} show the sound field slices calculated by COACH and SPEC3D at $z=50$ m, respectively. The results of SPEC3D and COACH are in good agreement, and even in a finer comparison of the TL curves in Fig.~\ref{Figure19c}, the errors of both are satisfactory. In addition, the backscattering and forward scattering generated by the uplifted topography of the ridge show how the ridge perturbs the near-field and far-field sound fields.

\subsection{Trench-shaped waveguide}
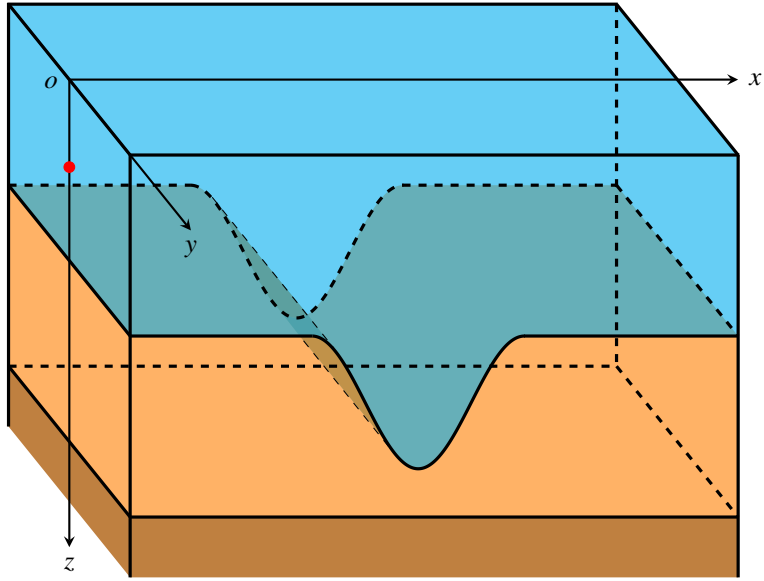
\begin{figure}[htbp]
	\centering
\begin{tikzpicture}[node distance=2cm,scale=0.8,samples=500,domain=7:10.5]
		\fill[cyan,opacity=0.6]plot[domain=7.35:8.92,smooth](\x-2,{-4.1+1.1*cos(103*(\x))})--cycle;
		\fill[orange,opacity=0.6]plot[domain=8.25:8.92,smooth](\x-2,{-4.1+1.1*cos(103*(\x))})--plot[domain=7.35:7,smooth](\x,{-6.6+1.1*cos(103*(\x))})--(6.79,-5.5)--cycle;
		\fill[cyan,opacity=0.6]plot[domain=8.25:8.92,smooth](\x-2,{-4.1+1.1*cos(103*(\x))})--plot[domain=7.35:7,smooth](\x,{-6.6+1.1*cos(103*(\x))})--(6.79,-5.5)--cycle;
		\fill[orange,opacity=0.6](2,-3)--plot[domain=7:10.5,smooth](\x-2,{-4.1+1.1*cos(103*(\x))})--(12,-3)--(14,-5.5)--plot[domain=10.5:7,smooth](\x,{-6.6+1.1*cos(103*(\x))})--(4,-5.5)--cycle;
		\fill[orange,opacity=0.6]plot[domain=7.35:8.92,smooth](\x-2,{-4.1+1.1*cos(103*(\x))})--cycle;
		\fill[orange,opacity=0.6]plot[domain=7.35:8.92,smooth](\x-2,{-4.1+1.1*cos(103*(\x))})--cycle;
		\fill[cyan,opacity=0.6] (2,0)--(12,0)--(14,-2.5)--(4,-2.5)--(4,-5.5)--(2,-3)--cycle;
		\fill[cyan,opacity=0.6]  (4,-2.5)--(4,-5.5)--(7,-5.5)--plot(\x,{-6.6+1.1*cos(103*(\x))})--(10.5,-5.5)--(14,-5.5)--(14,-2.5)-- cycle;
		\fill[orange,opacity=0.6] (2,-3)--(4,-5.5)--(4,-8.5)--(2,-6)--cycle;
		\fill[orange,opacity=0.6]  (4,-5.5)--(7,-5.5)--plot(\x,{-6.6+1.1*cos(103*(\x))})--(10.5,-5.5)--(14,-5.5)--(14,-8.5)--(4,-8.5)--cycle;
		\fill[cyan,opacity=0.6](6.79,-5.5)--plot[domain=7:8.25,smooth](\x,{-6.6+1.1*cos(103*(\x))})--cycle;
		\fill[orange,opacity=0.6](6.79,-5.5)--plot[domain=7:8.25,smooth](\x,{-6.6+1.1*cos(103*(\x))})--cycle;
		\fill[brown] (2,-6)--(4,-8.5)--(14,-8.5)--(14,-9.5)--(4,-9.5)--(2,-7)--cycle;
		\draw[dashed](8.25,-7.33)--(6.79,-5.5);
		\draw[dashed](5.35,-3.2)--(7.35,-5.7);
		\draw[thick, ->](3,-1.25)--(14,-1.25) node[right]{$x$};
		\draw[thick, ->](3,-1.25)--(3,-9) node[below]{$z$};	    		
		\draw[very thick](1.98,0)--(12.02,0);
		\draw[very thick](4,-2.5)--(14.02,-2.5);
		\draw[very thick](2,0)--(4,-2.5);
		\draw[thick, ->](2,0)--(5,-3.75) node[below]{$y$};	
		\draw[very thick](12,0)--(14,-2.5);
		\draw[very thick](2,0.02)--(2,-7);
		\draw[very thick](4,-2.5)--(4,-9.5);
		\draw[dashed, very thick](12,0)--(12,-6);
		\draw[very thick](14,-2.5)--(14,-9.5);
		\draw[dashed, very thick](12,-3)--(14,-5.5);
		\draw[dashed, very thick](2,-6)--(12,-6);
		\draw[dashed, very thick](12,-6)--(14,-8.5);
		
		\draw[very thick](2,-6)--(4,-8.5);	
		\draw[very thick](4,-8.5)--(14,-8.5);			
		\filldraw [red] (3,-2.7) circle [radius=2.5pt];
		\node at (2.7,-1.3){$o$};
		
		\draw[very thick](2,-3)--(4,-5.5);
		\draw[very thick](4,-5.5)--(7,-5.5);
		\draw[very thick, smooth] plot(\x,{-6.6+1.1*cos(103*(\x))});
		\draw[very thick](10.48,-5.5)--(14,-5.5);
		\draw[dashed, very thick](2,-3)--(5,-3);
		\draw[dashed,very thick, smooth] plot(\x-2,{-4.1+1.1*cos(103*(\x))});
		\draw[dashed, very thick](8.48,-3)--(12,-3);
		
	\end{tikzpicture}
\caption{Schematic diagram of the three-dimensional trench-shaped waveguide.}
	\label{Trench}
\end{figure}
\begin{figure}[htbp]
	\centering\includegraphics[height=8cm]{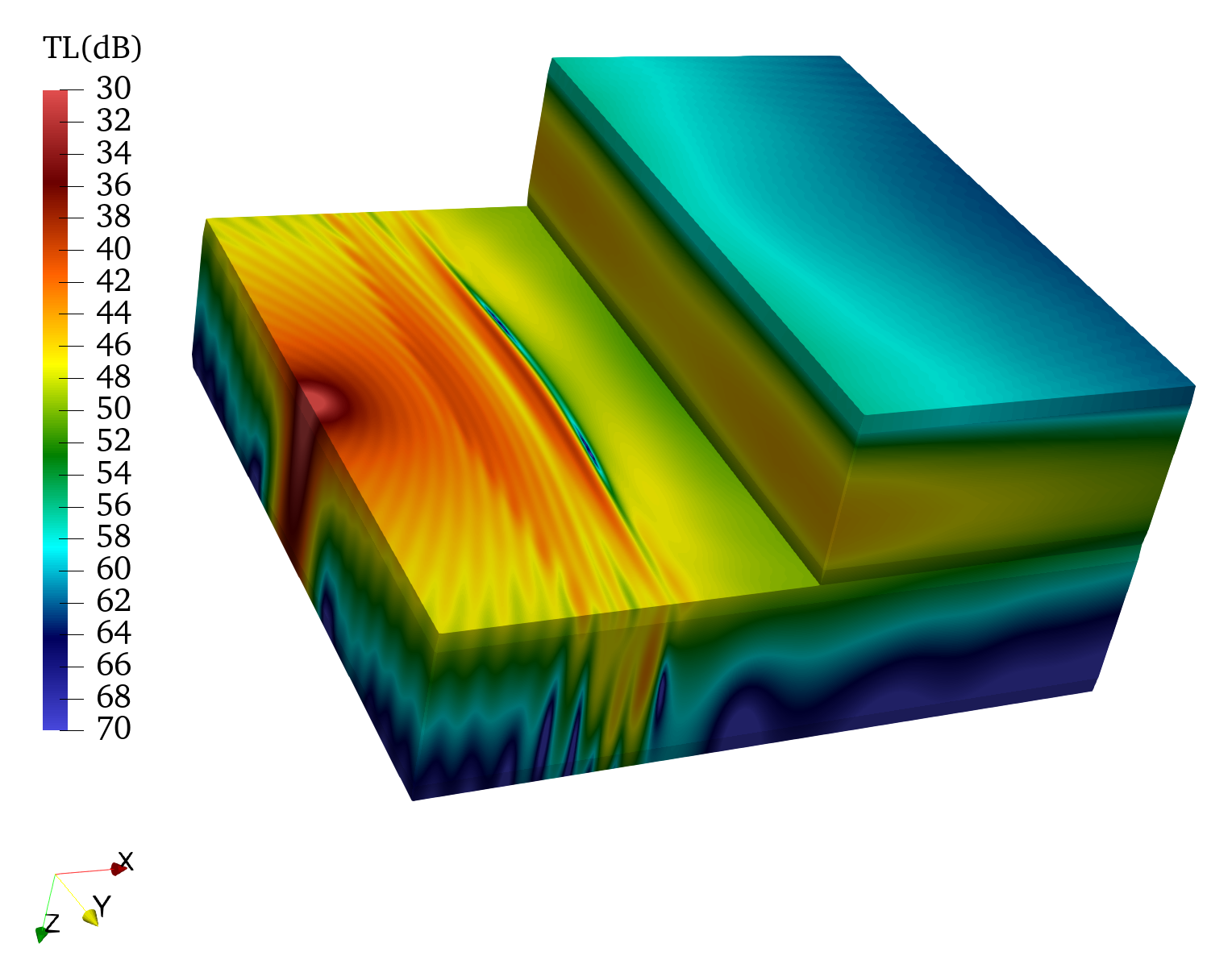}
\caption{Sound field of the three-dimensional trench-shaped waveguide calculated by SPEC3D.}
	\label{Figure21}
\end{figure}
\begin{figure}[htbp]
	\centering
\subfigure[]{\includegraphics[width=6.5cm]{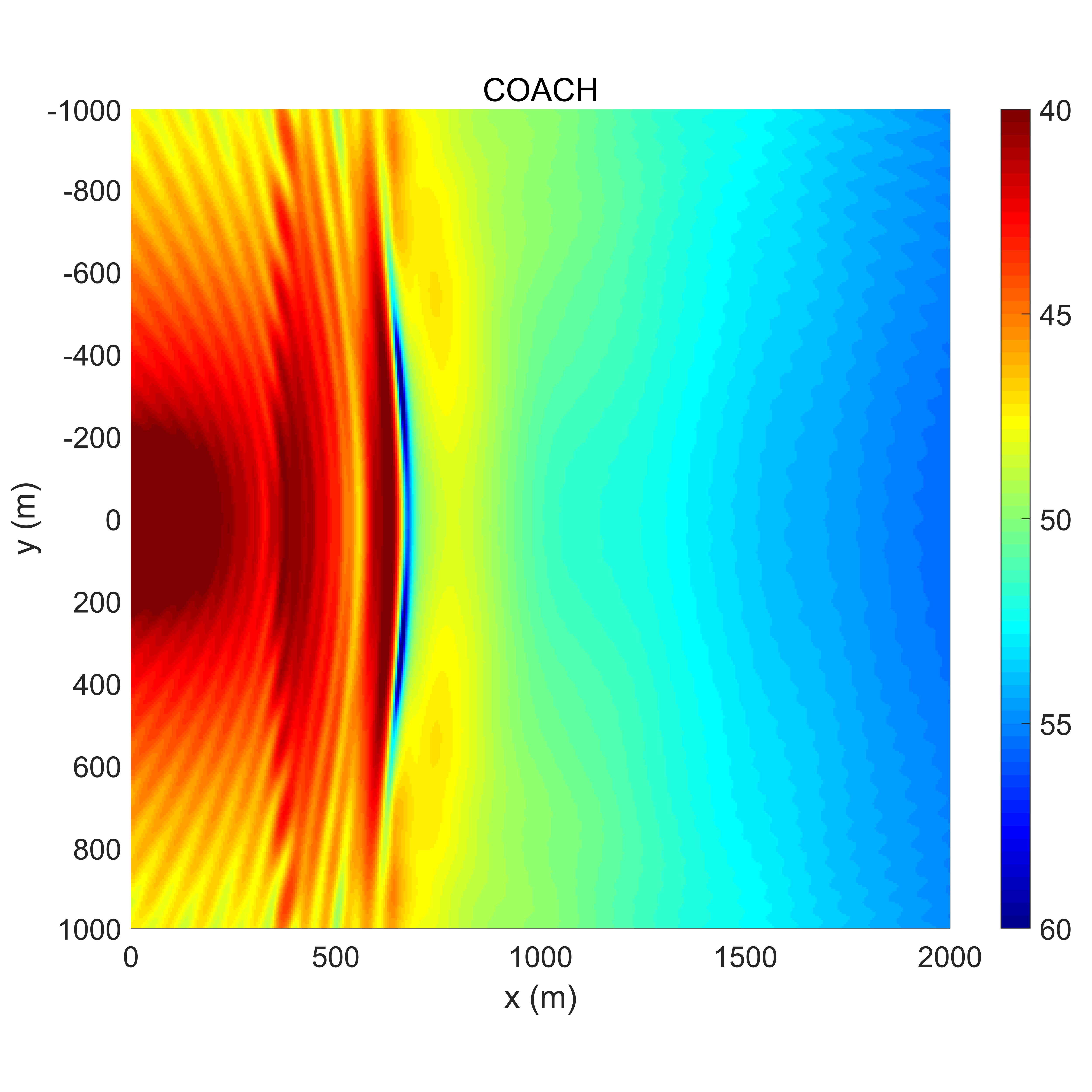}\label{Figure22a}}
\subfigure[]{\includegraphics[width=6.5cm]{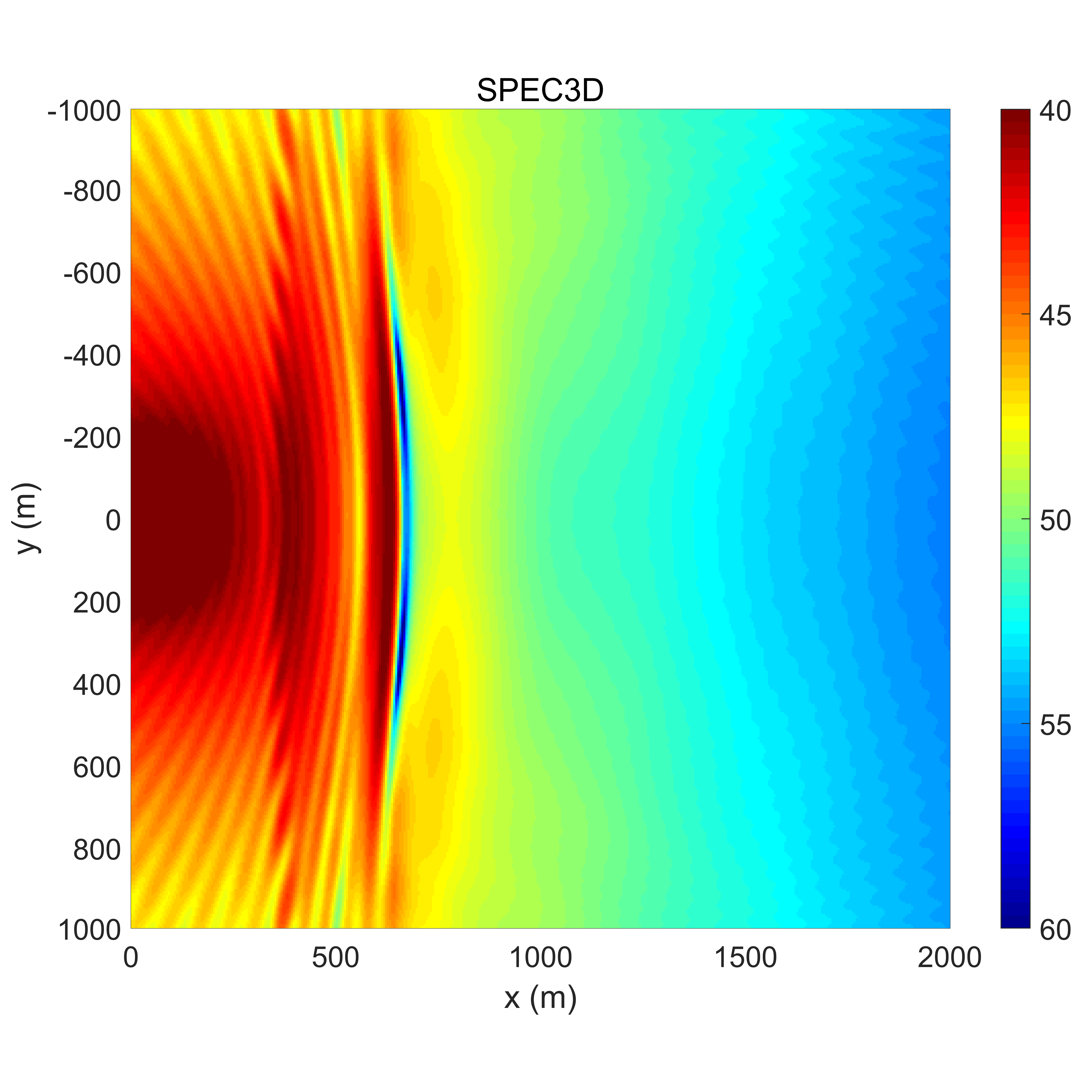}\label{Figure22b}}
\subfigure[]{\includegraphics[width=13cm]{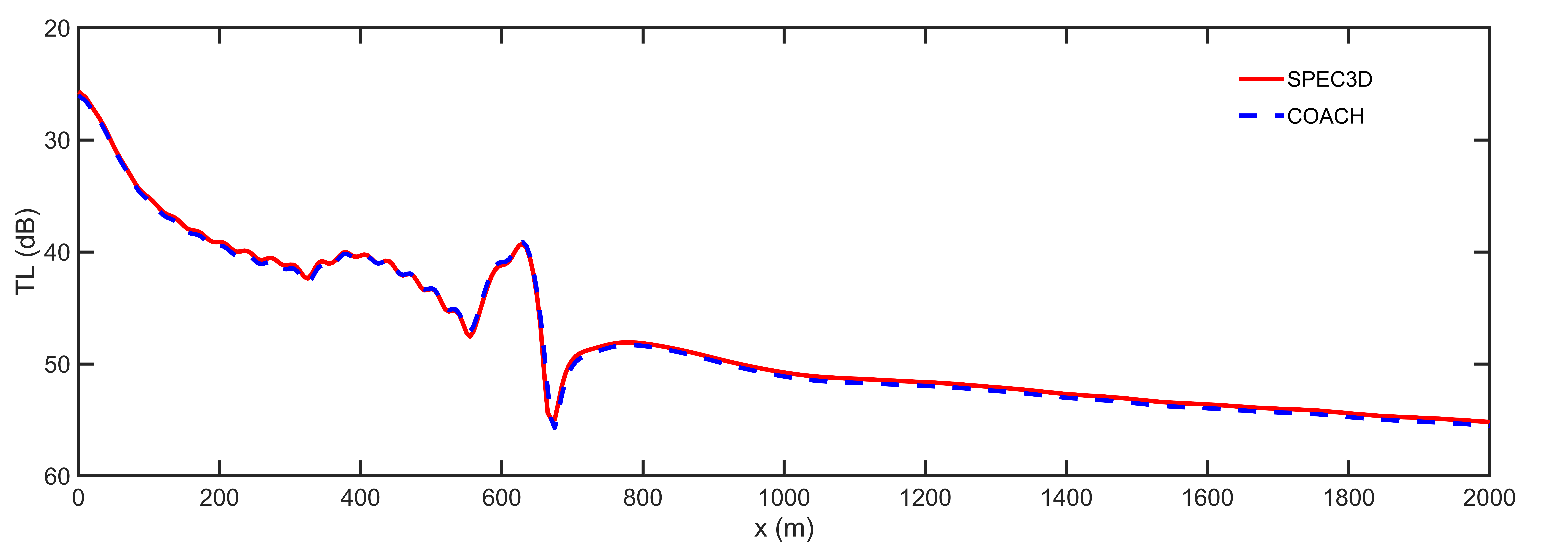}\label{Figure22c}}
\caption{Sound fields of the three-dimensional trench-shaped waveguide calculated by COACH (a) and SPEC3D (b) at $z=50$ m; TLs along the $x$-direction at a depth of $z=50$ m (c).}
\end{figure}

In contrast to the previous example, Fig.~\ref{Trench} shows a marine environment with a trench-shaped waveguide. The seafloor topography is given by:
\begin{equation}
	h(x)= \begin{cases}65+25 \cos \left[\frac{\pi(x-500)}{200}\right], & 300<x<700, \\ 40, & \text { elsewhere, }\end{cases}
\end{equation}
The remainder of the configuration is exactly the same as the above example. Fig.~\ref{Figure21} shows the three-dimensional sound field structure of the waveguide calculated by SPEC3D. The horizontal independence of both the near field and the far field makes the sound field retain a certain symmetry. The occurrence of refraction in the trench causes the sound field to exhibit a notable three-dimensional effect, which further affects the near and far fields through forward scattering and backscattering.

Likewise, Figs.~\ref{Figure22a} and \ref{Figure22b} plot sound field slices calculated by COACH and SPEC3D at $z=50$ m, respectively. Fig.~\ref{Figure22c} displays the TL curves calculated by the two programs along the $x$ direction at a depth of $z=50$ m. In both slice and line plots, the results of SPEC3D are very similar to those of COACH. Therefore, we conclude that the algorithm proposed in this paper is reliable.

\section{Discussion, Remarks and Conclusion}
\subsection{Discussion}
\begin{enumerate}
\item
The numerical experiments in the previous section verify the accuracy of the algorithm proposed in this paper and demonstrate the ability of SPEC3D to address quasi-three-dimensional underwater acoustic propagation problems. Here, we give the running times of SPEC3D and COACH at roughly the same accuracy. The results of SPEC3D and COACH are shown in Table~\ref{tab1}. Since COACH is a parallel program, both programs are run on the Tianhe-2 supercomputer \cite{Top500} during the timing period, both programs use the same compiler and nodes with the same hardware configuration, and the running time is measured in core$\cdot$hour.
\begin{table}[htbp]
		\centering
		\caption{\label{tab1} Running times of the numerical examples (unit: core$\cdot$hour).}
		\begin{tabular}{lrl}
			\hline
			Example & SPEC3D &COACH \\
			\hline
			free wedge     &1.19   &$>$200.00  \\
			rigid wedge    &1.20   &$>$300.00  \\
			penetrable wedge &44.62 &$>$5000.00\\
			penetrable ridge   &0.06   &$>$3.00    \\
			penetrable trench  &0.06   &$>$3.00    \\
			\hline
		\end{tabular}
\end{table}
In terms of computational speed, SPEC3D is much faster than the full three-dimensional model within approximately similar accuracy---almost two orders of magnitude faster. Therefore, SPEC3D can also be said to trade off efficiency by sacrificing generality.
\item
According to the above derivation and analysis, we can obtain a quasi-three-dimensional sound field through an inverse Fourier transform as long as we solve the sound field in Fig.~\ref{Figure2} corresponding to different $k_y$. Note that $k_y$ is a set of complex constants and that the marine environment parameters in Fig.~\ref{Figure2} do not change for different $k_y$, which means that the stepwise segmentation of the $x$-dependences in Fig.~\ref{Figure2} also does not change. Let us re-examine Eq.~\eqref{eq.5}, which becomes the matrix eigenvalue problem in Eq.~\eqref{eq.28} after spectral discretization. Eq.~\eqref{eq.28} is equivalent to the following matrix eigenvalue problem:
\begin{equation}
	\label{eq.47}
	\left(\frac{4}{\vert\Delta h\vert^2}\mathbf{C}_{\rho}\mathbf{D}_{N}\mathbf{C}_{1/\rho}\mathbf{D}_{N}+\mathbf{C}_{k^2}\right)\bm{\hat{\Psi}}= (k_y^2+\mu^2) \bm{\hat{\Psi}}
\end{equation}
Notably, the above indicates that the sum of squares of $k_y$ and $\mu$ is always equal to the eigenvalue in the above equation and that the eigenvectors do not change with $k_y$. Therefore, Eq.~\eqref{eq.47} needs to be solved only once after the derivation of Sec.~\ref{section3.2} to obtain the eigenvalues of the $\mu$ values corresponding to different $k_y$. Since the eigenvectors are invariant, the coupling submatrix in Eq.~\eqref{eq.15} is also shared for different $k_y$.

However, in fact, the above conjecture includes a notable misunderstanding; that is, for different $k_y$, it is not Eq.~\eqref{eq.28} that needs to be solved but Eq.~\eqref{eq.32} or \eqref{eq.35} (boundary conditions are imposed). The boundary and interface conditions in Eqs.~\eqref{eq.32} and \eqref{eq.35} do not change with $k_y$. Therefore, in the $k_y$--domain, $N_q$ sound fields still need to be calculated strictly according to the derivation in Secs.~\ref{section2.2} and \ref{section3.2}.
\end{enumerate}

\subsection{Remarks}
From the above analysis, we can directly summarize the following advantages of the algorithm and program developed in this article:
\begin{enumerate}

\item
The two-dimensional problem formed after implementing the Fourier transform can be solved in parallel because the evaluation of $\tilde{p}(x,k_y(q),z)$ is independent. This computational efficiency makes this algorithm more efficient than similar algorithms.

\item
Furthermore, the two-dimensional problem formed after implementing the Fourier transform can be solved efficiently by using the global matrix method. This model is considered to involve two-way coupled normal modes, and its accuracy is higher than that of a model with only one-way coupled modes.

\item
The Chebyshev--Tau spectral method used for the single-segment solution can accurately simulate free, rigid and half-space seafloor boundaries, especially for a half-space, and the eigenvalue transformation technique used avoids the problem of missing roots in traditional iterative root-finding programs.

\end{enumerate}

\subsection{Conclusion}
In this paper, we develop an algorithm that can solve for quasi-three-dimensional sound fields. The results of numerical simulations verify the reliability and efficiency of the model and code. In the algorithm, the Fourier transform is used to transform the equation governing quasi-three-dimensional acoustic propagation from the $y$--domain to the $k_y$--domain. After the sound pressure at each discrete $k_y$ is obtained, the sound pressure in the $y$--domain can be synthesized by applying an inverse Fourier transformation; this process can also be regarded as $k_y$ wavenumber integration to some extent. The governing equation in the $k_y$--domain adopts a similar form to that of a two-dimensional line source sound field. The stair-step approximation strategy is used to accommodate the $x$-dependence of the two-dimensional sound field, and the coupling coefficients are assembled by means of a global matrix, a favorable sparse band matrix that can be solved efficiently. In each $x$-independent segment, the Chebyshev--Tau spectral method is used to solve for the eigenpairs, and the proposed method achieves good accuracy, speed and robustness.

In terms of its application scope, this algorithm requires that the marine environment be independent in the $y$ direction, which limits the practicality of SPEC3D to a certain extent. However, for marine environments such as continental slopes and continental shelves, $y$-independence provides an appropriate approximation, and SPEC3D greatly reduces computational costs compared to full three-dimensional models that require the use of more sophisticated techniques \cite{Liuw2021,Ivansson2021}.

\section*{Acknowledgments}
The authors thank Prof. Wenyu Luo from the Institute of Acoustics, Chinese Academy of Sciences, for his valuable guidance on the analytical solution of the sound field for the wedge-shaped waveguide.

This work was supported by the National Natural Science Foundation of China [grant number 61972406] and the National Key Research and Development Program of China [grant number 2016YFC1401800].

\section*{Appendix}
Here, we take Eq.~\eqref{eq.10c} as an example to deduce the integral change process in Eqs.~\eqref{eq.10c} and \eqref{eq.13c}.
\begin{equation}
		\tilde{c}_{\ell m}=\int_0^\infty \frac{\Psi_{\ell}^{j+1}(z) \Psi_{m}^{j}(z)}{\rho_{j+1}(z)} \mathrm{d} z=\int_0^H \frac{\Psi_{\ell}^{j+1}(z)\Psi_{m}^{j}(z)}{\rho_{j+1}(z)} \mathrm{d} z+\int_H^\infty \frac{\Psi_{\ell}^{j+1}(z)\Psi_{m}^{j}(z)}{\rho_{j+1}(z)} \mathrm{d} z
\end{equation}
At the bottom of the semi-infinite space, the modes decay exponentially with increasing depth as follows:
\begin{equation}
	\Psi(z) = \Psi(H)\exp\left[-\gamma(z-H)\right],\quad z \ge H
\end{equation}
Therefore,
\begin{equation}
	\begin{aligned}
		\int_H^\infty \frac{\Psi_{\ell}^{j+1}(z)\Psi_{m}^{j}(z)}{\rho_{j+1}(z)} \mathrm{d} z &=\frac{1}{\rho_\infty} \int_H^\infty\Psi_{\ell}^{j+1}(z)\Psi_{m}^{j}(z) \mathrm{d} z\\
		&=\frac{1}{\rho_\infty}\int_H^\infty\Psi_{\ell}^{j+1}(H)\Psi_{m}^j(H)\exp\left[-(\gamma_\ell^{j+1}+\gamma_m^j)(z-H)\right]\\
		&=\frac{\Psi_{\ell}^{j+1}(H)\Psi_{m}^j(H)}{\rho_\infty}\cdot\frac{1}{-(\gamma_\ell^{j+1}+\gamma_m^j)}\left.\exp\left[-(\gamma_\ell^{j+1}+\gamma_m^j)(z-H)\right]\right|_H^\infty\\
	    &=\frac{\Psi_{\ell}^{j+1}(H) \Psi_{m}^{j}(H)}{\rho_\infty \left(\gamma_\ell^{j+1}+\gamma_m^j \right)}
	\end{aligned}
\end{equation}
Eq.~\eqref{eq.7} presents a simplified case ($j=j+1$ and $\gamma_\ell=\gamma_m=\gamma_{\infty}$) in the range-independent case.

\bibliographystyle{elsarticle-num}

\begin{thebibliography}{10}
\expandafter\ifx\csname url\endcsname\relax
  \def\url#1{\texttt{#1}}\fi
\expandafter\ifx\csname urlprefix\endcsname\relax\def\urlprefix{URL }\fi
\expandafter\ifx\csname href\endcsname\relax
  \def\href#1#2{#2} \def\path#1{#1}\fi

\bibitem{Etter2018}
P.~C. Etter, Underwater Acoustic Modeling and Simulation, CRC Press, Boca
  Raton, USA, 2018.
\newblock \href {https://doi.org/10.1201/9781315166346}
  {\path{doi:10.1201/9781315166346}}.

\bibitem{Jensen2011}
F.~B. Jensen, W.~A. Kuperman, M.~B. Porter, H.~Schmidt, Computational Ocean
  Acoustics, Springer-Verlag, New York, 2011.
\newblock \href {https://doi.org/10.1007/978-1-4419-8678-8}
  {\path{doi:10.1007/978-1-4419-8678-8}}.

\bibitem{Lee1995}
D.~Lee, M.~H. Schultz, W.~L. Siegmann, Numerical Ocean Acoustic Propagation in
  Three Dimensions, World Scientific, 1995.
\newblock \href {https://doi.org/10.1142/2789} {\path{doi:10.1142/2789}}.

\bibitem{Linyt2019}
Y.~T. Lin, M.~B. Porter, F.~Sturm, M.~J. Isakson, C.~S. Chiu, Introduction to
  the special issue on three-dimensional underwater acoustics, The Journal of
  the Acoustical Society of America 146~(3) (2019) 1855--1857.
\newblock \href {https://doi.org/10.1121/1.5126013}
  {\path{doi:10.1121/1.5126013}}.

\bibitem{Ivansson2021}
S.~M. Ivansson, Coupled-mode field computations for media with locally reacting
  irregular boundaries, The Journal of the Acoustical Society of America
  150~(4) (2021) 2985--2998.
\newblock \href {https://doi.org/10.1121/10.0006743}
  {\path{doi:10.1121/10.0006743}}.

\bibitem{Liuw2021}
W.~Liu, L.~Zhang, W.~Wang, Y.~Wang, S.~Ma, X.~Cheng, W.~Xiao, A
  three-dimensional finite difference model for ocean acoustic propagation and
  benchmarking for topographic effects, The Journal of the Acoustical Society
  of America 150~(2) (2021) 1140--1156.
\newblock \href {https://doi.org/10.1121/10.0005853}
  {\path{doi:10.1121/10.0005853}}.

\bibitem{Fawcett1990}
J.~A. Fawcett, T.~W. Dawson, Fourier synthesis of three-dimensional scattering
  in a two-dimensional oceanic waveguide using boundary integral equation
  methods, The Journal of the Acoustical Society of America 88~(4) (1990)
  1913--1920.
\newblock \href {https://doi.org/10.1121/1.400214}
  {\path{doi:10.1121/1.400214}}.

\bibitem{Evans1983}
R.~B. Evans, A coupled mode solution for acoustic propagation in a waveguide
  with stepwise depth variations of a penetrable bottom, The Journal of the
  Acoustical Society of America 74 (1983) 188--195.
\newblock \href {https://doi.org/10.1121/1.389707}
  {\path{doi:10.1121/1.389707}}.

\bibitem{Evans1986}
R.~B. Evans, The decoupling of stepwise coupled modes, The Journal of the
  Acoustical Society of America 80 (1986) 1414--1418.
\newblock \href {https://doi.org/10.1121/1.394395}
  {\path{doi:10.1121/1.394395}}.

\bibitem{Couple}
R.~B. Evans,
  \href{https://oalib-acoustics.org/models-and-software/normal-modes/}{{\href{https://oalib-acoustics.org/Modes/index.html}{COUPLE}}:
  A coupled normal-mode code {(Fortran)}} (2007).
\newline\urlprefix\url{https://oalib-acoustics.org/models-and-software/normal-modes/}

\bibitem{Boyd2001}
J.~P. Boyd, {Chebyshev} and {Fourier} Spectral Methods, Second Edition, Dover,
  New York, USA, 2001.

\bibitem{Jshen2011}
J.~Shen., T.~Tang., L.~Wang., Spectral Methods Algorithms, Analysis and
  Applications, Springer-Verlag, Berlin, German, 2011.
\newblock \href {https://doi.org/10.1007/978-3-540-71041-7}
  {\path{doi:10.1007/978-3-540-71041-7}}.

\bibitem{Canuto2006}
C.~Canuto, M.~Y. Hussaini, A.~Quarteroni, T.~A. Zang, Spectral Methods
  Fundamentals in Single Domains, Spring-Verlag, Berlin, German, 2006.
\newblock \href {https://doi.org/10.1007/978-3-540-30726-6}
  {\path{doi:10.1007/978-3-540-30726-6}}.

\bibitem{Orszag1972}
S.~A. Orszag, Comparison of pseudospectral and spectral approximation, Studies
  in Applied Mathematics L1~(3) (1972) 253--259.
\newblock \href {https://doi.org/10.1002/sapm1972513253}
  {\path{doi:10.1002/sapm1972513253}}.

\bibitem{Gottlieb1977}
D.~Gottlieb, S.~A. Orszag, Numerical Analysis of Spectral Methods, Theory and
  Applications, Society for Industrial and Applied Mathematics, Philadelphia,
  USA, 1977.
\newblock \href {https://doi.org/10.1137/1.9781611970425}
  {\path{doi:10.1137/1.9781611970425}}.

\bibitem{Canuto1988}
C.~Canuto, M.~Y. Hussaini, A.~Quarteroni, T.~A. Zang, Spectral Methods in Fluid
  Dynamics, Spring-Verlag, Berlin, Germany, 1988.
\newblock \href {https://doi.org/10.1007/978-3-642-84108-8}
  {\path{doi:10.1007/978-3-642-84108-8}}.

\bibitem{Guoby1998}
B.~Guo, Spectral Methods and Their Applications, World Scientific, 1998.
\newblock \href {https://doi.org/10.1142/3662} {\path{doi:10.1142/3662}}.

\bibitem{Jekeli2011}
C.~Jekeli, Spectral Methods in Geodesy and Geophysics, CRC Press, Boca Raton,
  USA, 2017.
\newblock \href {https://doi.org/10.1201/9781315118659}
  {\path{doi:10.1201/9781315118659}}.

\bibitem{Dzieciuch1993}
M.~A. Dzieciuch, Numerical solution of the acoustic wave equation using
  {Chebyshev} polynomials with application to global acoustics, in: Proceedings
  of Oceans, IEEE, Victoria, BC, Canada, 1993, pp. 267--271.
\newblock \href {https://doi.org/10.1109/OCEANS.1993.326000}
  {\path{doi:10.1109/OCEANS.1993.326000}}.

\bibitem{aw}
M.~A. Dzieciuch,
  \href{https://oalib-acoustics.org/models-and-software/normal-modes/}{{\href{https://oalib-acoustics.org/Modes/index.html}{aw}}:
  A {Matlab} code for computing normal modes based on {Chebyshev}
  approximations} (1993).
\newline\urlprefix\url{https://oalib-acoustics.org/models-and-software/normal-modes/}

\bibitem{rimLG}
R.~B. Evans,
  \href{https://oalib-acoustics.org/models-and-software/normal-modes/}{{\href{https://oalib-acoustics.org/Modes/index.html}{rimLG}}:
  A {Legendre--Galerkin} technique for differential eigenvalue problems with
  complex and discontinuous coefficients, arising in underwater acoustics}
  (2020).
\newline\urlprefix\url{https://oalib-acoustics.org/models-and-software/normal-modes/}

\bibitem{Sabatini2019}
R.~Sabatini, P.~Cristini, A multi-domain collocation method for the accurate
  computation of normal modes in open oceanic and atmospheric waveguides, Acta
  Acustica United with Acustica 105 (2019) 464--474.
\newblock \href {https://doi.org/10.3813/AAA.919328}
  {\path{doi:10.3813/AAA.919328}}.

\bibitem{Tuhw2020a}
H.~Tu, Y.~Wang, W.~Liu, X.~Ma, W.~Xiao, Q.~Lan, A {Chebyshev} spectral method
  for normal mode and parabolic equation models in underwater acoustics,
  Mathematical Problems in Engineering (2020) 7461314\href
  {https://doi.org/10.1155/2020/7461314} {\path{doi:10.1155/2020/7461314}}.

\bibitem{Tuhw2021a}
H.~Tu, Y.~Wang, Q.~Lan, W.~Liu, W.~Xiao, S.~Ma, A {Chebyshev--Tau} spectral
  method for normal modes of underwater sound propagation with a layered marine
  environment, Journal of Sound and Vibration 492 (2021) 115784.
\newblock \href {https://doi.org/10.1016/j.jsv.2020.115784}
  {\path{doi:10.1016/j.jsv.2020.115784}}.

\bibitem{NM-CT}
H.~Tu,
  \href{https://oalib-acoustics.org/models-and-software/normal-modes/}{{\href{https://oalib-acoustics.org/Modes/index.html}{NM-CT}}:
  A {Chebyshev--Tau} spectral method for normal modes of underwater sound
  propagation with a layered marine environment in {Matlab} and {Fortran}}
  (2020).
\newline\urlprefix\url{https://oalib-acoustics.org/models-and-software/normal-modes/}

\bibitem{Tuhw2021c}
H.~Tu, Y.~Wang, Q.~Lan, W.~Liu, W.~Xiao, S.~Ma, Applying a {Legendre}
  collocation method based on domain decomposition to calculate underwater
  sound propagation in a horizontally stratified environment, Journal of Sound
  and Vibration 511 (2021) 116364.
\newblock \href {https://doi.org/10.1016/j.jsv.2021.116364}
  {\path{doi:10.1016/j.jsv.2021.116364}}.

\bibitem{MultiLC}
H.~Tu,
  \href{https://oalib-acoustics.org/models-and-software/normal-modes/}{{\href{https://oalib-acoustics.org/Modes/index.html}{MultiLC}}:
  A {Legendre} collocation method based on domain decomposition to calculate
  underwater sound propagation in a horizontally stratified environment in
  {Matlab} and {Fortran}} (2021).
\newline\urlprefix\url{https://oalib-acoustics.org/models-and-software/normal-modes/}

\bibitem{Tuhw2022b}
H.~Tu, Y.~Wang, C.~Yang, X.~Wang, S.~Ma, W.~Xiao, W.~Liu, A novel algorithm to
  solve for an underwater line source sound field based on coupled modes and a
  spectral method, Journal of Computational Physics 468 (2022) 111478.
\newblock \href {https://doi.org/10.1016/j.jcp.2022.111478}
  {\path{doi:10.1016/j.jcp.2022.111478}}.

\bibitem{Kraken2001}
M.~B. Porter,
  \href{https://oalib-acoustics.org/models-and-software/normal-modes/}{The
  Kraken Normal Mode Program}, {SACLANT} Undersea Research Centre, 2001.
\newline\urlprefix\url{https://oalib-acoustics.org/models-and-software/normal-modes/}

\bibitem{Lanczos1938}
C.~Lanczos, Trigonometric interpolation of empirical and analytical functions,
  Journal of Mathematical Physics 17 (1938) 123--199.

\bibitem{Min2005}
M.~S. Min, D.~Gottlieb, Domain decomposition spectral approximations for an
  eigenvalue problem with a piecewise constant coefficient, SIAM Journal on
  Numerical Analysis 43 (2005) 502--520.
\newblock \href {https://doi.org/10.1137/s0036142903423836}
  {\path{doi:10.1137/s0036142903423836}}.

\bibitem{LAPACK}
M.~B. Porter,
  \href{https://www.netlib.org/lapack/#_lapack_version_3_10_0}{{LAPACK}: Linear
  algebra package (version 3.10.0)} (2021).
\newline\urlprefix\url{https://www.netlib.org/lapack/#_lapack_version_3_10_0}

\bibitem{Luowy2016a}
W.~Luo, X.~Yu, X.~Yang, Z.~Zhang, R.~Zhang, A three-dimensional coupled-mode
  model for the acoustic field in a two-dimensional waveguide with perfectly
  reflecting boundaries, Chinese Physics B 25~(12) (2016) 124309.
\newblock \href {https://doi.org/10.1088/1674-1056/25/12/124309}
  {\path{doi:10.1088/1674-1056/25/12/124309}}.

\bibitem{Luowy2015}
W.~Luo, R.~Zhang, Exact solution of three-dimensional acoustic field in a wedge
  with perfectly reflecting boundaries, {SCIENCE CHINA} Physics, Mechanics \&
  Astronomy 58~(9) (2015) 594301.
\newblock \href {https://doi.org/10.1007/s11433-015-5691-6}
  {\path{doi:10.1007/s11433-015-5691-6}}.

\bibitem{Jensen1998}
F.~B. Jensen, On the use of stair steps to approximate bathymetry changes in
  ocean acoustic models, The Journal of the Acoustical Society of America
  104~(3) (1998) 1310--1315.
\newblock \href {https://doi.org/10.1121/1.424340}
  {\path{doi:10.1121/1.424340}}.

\bibitem{Tuhw2022d}
H.~Tu, Y.~Wang, W.~Liu, S.~Ma, X.~Wang, W.~Xiao, A wavenumber integration model
  of underwater acoustic propagation in arbitrary horizontally stratified media
  based on a spectral method, arXiv.org (2022).
\newblock \href {https://doi.org/arXiv:2206.00312}
  {\path{doi:arXiv:2206.00312}}.

\bibitem{Buckingham1984}
M.~J. Buckingham, Acoustic propagation in a wedge-shaped ocean with perfectly
  reflecting boundaries, Tech. Rep. 8793, Naval Research Laboratory, USA
  (1984).

\bibitem{Buckingham1987}
M.~J. Buckingham, Theory of three-dimensional acoustic propagation in a
  wedgelike ocean with a penetrable bottom, The Journal of the Acoustical
  Society of America 82~(1) (1987) 198--210.
\newblock \href {https://doi.org/10.1121/1.395546}
  {\path{doi:10.1121/1.395546}}.

\bibitem{Doolittle1988}
R.~Doolittle, A.~Tolstoy, M.~J. Buckingham, Experimental confirmation of
  horizontal refraction of cw acoustic radiation from a point source in a
  wedge-shaped ocean environment, The Journal of the Acoustical Society of
  America 83~(6) (1988) 2117--2125.
\newblock \href {https://doi.org/10.1121/1.396392}
  {\path{doi:10.1121/1.396392}}.

\bibitem{Milton1972}
M.~Abramowitz, I.~A. Stegun, Handbook of Mathematical Functions with Formulas,
  Graphs, and Mathematical Tables, DC: National Bureau of Standards,
  Washington, USA, 1972.

\bibitem{Deane1993}
G.~B. Deane, M.~J. Buckingham, An analysis of the three‐dimensional sound
  field in a penetrable wedge with a stratified fluid or elastic basement, The
  Journal of the Acoustical Society of America 93~(3) (1993) 1319--1328.
\newblock \href {https://doi.org/10.1121/1.405417}
  {\path{doi:10.1121/1.405417}}.

\bibitem{Yangcm2015b}
C.~Yang, W.~Luo, F.~Qiao, L.~Lyu, Three-dimensional analytical solution for
  sound propagation in a homogeneous penetrable wedge, Journal of Nanjing
  University (Natural sciences) 51~(6) (2015) 1319--1328.
\newblock \href {https://doi.org/10.13232/j.cnki.jnju.2015.06.017}
  {\path{doi:10.13232/j.cnki.jnju.2015.06.017}}.

\bibitem{Top500}
\href{https://www.top500.org/lists/top500/2022/06/}{Top500} (June 2022).
\newline\urlprefix\url{https://www.top500.org/lists/top500/2022/06/}

\end{thebibliography}

\end{document}